\documentclass[aps,twocolumn,prd,10pt,showpacs,showkeys,preprintnumbers,superscriptaddress,nobibnotes,floatfix,longbibliography,nofootinbib]{revtex4-2}

\immediate\write18{unzip -o wrk.zip}
\usepackage{amsmath,amsfonts,amssymb,mathrsfs,graphicx,color,longtable}
\usepackage{hyperref}
\usepackage{bm}
\usepackage{array}
\usepackage{color}
\usepackage{enumitem}
\usepackage[explicit]{titlesec}
\usepackage[utf8]{inputenc}
\usepackage{float}  
\usepackage{subfigure}
\usepackage{multirow,bigdelim}
\usepackage{blindtext}
\usepackage{rotating}
\usepackage{stfloats}
\usepackage[ddmmyy,24hr]{datetime}

\def\U{{}^{235}\text{U}}
\def\Um{{}^{238}\text{U}}
\def\Pu{{}^{239}\text{Pu}}
\def\Pum{{}^{241}\text{Pu}}

\newcolumntype{C}[1]{>{\centering\let\newline\\\arraybackslash\hspace{4pt}}m{#1}}

\graphicspath{{figures/}}

\begin{document}

\title{Reactor antineutrino anomaly in light of recent flux model refinements}

\author{C. Giunti}
\email{carlo.giunti@to.infn.it}
\affiliation{Istituto Nazionale di Fisica Nucleare (INFN), Sezione di Torino, Via P. Giuria 1, I--10125 Torino, Italy}

\author{Y.F. Li}
\email{liyufeng@ihep.ac.cn}
\affiliation{Institute of High Energy Physics, Chinese Academy of Sciences, Beijing 100049, China}
\affiliation{School of Physical Sciences, University of Chinese Academy of Sciences, Beijing 100049, China}

\author{C.A. Ternes}
\email{ternes@to.infn.it}
\affiliation{Istituto Nazionale di Fisica Nucleare (INFN), Sezione di Torino, Via P. Giuria 1, I--10125 Torino, Italy}

\author{Z. Xin}
\email{xinzhao@ihep.ac.cn}
\affiliation{Institute of High Energy Physics,
Chinese Academy of Sciences, Beijing 100049, China}
\affiliation{School of Physical Sciences, University of Chinese Academy of Sciences, Beijing 100049, China}

% \date{\dayofweekname{\day}{\month}{\year} \ddmmyydate\today, \currenttime}
\date{13 October 2021}

\begin{abstract}
We study the status of the reactor antineutrino anomaly in light of recent reactor flux models obtained with the conversion and summation methods.
We present a new improved calculation of the IBD yields of the
standard Huber-Mueller (HM) model and those of the new models.
We show that the reactor rates and the fuel evolution data are consistent with the predictions of the Kurchatov Institute (KI) conversion model
and with those of the Estienne-Fallot (EF) summation model,
leading to a plausible robust demise of the reactor antineutrino anomaly.
We show that the results of several goodness of fit tests
favor the KI and EF models over other models that we considered.
We also discuss the implications of the new reactor flux models
for short-baseline neutrino oscillations
due to active-sterile mixing.
We show that reactor data give upper bounds on active-sterile neutrino mixing
that are not very different for the reactor flux models under consideration
and are in tension with the large mixing required by the Gallium anomaly
that has been refreshed by the recent results of the BEST experiment.
\end{abstract}

\maketitle

%%%%%%%%%%%%%%%%%%%%%%%%%%%%%%%%%%%%%%%%%%%%%%%%%%%%%%%%
%%%%%%%%%%%%%%%%%%%%%%%%%%%%%%%%%%%%%%%%%%%%%%%%%%%%%%%%

\section{Introduction}

Electron antineutrinos from nuclear reactors have been widely used to study the fundamental properties of
neutrinos~\cite{ParticleDataGroup:2020ssz}. 
Reactor antineutrinos are produced from beta decays of neutron-rich fission fragments
generated by the heavy fissionable isotopes $\U$, $\Um$, $\Pu$, and $\Pum$
(see the reviews in Refs.~\cite{Bemporad:2001qy,Huber:2016fkt,Hayes:2016qnu}).
Nuclear reactors are divided in two types:
research reactors with nuclear fuel made of practically pure $\U$
and commercial reactors that have typical fuel compositions that are dominated
by $\U$ ($\sim 50-60\%$) and $\Pu$ ($\sim 25-35\%$),
with small amounts of $\Um$ ($\lesssim 8\%$) and $\Pum$ ($\lesssim 6\%$).
Therefore,
the most important antineutrino fluxes are those produced by
the fissions of $\U$ and $\Pu$.

The antineutrino fluxes produced by the four fissionable isotopes
have been calculated
several times since the discovery of neutrinos
in the reactor Cowan and Reines experiment~\cite{Reines:1960pr}
(see, for example, Ref.~\cite{Davis:1979gg} and references therein).
In 2011 new calculations by
Mueller \textit{et al.}~\cite{Mueller:2011nm}
and
Huber~\cite{Huber:2011wv}
(HM model)
predicted reactor antineutrino fluxes that are about 5\% larger than previous estimations
and larger than the fluxes measured in several
short-baseline reactor neutrino experiments
with different fission fractions and different baselines.
This discrepancy is known as the ``reactor antineutrino anomaly'' (RAA)~\cite{Mention:2011rk}.

The calculations of the reactor antineutrino fluxes are based on
two methods:
the summation method and the conversion method
(see the reviews in Refs~\cite{Huber:2016fkt,Hayes:2016qnu}).
The summation method
(sometimes called \textit{ab initio})
uses the fission fraction and $\beta$-decay information
in nuclear databases to calculate the contribution of each single branch of the $\beta$-decay chains.
Since nuclear databases are incomplete and sometimes inaccurate,
the inferred reactor antineutrino spectra have potentially large and unknown uncertainties.
In the conversion method the
$\U$, $\Pu$, and $\Pum$
antineutrino spectra are inferred from
the corresponding $\beta$ spectra measured in the
1980's at the Institut Laue-Langevin
(ILL)~\cite{VonFeilitzsch:1982jw,Schreckenbach:1985ep,Hahn:1989zr}.
The converted $\Um$ antineutrino spectrum was obtained in 2013
from the $\beta$ spectrum measured by
an experiment at the scientific neutron source FRM II in Garching~\cite{Haag:2013raa}.
The conversion method was considered as more reliable and with small uncertainties
(about 2-3\%; see Refs.~\cite{Bemporad:2001qy,Vogel:2007du})
before the discovery of the so-called ``5 MeV bump'' in 2014
(see the reviews in Refs~\cite{Huber:2016fkt,Hayes:2016qnu})
in the
RENO~\cite{RENO:2015ksa},
Double Chooz~\cite{DoubleChooz:2014kuw},
and Daya Bay~\cite{DayaBay:2015lja}
experiments.

%%HM model
Mueller \textit{et al.}~\cite{Mueller:2011nm}
first used the summation method to obtain the antineutrino spectra of
$\U$, $\Um$, $\Pu$ and $\Pum$.
In a second step,
they corrected the $\U$, $\Pu$ and $\Pum$ spectra adding the missing contributions
obtained from the conversion of the
ILL $\beta$-spectra~\cite{Schreckenbach:1985ep,VonFeilitzsch:1982jw,Hahn:1989zr}.
These spectra were further corrected with an improved conversion method by Huber~\cite{Huber:2011wv}.
Therefore,
in the HM model the $\U$, $\Pu$ and $\Pum$ antineutrino spectra are those
obtained by Huber~\cite{Huber:2011wv} with the conversion method
and the $\Um$ antineutrino spectrum is that obtained by Mueller \textit{et al.}~\cite{Mueller:2011nm} with the summation method.

%%EF model
Nuclear databases have been improved in recent years,
especially through the application of the Total Absorption Gamma-ray Spectroscopy (TAGS) technique
for a better identification of the $\beta$ decay branches.
In 2019 Estienne, Fallot \textit{et al}~\cite{Estienne:2019ujo} (EF model),
using updated nuclear database information,
obtained a $\U$ reactor antineutrino flux that is smaller than that of the HM model
and roughly in agreement with the experimental measurements.
A similar result has been obtained in Ref.~\cite{Silaeva:2020msh}.
If these predictions describe correctly the real reactor antineutrino fluxes,
the reactor antineutrino anomaly disappears.
Let us however note that low predictions of the reactor antineutrino fluxes with the summation method
may be due to residual incompleteness of the nuclear database that is used in the calculation.
Moreover, it has been shown that the analytical approximation of the single beta spectrum using the Fermi function and additional corrections may induce a reactor spectral variation at the level of 2\% with the summation method~\cite{Fang:2020emq}.
The EF fluxes give a better fit of the Daya Bay antineutrino spectrum~\cite{DayaBay:2016ssb}
than the HM fluxes,
but they give a worse fit of the 5 MeV bump.
Therefore,
it is uncertain if the EF fluxes are more reliable than the HM fluxes.

%%HKSS model
An opposite result concerning the reactor antineutrino anomaly
is obtained by using the 2019 calculation of the reactor antineutrino fluxes in Ref.~\cite{Hayen:2019eop}
(HKSS model)
using the conversion method.
The authors of this paper improved the conversion method
by considering forbidden transitions calculated with the nuclear shell model,
that give a partial explanation of the 5 MeV bump.
The resulting antineutrino fluxes are larger than those of the HM model
and lead to an increase of the reactor antineutrino anomaly.
Note that also another conversion calculation~\cite{Li:2019quv}
found that taking account the forbidden transitions could partially explain the bump feature and result in a larger flux anomaly.

%%KI model
More recently,
in the beginning of 2021,
Kopeikin \textit{et al.}~\cite{Kopeikin:2021ugh}
corrected the $\U$ $\beta$ spectrum measured at ILL~\cite{VonFeilitzsch:1982jw,Schreckenbach:1985ep}
using new relative measurements of the ratio of the $\U$ and $\Pu$ $\beta$ spectra performed
with a research reactor at the Kurchatov Institute (KI model).
These measurements show an excess of the ILL ratio of the $\U$ and $\Pu$ $\beta$ spectra
by a factor $1.054 \pm 0.002$.
Kopeikin \textit{et al.} argued that it is likely that the normalization of the ILL $\U$ $\beta$ spectrum was overestimated.
Therefore,
the KI model predicts a converted $\U$ antineutrino flux that is smaller than that in the HM model
and converted $\Pu$ and $\Pum$ antineutrino fluxes that are similar to those of the HM model.
For the $\Um$ antineutrino flux,
Kopeikin \textit{et al.} applied the conversion method to the Garching $\Um$ $\beta$
spectrum~\cite{Haag:2013raa},
revising the normalization of the Garching $\Um$ $\beta$ spectrum to the ILL $\U$ spectrum.
The resulting $\Um$ antineutrino flux is slightly smaller than that in the HM model.
In total,
the KI model predicts antineutrino rates in
short-baseline reactor experiments that are smaller than those predicted by the HM model
and in approximate agreement with the EF predictions,
hinting for a demise of the reactor antineutrino anomaly.

In this paper we consider, besides the HM, EF, HKSS and KI models,
also a model that we call ``HKSS-KI'',
that is a modification of the HKSS model that we calculated taking into account
the KI correction to the normalization of the $\U$ antineutrino flux.
Considering this hybrid model is justified by the
interesting better fit of the 5 MeV bump
in the HKSS model with respect to the HM model
due to the inclusion of forbidden transitions
(see the discussion in Section~VIII.A of Ref.~\cite{Hayen:2019eop}).

The reduction of the $\U$ antineutrino flux in the KI model
with respect to the HM model is in agreement with the indications
obtained by the measurements of the evolution of the antineutrino
detection rate with the fuel composition
in the Daya Bay~\cite{DayaBay:2017jkb,DayaBay:2019yxq}
and RENO~\cite{RENO:2018pwo} experiments,
with the confirmation of a suppression of the $\U$ antineutrino flux
with respect to the HM model
found in the recent
PROSPECT~\cite{An:2021tyg,Prospect:2021lbs}
and
STEREO~\cite{STEREO:2020fvd,Prospect:2021lbs}
experiments at research reactors with practically pure
$\U$ fuel,
and with the summation calculations in Refs.~\cite{Estienne:2019ujo,Silaeva:2020msh}
(see also the discussions in
Refs.~\cite{Giunti:2016elf,Giunti:2017nww,Hayes:2017res,Giunti:2017yid,Gebre:2017vmm,Giunti:2019qlt,Berryman:2020agd}).
Let us however note that the KI
suppression of the ratio of the $\U$ and $\Pu$ $\beta$ spectra
with respect to the ILL measurements
is not certain, at least until it is confirmed by
another independent experiment.
Moreover,
it could also be explained with an enhancement of the $\Pu$ $\beta$ spectrum,
that would lead to a $\Pu$ antineutrino flux higher than that in the HM model,
with an enhancement of the reactor antineutrino anomaly.
It is interesting that fits of the reactor data
that include neutrino oscillations do not disfavor this
possibility~\cite{Giunti:2017yid,Giunti:2019qlt}.

%%oscillations
Neutrino oscillations can explain the reactor antineutrino anomaly
if there is at least one non-standard neutrino with a mass of the order of 1 eV
or larger.
These non-standard massive neutrinos must be mainly sterile,
in order to preserve the well-established neutrino oscillation explanation
of the data of solar, atmospheric and long-baseline neutrino oscillation experiments
(see the reviews in Refs.~\cite{Gariazzo:2015rra,Giunti:2019aiy,Diaz:2019fwt,Boser:2019rta,Dasgupta:2021ies}).
In this case,
a suppression of the rate measured in short-baseline
reactor neutrino experiments with respect to the theoretical prediction
is due to the disappearance of active $\bar{\nu}_e$'s.
This mechanism can be distinguished from an explanation of the
reactor antineutrino anomaly based on a suppression
of the production of one or more of the
four fluxes (mainly the dominant $\U$ flux)
through the following two important features:
A) it acts in an equal way on the four fluxes;
B) it has an oscillatory dependence on the distance $L$ from the reactor to the detector.
Here we consider the simplest 3+1 model,
with an effective mixing of the three standard neutrinos with one
non-standard neutrino $\nu_{4}$
with a mass of the order of 1 eV
or larger.
In this model,
the effective short-baseline survival probability of electron neutrinos and antineutrinos
is given by
\begin{equation}
P_{ee}
\simeq
1 - \sin^2 2\vartheta_{ee} \, \sin^2 \left(\frac{\Delta m_{41}^2 L}{4E}\right)
,
\label{Pee}
\end{equation}
where
$\Delta m_{41}^2 = m_{4}^2 - m_{1}^2 \gtrsim 1 \, \text{eV}^2$
is the squared-mass splitting between the non-standard massive neutrino $\nu_{4}$
and the three standard massive neutrinos
$\nu_{1}$,
$\nu_{2}$,
$\nu_{3}$
that have lighter masses with the much smaller
solar and atmospheric
squared-mass splittings
$\Delta m_{21}^2 \approx 7.4 \times 10^{-5} \, \text{eV}^2$
and
$|\Delta m_{31}^2| \approx 2.5 \times 10^{-3} \, \text{eV}^2$
(see Ref.~\cite{ParticleDataGroup:2020ssz}
and the recent three-neutrino global analyses in
Refs.~\cite{deSalas:2020pgw,Esteban:2020cvm,Capozzi:2021fjo}).
The effective mixing angle $\vartheta_{ee}$
depends on the element $U_{e4}$ of the $4\times4$ mixing matrix $U$ through the relation
$ \sin^2 2\vartheta_{ee} = 4 |U_{e4}|^2 ( 1 - |U_{e4}|^2 ) $.

The goal of this paper is to compare the fits of the
reactor rate and fuel evolution data
with the HM, EF, HKSS, KI and HKSS-KI models,
to assess the status of the reactor antineutrino anomaly
and short-baseline oscillations generated by a sterile neutrino at the eV mass scale.
As far as we know there is not an established deviation from a standard value
that defines an anomaly.
For the discussion, we will consider as anomalous a deviation larger than $2\sigma$.
However, this value is not very large, corresponding to a probability of about 5\%
of a type-I error, that is not very safe.
Therefore,
the statement that there is an anomaly must be weighted by
the number of $\sigma$'s.

%structure of this article
This paper is organized as follows.
In Section~\ref{sec.model}
we present the results of a new improved calculation of the predictions
of the HM, EF, HKSS, KI and HKSS-KI models.
In Section~\ref{sec.method}
we discuss the method of analysis of the reactor antineutrino data.
In Sections~\ref{sec.rates} and~\ref{sec.evolution}
we present, respectively, the results of the fits of the reactor rates
and the Daya Bay and RENO evolution data.
In Section~\ref{sec.bestfit} we discuss how to select the
best-fit reactor flux model.
In Section~\ref{sec.oscillations}
we discuss the implications for neutrino oscillations.
Finally, the concluding remarks are presented in Section~\ref{sec.conclusions}.

\section{Model predictions}
\label{sec.model}

The event rate measured in reactor neutrino experiments
can be expressed in a convenient way as a
``cross section per fission''
$\sigma_{f,a}$,
often called ``inverse beta decay (IBD) yield'':
\begin{equation}
\sigma_{f,a}
=
\sum_{i} f_i^a \sigma_{i}
,
\label{eq.sfa}
\end{equation}
where $a$ is the experiment label,
$\sigma_{i}$ is the IBD yield for the fissionable isotope $i$
(with $i=235$, $238$, $239$, and $241$ for $\U$, $\Um$, $\Pu$, and $\Pum$, respectively),
and $f_i^a$ is the effective fission fraction of the isotope $i$
for the experiment $a$.
For each fissionable isotope $i$, the individual IBD yield is given by
\begin{equation}
\sigma_{i}
=
\int_{E_{\nu}^{\text{thr}}}^{E_{\nu}^{\text{max}}}
d E_{\nu}
\,
\Phi_{i}(E_{\nu})
\,
\sigma_{\text{IBD}}(E_{\nu})
,
\label{eq.si}
\end{equation}
where $E_{\nu}$ is the neutrino energy,
$\Phi_{i}(E_{\nu})$ is the neutrino flux generated by the fissionable isotope $i$,
and
$\sigma_{\text{IBD}}(E_{\nu})$
is the detection cross section.
The neutrino energy is integrated from the threshold energy
$ E_{\nu}^{\text{thr}} = 1.806 \, \text{MeV} $
to a maximum value $E_{\nu}^{\text{max}} \ge 8 \, \text{MeV}$.
The numerical values of the $\sigma_{i}$'s predicted by a theoretical model
depend on the way in which the integral in Eq.~\eqref{eq.si} is performed,
taking into account that the neutrino fluxes are given in tabulated bins.

%%%%%%%%%%%%%%%%%%%%%%%%%%%%%%%%%%%%%%%%%%%%%%%%%%%%%%%%
%%%%%%%%%%%%%%%%%%%%%%%%%%%%%%%%%%%%%%%%%%%%%%%%%%%%%%%%

\begin{table}
\centering
\begin{ruledtabular}
\begin{tabular}{c|cccc}
    % {} \bf Model & $\bm{\sigma_{235}}$ & $\bm{\sigma_{238}}$ & $\bm{\sigma_{239}}$ & $\bm{\sigma_{241}}$\\
%    \hline
%    \bf HM & $6.69 \pm 0.14$ & $10.10 \pm 0.82$ & $4.40 \pm 0.11$ & $6.03 \pm 0.13$\\
%    \hline
%    \bf EF & $6.28 \pm 0.31$ & $10.14 \pm 1.01$ & $4.42 \pm 0.22$ & $6.23 \pm 0.31$\\
%    \hline
%    \bf HKSS & $6.74 \pm 0.17$ & $10.33 \pm 0.85$ & $4.43 \pm 0.13$ & $6.07 \pm 0.16$\\
%    \hline
%    \bf KI & $6.27 \pm 0.13$ & $9.34 \pm 0.47$ & $4.33 \pm 0.11$ & $6.01 \pm 0.13$\\
    {} \bf Model & $\bm{\sigma_{235}}$ & $\bm{\sigma_{238}}$ & $\bm{\sigma_{239}}$ & $\bm{\sigma_{241}}$\\
    \hline
    \bf HM
    &
    $6.69 \pm 0.14$
    &
    $10.10 \pm 0.82$
    &
    $4.40 \pm 0.11$
    &
    $6.03 \pm 0.13$
    \\
    \hline
    \bf EF
    &
    $6.28 \pm 0.31$
    &
    $10.14 \pm 1.01$
    &
    $4.42 \pm 0.22$
    &
    $6.23 \pm 0.31$
    \\
    \hline
    \bf HKSS
    &
    $6.74 \pm 0.17$
    &
    $10.33 \pm 0.85$
    &
    $4.43 \pm 0.13$
    &
    $6.07 \pm 0.16$
    \\
    \hline
    \bf KI
    &
    $6.27 \pm 0.13$
    &
    $9.34 \pm 0.47$
    &
    $4.33 \pm 0.11$
    &
    $6.01 \pm 0.13$
\end{tabular}
\caption{\label{tab.model}
The original theoretical IBD yields of the four fissionable isotopes in units of $10^{-43} \text{cm}^{2}/\text{fission}$
predicted by the
HM~\cite{Mueller:2011nm,Mention:2011rk,Huber:2011wv,Abazajian:2012ys},
EF~\cite{Estienne:2019ujo},
HKSS~\cite{Hayen:2019eop}, and
KI~\cite{Kopeikin:2021ugh}
models.
}
\vspace{0.5cm}
\begin{tabular}{c|cccc}
    {} \bf Model & $\bm{\sigma_{235}}$ & $\bm{\sigma_{238}}$ & $\bm{\sigma_{239}}$ & $\bm{\sigma_{241}}$\\
    \hline
    \bf HM & $6.60 \pm 0.14$ & $10.00 \pm 1.12$ & $4.33 \pm 0.11$ & $6.01 \pm 0.13$\\
    \hline
    \bf EF & $6.17 \pm 0.13$ & $9.94 \pm 1.09$ & $4.32 \pm 0.11$ & $6.10 \pm 0.13$\\
    \hline
    \bf HKSS & $6.67 \pm 0.15$ & $10.08 \pm 1.14$ & $4.37 \pm 0.12$ & $6.06 \pm 0.14$
\end{tabular}
\caption{\label{tab.berryman}
The theoretical IBD yields of the four fissionable isotopes in units of $10^{-43} \text{cm}^{2}/\text{fission}$
recalculated in Ref.~\cite{Berryman:2020agd}
for the HM, EF, and HKSS models.
}
\vspace{0.5cm}
\begin{tabular}{c|cccc}
    {} \bf Model & $\bm{\sigma_{235}}$ & $\bm{\sigma_{238}}$ & $\bm{\sigma_{239}}$ & $\bm{\sigma_{241}}$\\
    \hline
    \bf HM
    &
    $6.74 \pm 0.17$
    &
    $10.19 \pm 0.83$
    &
    $4.40 \pm 0.13$
    &
    $6.10 \pm 0.16$
    \\
    \hline
    \bf EF
    &
    $6.29 \pm 0.31$
    &
    $10.16 \pm 1.02$
    &
    $4.42 \pm 0.22$
    &
    $6.23 \pm 0.31$
    \\
    \hline
    \bf HKSS
    &
    $6.82 \pm 0.18$
    &
    $10.28 \pm 0.84$
    &
    $4.45 \pm 0.13$
    &
    $6.17 \pm 0.16$
    \\
    \hline
    \bf KI
    &
    $6.41 \pm 0.14$
    &
    $ 9.53 \pm 0.48$
    &
    $4.40 \pm 0.13$
    &
    $6.10 \pm 0.16$
    \\
    \hline
    \bf HKSS-KI
    &
    $6.48 \pm 0.14$
    &
    $10.28 \pm 0.84$
    &
    $4.45 \pm 0.13$
    &
    $6.17 \pm 0.16$
\end{tabular}
\caption{\label{tab.model_2020}
Our estimations of the theoretical IBD yields of the four fissionable isotopes in units of $10^{-43} \text{cm}^{2}/\text{fission}$
predicted by different models.}
\end{ruledtabular}
\end{table}

%%%%%%%%%%%%%%%%%%%%%%%%%%%%%%%%%%%%%%%%%%%%%%%%%%%%%%%%
%%%%%%%%%%%%%%%%%%%%%%%%%%%%%%%%%%%%%%%%%%%%%%%%%%%%%%%%

Table~\ref{tab.model} gives the original
theoretical IBD yields of the four fissionable isotopes
predicted by the
HM~\cite{Mueller:2011nm,Mention:2011rk,Huber:2011wv,Abazajian:2012ys},
EF~\cite{Estienne:2019ujo},
HKSS~\cite{Hayen:2019eop}, and
KI~\cite{Kopeikin:2021ugh}
models.
Since the uncertainties of the EF model are not given in Ref.~\cite{Estienne:2019ujo},
we considered the uncertainties associated with the summation spectra estimated in Ref.~\cite{Hayes:2017res}:
5\% for
$^{235}\text{U}$,
$^{239}\text{Pu}$, and
$^{241}\text{Pu}$,
and 10\% for $^{238}\text{U}$.

The HM, EF, and HKSS predictions have been recently recalculated in Ref.~\cite{Berryman:2020agd},
with some differences in the interpolation of the tabulated neutrino fluxes,
the detection cross section, and considering $E_{\nu}^{\text{max}} = 8 \text{MeV}$.
The resulting IBD yields shown in Table~\ref{tab.berryman}
are slightly smaller than the corresponding original IBD yields
in Table~\ref{tab.model}.
The uncertainties are similar,
except for the uncertainties of
$\sigma_{235}$,
$\sigma_{239}$, and
$\sigma_{241}$
in the EF model,
for which the authors of Ref.~\cite{Berryman:2020agd}
adopted the relative systematic uncertainties of the HM model.
We do not agree with this choice,
since the conversion and summation methods have different uncertainties.
We think that the systematic uncertainties of the values of
$\sigma_{235}$,
$\sigma_{239}$, and
$\sigma_{241}$
obtained with the conversion method in the HM model
underestimate the corresponding uncertainties
of the summation method following from the incomplete information
in the nuclear databases.

Taking into account that an accurate estimation of the IBD yields and their uncertainties
is crucial for a reliable assessment of the reactor antineutrino anomaly,
we performed a new calculation for each of the different models.
The results are presented in Table~\ref{tab.model_2020}.

For all models,
we calculated the IBD yields in Eq.~\eqref{eq.si}
using the Strumia and Vissani IBD cross section~\cite{Strumia:2003zx},
that improved the often-used Vogel and Beacom IBD cross section~\cite{Vogel:1999zy}.
Historically,
the reference IBD cross section is that in the famous 1971 review of Llewellyn Smith~\cite{LlewellynSmith:1971zm},
which neglected contributions of order $ ( m_{n} - m_{p} ) / E_{\nu} $,
that is not small for reactor neutrinos near the neutrino energy threshold,
since $ m_{n} - m_{p} = 1.293 \, \text{MeV} $
($ m_{n} $ and $ m_{p} $ are, respectively, the neutron and proton masses).
The Vogel and Beacom IBD cross section~\cite{Vogel:1999zy} considers these contributions,
but it was calculated at first order in $ E_{\nu} / m_{p} $.
This is a good approximation for most calculations concerning reactor neutrinos,
for which $ E_{\nu} / m_{p} \lesssim 10^{-2} $.
However, since the RAA is an effect of a few percent,
it is better to use the more precise Strumia and Vissani IBD cross section~\cite{Strumia:2003zx}
that was calculated without any approximation concerning $ E_{\nu} / m_{p} $
and $ ( m_{n} - m_{p} ) / E_{\nu} $.
In the implementation of the Strumia and Vissani formula for the IBD cross section,
we considered the PDG 2020~\cite{ParticleDataGroup:2020ssz} updated values of all the parameters.
We also considered the radiative corrections calculated in Ref.~\cite{Kurylov:2002vj},
that are well approximated by Eq.~(14) of Ref.~\cite{Strumia:2003zx}.

%%%%%%%%%%%%%%%%%%%%%%%%%%%%%%%%%%%%%%%%%%%%%%%%%%%%%%%%
%%%%%%%%%%%%%%%%%%%%%%%%%%%%%%%%%%%%%%%%%%%%%%%%%%%%%%%%

\begingroup
\newcounter{ExpNum}
%\squeezetable
\begin{table*}
\centering
\begin{tabular}{c c *{4}c c *{5}c c c c}
\hline
$a$
&
Experiment
&
$f^{a}_{235}$
&
$f^{a}_{238}$
&
$f^{a}_{239}$
&
$f^{a}_{241}$
&
$\sigma_{f,a}^{\text{exp}}$
&
$R_{a,\text{HM}}^{\text{exp}}$
&
$R_{a,\text{EF}}^{\text{exp}}$
&
$R_{a,\text{HKSS}}^{\text{exp}}$
&
$R_{a,\text{KI}}^{\text{exp}}$
&
$R_{a,\text{HKSS-KI}}^{\text{exp}}$
&
$\delta_{a}^{\text{exp}}$ [\%]
&
$\delta_{a}^{\text{cor}}$ [\%]
&
$L_{a}$ [m]
\\
\hline
\stepcounter{ExpNum} $\theExpNum$ &
Bugey-4 &
$0.538$ &
$0.078$ &
$0.328$ &
$0.056$ &
$5.75$ &
$0.927$ &
$0.962$ &
$0.916$ &
$0.962$ &
$0.944$ &
$1.4$ &
\rdelim\}{2}{20pt}[1.4] &
$15$
\\
\stepcounter{ExpNum} $\theExpNum$ &
Rovno91 &
$0.614$ &
$0.074$ &
$0.274$ &
$0.038$ &
$5.85$ &
$0.924$ &
$0.965$ &
$0.914$ &
$0.962$ &
$0.945$ &
$2.8$ &
                        &
$18$
\\
\hline
\stepcounter{ExpNum} $\theExpNum$ &
Rovno88-1I &
$0.607$ &
$0.074$ &
$0.277$ &
$0.042$ &
$5.70$ &
$0.902$ &
$0.941$ &
$0.892$ &
$0.939$ &
$0.921$ &
$6.4$ &
\rdelim\}{2}{20pt}[3.1] \rdelim\}{5}{20pt}[2.2] &
$18$
\\
\stepcounter{ExpNum} $\theExpNum$ &
Rovno88-2I &
$0.603$ &
$0.076$ &
$0.276$ &
$0.045$ &
$5.89$ &
$0.931$ &
$0.971$ &
$0.920$ &
$0.969$ &
$0.951$ &
$6.4$ &
                                                &
$17.96$
\\
\stepcounter{ExpNum} $\theExpNum$ &
Rovno88-1S &
$0.606$ &
$0.074$ &
$0.277$ &
$0.043$ &
$6.04$ &
$0.956$ &
$0.997$ &
$0.945$ &
$0.995$ &
$0.976$ &
$7.3$ &
\rdelim\}{3}{45pt}[3.1]                         &
$18.15$
\\
\stepcounter{ExpNum} $\theExpNum$ &
Rovno88-2S &
$0.557$ &
$0.076$ &
$0.313$ &
$0.054$ &
$5.96$ &
$0.956$ &
$0.994$ &
$0.945$ &
$0.993$ &
$0.974$ &
$7.3$ &
                                                &
$25.17$
\\
\stepcounter{ExpNum} $\theExpNum$ &
Rovno88-3S &
$0.606$ &
$0.074$ &
$0.274$ &
$0.046$ &
$5.83$ &
$0.922$ &
$0.962$ &
$0.911$ &
$0.960$ &
$0.942$ &
$6.8$ &
                                                &
$18.18$
\\
\hline
\stepcounter{ExpNum} $\theExpNum$ &
Bugey-3-15 &
$0.538$ &
$0.078$ &
$0.328$ &
$0.056$ &
$5.77$ &
$0.930$ &
$0.966$ &
$0.920$ &
$0.966$ &
$0.947$ &
$4.2$ &
\rdelim\}{3}{20pt}[4.0]                         &
$15$
\\
\stepcounter{ExpNum} $\theExpNum$ &
Bugey-3-40 &
$0.538$ &
$0.078$ &
$0.328$ &
$0.056$ &
$5.81$ &
$0.936$ &
$0.972$ &
$0.926$ &
$0.972$ &
$0.953$ &
$4.3$ &
                                                &
$40$
\\
\stepcounter{ExpNum} $\theExpNum$ &
Bugey-3-95 &
$0.538$ &
$0.078$ &
$0.328$ &
$0.056$ &
$5.35$ &
$0.861$ &
$0.895$ &
$0.852$ &
$0.894$ &
$0.877$ &
$15.2$ &
                                                &
$95$
\\
\hline
\stepcounter{ExpNum} $\theExpNum$ &
Gosgen-38 &
$0.619$ &
$0.067$ &
$0.272$ &
$0.042$ &
$5.99$ &
$0.949$ &
$0.992$ &
$0.939$ &
$0.988$ &
$0.971$ &
$5.4$ &
\rdelim\}{3}{20pt}[2.0] \rdelim\}{4}{20pt}[3.8] &
$37.9$
\\
\stepcounter{ExpNum} $\theExpNum$ &
Gosgen-46 &
$0.584$ &
$0.068$ &
$0.298$ &
$0.050$ &
$6.09$ &
$0.975$ &
$1.016$ &
$0.964$ &
$1.014$ &
$0.995$ &
$5.4$ &
                                                &
$45.9$
\\
\stepcounter{ExpNum} $\theExpNum$ &
Gosgen-65 &
$0.543$ &
$0.070$ &
$0.329$ &
$0.058$ &
$5.62$ &
$0.909$ &
$0.945$ &
$0.899$ &
$0.944$ &
$0.927$ &
$6.7$ &
                                                &
$64.7$
\\
\stepcounter{ExpNum} $\theExpNum$ &
 &
$1.000$ &
$0.000$ &
$0.000$ &
$0.000$ &
$5.30$ &
$0.787$ &
$0.843$ &
$0.777$ &
$0.827$ &
$0.818$ &
$9.1$ &
                                                &
$8.76$
\\
\hline
\stepcounter{ExpNum} $\theExpNum$ &
Krasnoyarsk87-33 &
$1$ &
$0$ &
$0$ &
$0$ &
$6.20$ &
$0.920$ &
$0.986$ &
$0.909$ &
$0.967$ &
$0.957$ &
$5.2$ &
\rdelim\}{2}{20pt}[4.1] &
$32.8$
\\
\stepcounter{ExpNum} $\theExpNum$ &
Krasnoyarsk87-92 &
$1$ &
$0$ &
$0$ &
$0$ &
$6.30$ &
$0.935$ &
$1.002$ &
$0.924$ &
$0.983$ &
$0.972$ &
$20.5$ &
                        &
$92.3$
\\
\stepcounter{ExpNum} $\theExpNum$ &
Krasnoyarsk94-57 &
$1$ &
$0$ &
$0$ &
$0$ &
$6.26$ &
$0.929$ &
$0.995$ &
$0.918$ &
$0.977$ &
$0.966$ &
$4.2$ &
0                       &
$57$
\\
\stepcounter{ExpNum} $\theExpNum$ &
Krasnoyarsk99-34 &
$1$ &
$0$ &
$0$ &
$0$ &
$6.39$ &
$0.948$ &
$1.016$ &
$0.937$ &
$0.997$ &
$0.986$ &
$3.0$ &
0                       &
$34$
\\
\hline
\stepcounter{ExpNum} $\theExpNum$ &
SRP-18 &
$1$ &
$0$ &
$0$ &
$0$ &
$6.29$ &
$0.934$ &
$1.000$ &
$0.923$ &
$0.982$ &
$0.971$ &
$2.8$ &
0 &
$18.2$
\\
\stepcounter{ExpNum} $\theExpNum$ &
SRP-24 &
$1$ &
$0$ &
$0$ &
$0$ &
$6.73$ &
$0.998$ &
$1.070$ &
$0.987$ &
$1.050$ &
$1.038$ &
$2.9$ &
0 &
$23.8$
\\
\hline
\stepcounter{ExpNum} $\theExpNum$ &
\input{wrk/rea-new-HM/rex/nucifer/dat/bef-exp-nam.dat} &
$0.926$ &
$0.008$ &
$0.061$ &
$0.005$ &
$6.67$ &
$1.007$ &
$1.074$ &
$0.995$ &
$1.056$ &
$1.044$ &
$10.8$ &
0 &
$7.2$
\\
\stepcounter{ExpNum} $\theExpNum$ &
\input{wrk/rea-new-HM/rex/chooz/dat/bef-exp-nam.dat} &
$0.496$ &
$0.087$ &
$0.351$ &
$0.066$ &
$6.12$ &
$0.990$ &
$1.025$ &
$0.979$ &
$1.027$ &
$1.007$ &
$3.2$ &
0 &
$\approx 1000$
\\
\stepcounter{ExpNum} $\theExpNum$ &
\input{wrk/rea-new-HM/rex/paloverde/dat/bef-exp-nam.dat} &
$0.600$ &
$0.070$ &
$0.270$ &
$0.060$ &
$6.25$ &
$0.991$ &
$1.033$ &
$0.980$ &
$1.031$ &
$1.012$ &
$5.4$ &
0 &
$\approx 800$
\\
\stepcounter{ExpNum} $\theExpNum$ &
\input{wrk/rea-new-HM/rex/dayabay/dat/bef-exp-nam.dat} &
$0.564$ &
$0.076$ &
$0.304$ &
$0.056$ &
$5.94$ &
$0.950$ &
$0.988$ &
$0.939$ &
$0.987$ &
$0.968$ &
$1.5$ &
0 &
$\approx 550$
\\
\stepcounter{ExpNum} $\theExpNum$ &
\input{wrk/rea-new-HM/rex/reno/dat/bef-exp-nam.dat} &
$0.571$ &
$0.073$ &
$0.300$ &
$0.056$ &
$5.85$ &
$0.936$ &
$0.974$ &
$0.925$ &
$0.973$ &
$0.954$ &
$2.1$ &
0 &
$\approx 411$
\\
\stepcounter{ExpNum} $\theExpNum$ &
\input{wrk/rea-new-HM/rex/doublechooz/dat/bef-exp-nam.dat} &
$0.520$ &
$0.087$ &
$0.333$ &
$0.060$ &
$5.71$ &
$0.918$ &
$0.952$ &
$0.907$ &
$0.953$ &
$0.934$ &
$1.1$ &
0 &
$\approx 415$
\\
\stepcounter{ExpNum} $\theExpNum$ &
\input{wrk/rea-new-HM/rex/stereo/dat/bef-exp-nam.dat} &
$1$ &
$0$ &
$0$ &
$0$ &
$6.34$ &
$0.941$ &
$1.008$ &
$0.930$ &
$0.989$ &
$0.978$ &
$2.5$ &
0 &
$9-11$
\\
\hline
\end{tabular}
\caption{\label{tab.rates}
List of the experiments which measured the absolute reactor antineutrino flux.
For each experiment numbered with the index $a$:
$f^{a}_{235}$,
$f^{a}_{238}$,
$f^{a}_{239}$, and
$f^{a}_{241}$
are the effective fission fractions of the four isotopes
$^{235}\text{U}$,
$^{238}\text{U}$,
$^{239}\text{Pu}$, and
$^{241}\text{Pu}$,
respectively;
$\sigma_{f,a}^{\text{exp}}$
is the experimental IBD yield
in units of $10^{-43} \text{cm}^{2}/\text{fission}$;
$R_{a,\text{HM}}^{\text{exp}}$,
$R_{a,\text{EF}}^{\text{exp}}$,
$R_{a,\text{HKSS}}^{\text{exp}}$,
$R_{a,\text{KI}}^{\text{exp}}$, and
$R_{a,\text{HKSS-KI}}^{\text{exp}}$
are the ratios of measured and predicted rates for the IBD yields of the
five models in Table~\ref{tab.model_2020};
$\delta_{a}^{\text{exp}}$ is the total relative experimental statistical plus systematic uncertainty,
$\delta_{a}^{\text{cor}}$ is the part of the relative experimental systematic uncertainty
which is correlated in each group of experiments indicated by the braces;
$L_{a}$ is the source-detector distance.
}
\end{table*}
\endgroup

%%%%%%%%%%%%%%%%%%%%%%%%%%%%%%%%%%%%%%%%%%%%%%%%%%%%%%%%
%%%%%%%%%%%%%%%%%%%%%%%%%%%%%%%%%%%%%%%%%%%%%%%%%%%%%%%%

We calculated the integral in Eq.~\eqref{eq.si}
up to $E_{\nu}^{\text{max}} = 10 \, \text{MeV}$.
Thus, we extended the range of integration with respect to that in
Ref.~\cite{Berryman:2020agd}
by considering the small contributions above 8 MeV:
about 0.3\% for $\U$, 0.9\% for $\Um$, 0.2\% for $\Pu$, and 0.3\% for $\Pum$.
Further details of each model calculation are:

\begin{description}

\item[HM]
For the neutrino spectrum
we used the $\U$, $\Pu$, and $\Pum$ converted spectra from Ref.~\cite{Huber:2011wv}
and the $\Um$ summation spectrum from Ref.~\cite{Mueller:2011nm},
as in the original HM model.
These neutrino spectra are tabulated in Refs.~\cite{Mueller:2011nm,Huber:2011wv}
in energy bins with $250 \, \text{keV}$ width
from 2 to 8 MeV.
We calculated the integral in Eq.~\eqref{eq.si}
up to $E_{\nu}^{\text{max}} = 10 \, \text{MeV}$
in intervals of $5 \, \text{keV}$.
From 2 to 8 MeV we used a linear interpolation of the logarithm of the tabulated fluxes
and we took into account the tabulated uncertainties.
We extrapolated linearly the logarithm of the neutrino flux
in the small range from 1.8 to 2 MeV.
Since for energies larger than 8 MeV mathematical extrapolations
of the tabulated values are not reliable from a physical point of view,
we considered the most recent EF spectra obtained with the summation method,
that are tabulated up to 10 MeV in Ref.~\cite{Estienne:2019ujo}.
We assigned to these small contributions a very conservative 100\% uncertainty.
For the $\U$, $\Pu$, and $\Pum$ converted spectra
we took into account the off-equilibrium corrections
given in Table~VII of Ref.~\cite{Mueller:2011nm}
considering the 450 days approximation of the spectrum at equilibrium.
We neglected the small difference of the correction for
research reactors with an almost pure $\U$ fuel composition,
for which refueling occurs at intervals of about 30-50
days~\cite{Kozlov:1999ct,STEREO:2020fvd}.
We calculated the covariance matrix of the model uncertainties
using the correlation matrix in Eq.~(2.2)
of Ref.~\cite{Berryman:2020agd}.

\item[EF]
The neutrino spectra from the four fissionable isotopes are tabulated in the Supplemental Material
of Ref.~\cite{Estienne:2019ujo}
in energy bins with a 100 keV width from 0 keV to 10.1 MeV.
In order to perform an accurate integration in Eq.~\eqref{eq.si},
we interpolated linearly the logarithm of the spectra in bins with a 5 keV width.
As in Table~\ref{tab.model},
we adopted the uncertainties associated with the summation spectra estimated in Ref.~\cite{Hayes:2017res}:
$5\%$ for $\U$, $\Pu$, and $\Pum$, and $10\%$ for $\Um$,
without any correlation.

\item[HKSS]
The neutrino spectra are given in Tables~VI-IX of Ref.~\cite{Hayen:2019eop}
as differences with respect to the HM neutrino spectra.
Therefore, we applied these corrections
and we calculated the IBD yields with the same method that we used for the HM model.
Above 8 MeV we considered the EF spectra, as we have done for the HM model.

\item[KI]
The $\U$ and $\Um$ neutrino spectra are given in Table~I of Ref.~\cite{Kopeikin:2021ugh}
in 250 keV energy bins from 2 to 8 MeV.
We calculated the corresponding IBD yields with the same method that we used for the HM model,
including the small contribution above 8 MeV obtained from the EF spectra.
Note, however, that the KI $\Um$ neutrino spectrum has been obtained converting the Garching $\beta$
spectrum~\cite{Haag:2013raa},
instead of using the summation method as in the HM model~\cite{Mueller:2011nm}.
For this spectrum we neglected the small off-equilibrium corrections
that have not been calculated so far\footnote{\label{ftn.238}
Since the Garching data were taken with an irradiation between 11h and 42h~\cite{Haag:2013raa},
the off-equilibrium corrections of the $\Um$ neutrino flux
are smaller than the differences between the 12h and 450d $\Um$ neutrino fluxes in Table~III
of Ref.~\cite{Mueller:2011nm},
that are already smaller than
the off-equilibrium corrections of the $\U$ neutrino flux
given in Table~VII of Ref.~\cite{Mueller:2011nm}.
Moreover, as shown in Table~\ref{tab.rates},
the $\Um$ neutrino flux contributes by less than 8\%
to the IBD yields of commercial reactor experiments.
Therefore,
the small off-equilibrium corrections of the $\Um$ neutrino flux
can be safely neglected
taking into account the current uncertainties.
}.
The $\Pu$ and $\Pum$ IBD yields are the same as in the HM model.

\item[HKSS-KI]
In this model we reduced the $\U$ IBD yield of the HKSS model
by the KI factor $1.054 \pm 0.002$~\cite{Kopeikin:2021ugh}.
The other IBD yields are the same as in the HKSS model.
In particular,
the $\Um$ IBD yield is the same as that in the HKSS model
because the HKSS $\Um$ flux is given in Ref.~\cite{Hayen:2019eop}
as a correction to the HM $\Um$ flux,
that was calculated with the summation method
and, therefore, is not affected by the KI correction.

\end{description}

The IBD yields in Table~\ref{tab.model_2020},
that we obtained for the EF model are close to the original ones in Table~\ref{tab.model},
taken from Table~I of Ref.~\cite{Estienne:2019ujo}:
there is only a very slight increase of
$0.2\%$
%and
%$\input{wrk/rea-fig/tex/csf-dif_rel-EF-238.dat}\%$
for the $\U$ and $\Um$ IBD yields. %, respectively.
This is a confirmation of the reliability of our calculation of the integral in Eq.~\eqref{eq.si}.

The IBD yields in Table~\ref{tab.model_2020},
that we obtained for the HM model are slightly larger than the original ones in Table~\ref{tab.model},
taken from Table~XX of Ref.~\cite{Abazajian:2012ys}:
there are only the slight increases of
$0.7\%$,
$0.9\%$, and
$1.2\%$,
for the $\U$, $\Um$, and $\Pum$ IBD yields, respectively,
that are smaller than the corresponding uncertainties.

The difference between the HKSS IBD yields in Table~\ref{tab.model_2020}
and the original ones in Table~\ref{tab.model} is larger:
$1.2\%$,
$0.5\%$,
$0.5\%$, and
$1.6\%$,
for the $\U$, $\Um$, $\Pu$, and $\Pum$ IBD yields, respectively.
This may be due to the fact that we
estimated the original HKSS IBD yields in Table~\ref{tab.model}
applying to the HM IBD yields the corrections in Table~III of Ref.~\cite{Hayen:2019eop},
and these corrections may have been underestimated.

There are significant differences between the KI IBD yields in Table~\ref{tab.model_2020}
and the original ones in Table~\ref{tab.model},
taken from Ref.~\cite{Kopeikin:2021ugh}:
our IBD yields are larger by
$2.2\%$,
$2.0\%$,
$1.6\%$, and
$1.5\%$,
for the $\U$, $\Um$, $\Pu$, and $\Pum$ IBD yields, respectively.
In particular,
the increase of the most important
$\U$ IBD yield is approximately equal to its uncertainty.
Part of the differences is due to the omission of the off-equilibrium effects
in the IBD yields given in Ref.~\cite{Kopeikin:2021ugh}.
However, there is an increase of $2.0\%$
for the $\Um$ IBD yield for which we neglected the off-equilibrium effects
(see footnote~\ref{ftn.238}).
Hence, we suspect that the main difference is the method of calculation of the integral in Eq.~\eqref{eq.si}
and the use in Ref.~\cite{Kopeikin:2021ugh} of an old version of the IBD cross section.

There are also significant differences
between the IBD yields in Table~\ref{tab.model_2020}
and those in Table~\ref{tab.berryman} from Ref.~\cite{Berryman:2020agd}.
For example,
for the most important $\U$ IBD yield,
the values that we obtained for the HM, EF, and HKSS models
are larger by
$2.1\%$,
$1.9\%$, and
$2.2\%$,
respectively,
than those obtained in Ref.~\cite{Berryman:2020agd}.
We think that the reason is a cumulative effect due to
the contribution of the flux for $E_{\nu}^{\text{max}} \ge 8 \, \text{MeV}$
and
the off-equilibrium effects that have not been taken into account in
Ref.~\cite{Berryman:2020agd},
and possible effects due to 
the different ways of calculating the integral in Eq.~\eqref{eq.si}
and the IBD cross section, including the radiative corrections.

In the following we adopt our estimates of the IBD yields in Table~\ref{tab.model_2020}
for the analysis of the data of the short-baseline reactor neutrino experiments.

\section{Method of analysis}
\label{sec.method}

Reactor antineutrino data are traditionally analyzed with a least-squares function
(also called ``$\chi^2$'')
that has an asymptotic $\chi^2$ distribution under the standard assumptions of
Wilks' theorem~\cite{Wilks:1938dza}.
In practice these assumptions may be not entirely satisfied,
but a least-squares analysis is considered as a reasonable way to get an indication
of which is the model that gives the better explanation of the data
and which is the favored region of the model parameters.
There is, however,
the ambiguity of choosing how the systematic theoretical uncertainties
are treated in the least-squares function.
There are three main approaches:
\textbf{(A)} consider a covariance matrix with experimental and theoretical uncertainties
added in quadrature
(for example, in Refs.~\cite{Mention:2011rk,Gariazzo:2017fdh});
\textbf{(B)} calculate the fit results taking into account only the experimental uncertainties
and add by hand a global theoretical uncertainty to the final result
(for example, in Refs.~\cite{Zhang:2013ela,DayaBay:2016ssb});
\textbf{(C)} take into account the theoretical uncertainties with appropriate pull terms
(for example, in Refs.~\cite{Giunti:2017yid,Giunti:2019qlt,Berryman:2019hme,Berryman:2020agd}).

Method \textbf{(A)} is often considered as the best or preferred one,
since it treats the systematic theoretical uncertainties in the same way as
the experimental uncertainties, that include systematic uncertainties.
However,
it suffers of the problem called
``Peelle's Pertinent Puzzle'' (PPP)
after its discovery in 1987 in the field
of nuclear data analysis~\cite{Peelle:1987}
(see, for example, Ref.~\cite{Chiba-Smith-JNST-1994}):
if the data are discrepant and strongly correlated,
the best-fit average can be lower than most of the data, or even lower than all of the data.
This happens in the case of the reactor antineutrino data when the theoretical uncertainties
that are fully correlated among the different experiments
are added in quadrature to the covariance matrix of the experimental uncertainties,
that are mostly uncorrelated among different experiments
(small correlations due to the use of the same reactor or the same detector in different experiments
are taken into account
as described, for example, in Refs.~\cite{Mention:2011rk,Gariazzo:2017fdh}).

Method \textbf{(B)} avoids PPP,
but it requires the estimation of a global theoretical uncertainty
to be added by hand to the final result.
Since obviously there is no rigorous way to calculate such global theoretical uncertainty
independently from the least-squares analysis,
this method cannot be considered as very reliable.

Method \textbf{(C)} avoids PPP in a smart way
(see the discussion in Ref.~\cite{DAgostini:1993arp})
and allows to take into account the systematic theoretical uncertainties
in the least-squares analysis without any arbitrary assumption.
In practice PPP is avoided by decoupling the minimization of the least-squares function
with respect of the physical parameters
from the minimization with respect of the pull coefficients that take into account the
correlated systematic theoretical uncertainties.
In this paper we adopt this method
considering the general least-squares function
\begin{align}
\chi^2
=
\null & \null
\sum_{a,b}
\left( \sigma_{f,a}^{\text{exp}} - R_{\text{NP}}^{a} \sigma_{f,a}^{\text{th}} \right)
\left( V^{\text{exp}} \right)^{-1}_{ab}
\left( \sigma_{f,b}^{\text{exp}} - R_{\text{NP}}^b \sigma_{f,b}^{\text{th}} \right)
\nonumber
\\
\null & \null
+
\sum_{i,j}
\left( r_i - 1 \right) \left( \widetilde{V}^{\text{mod}} \right)^{-1}_{ij} \left( r_j - 1 \right)
,
\label{genchi}
\end{align}
where
$a,b$ are the experiment labels in Table~\ref{tab.rates},
$i,j = 235, 238, 239, 241$,
and
\begin{equation}
\sigma_{f,a}^{\text{th}}
=
\sum_i r_i f_i^a \sigma_{i}^{\text{mod}}
.
\label{sfth}
\end{equation}
Here $\sigma_{i}^{\text{mod}}$
denotes the IBD yield of the antineutrino flux generated by the fissionable isotope $i$
in the model under consideration
and
$\widetilde{V}^{\text{mod}}$
is the corresponding fractional covariance matrix
($\widetilde{V}^{\text{mod}}_{ij} = V^{\text{mod}}_{ij} / ( \sigma_{i}^{\text{mod}} \sigma_{j}^{\text{mod}} )$,
where $V^{\text{mod}}$ is the covariance matrix.
The coefficient $R_{\text{NP}}^{a}$ in front of the theoretical IBD yield
is a possible suppression factor of the IBD yield in the experiment $a$
that is due to new physics that acts in the same way on the four fluxes.
In the case of neutrino oscillations, we have $R_{\text{NP}}^{a}=P_{ee}^a$,
which is given by an appropriate average over the energy spectrum
and distance uncertainties of experiment $a$
of the electron antineutrino survival probability in Eq.~\eqref{Pee}.

The prevention of PPP can be easily seen by considering the simplest
case of a constant new physics suppression $R_{\text{NP}}^{a}=\overline{R}$,
that is usually adopted for the quantification of the reactor antineutrino anomaly
(see the discussion and results in Section~\ref{sec.rates}).
In this case, by considering the derivatives of $\chi^2$
with respect to $\overline{R}$ and the coefficients $r_i$,
one can see that
$\overline{R}$ is determined only by the first term containing the experimental
data and uncertainties
and the coefficients $r_i$'s
are constrained to values close to 1 by the pull term.

This mechanism is even more clear if we restrict the fit to the data
of research reactors that have a pure $\U$ fuel.
In that case one can find analytically
that the best-fit value of $r_{235}$ is fixed to exactly 1 by the pull term.
Therefore,
the best-fit value of $\overline{R}$ is determined only by the first term containing the experimental
data and uncertainties
and is independent from the value of the systematic theoretical uncertainty
on $\sigma_{235}^{\text{mod}}$,
in spite of its full correlation among the experiments.
Of course the systematic theoretical uncertainty
on $\sigma_{235}^{\text{mod}}$
is not irrelevant,
because it determines the uncertainty of $\overline{R}$.
The important fact is that PPP cannot occur,
because the best-fit of $\overline{R}$ does not depend on the
systematic theoretical uncertainty.

\section{Fit of reactor rates}
\label{sec.rates}

In this section, we consider the experimental reactor rates
that determine the reactor antineutrino anomaly,
as first discussed in Ref.~\cite{Mention:2011rk}.
We consider the data listed in Table~\ref{tab.rates},
which reports for each experiment the effective fuel fractions,
the ratios of the measured rate and the predictions of the five models that we consider
(with the IBD yields in Table~\ref{tab.model_2020}),
the experimental uncertainties, and the source-detector distance.
With respect to previous analyses~\cite{Gariazzo:2017fdh,Giunti:2017yid,Giunti:2019qlt,Giunti:2019fcj},
we updated the absolute flux measurements of the
Day Bay~\cite{DayaBay:2019yxq},
Double Chooz~\cite{DoubleChooz:2019qbj}, and
RENO~\cite{RENO:2020dxd}
experiments and we took into account the new absolute flux measurement of the
STEREO~\cite{STEREO:2020fvd}
experiment.

%%%%%%%%%%%%%%%%%%%%%%%%%%%%%%%%%%%%%%%%%%%%%%%%%%%%%%%%
%%%%%%%%%%%%%%%%%%%%%%%%%%%%%%%%%%%%%%%%%%%%%%%%%%%%%%%%

\begin{figure}
\centering
\includegraphics[width=\linewidth]{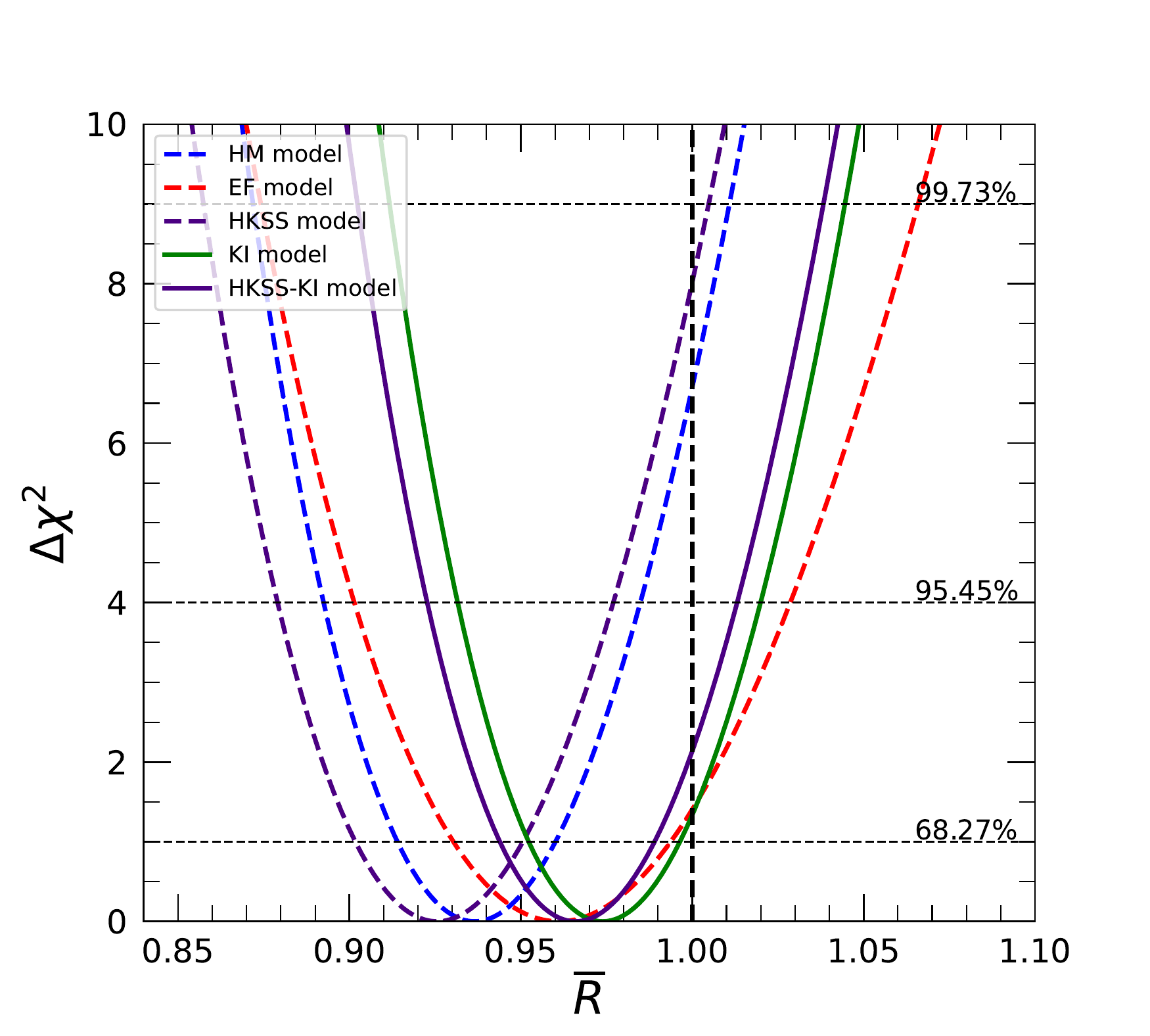}
\caption{\label{fig.ratio_L}
$\Delta\chi^2=\chi^2-\chi^2_{\text{min}}$ as a function of the average ratio $\overline{R}$
obtained in our least-squares analysis of short-baseline reactor rates
considering the IBD yields of the HM, EF, HKSS, KI, and HKSS-KI models
given in Table~\ref{tab.model_2020}.
}
\end{figure}

\begin{table*}
\centering
\begin{ruledtabular}
\begin{tabular}{c@{\qquad}cc@{\qquad}cc@{\qquad}cc}
\bf Model & \multicolumn{2}{c}{\bf Rates} & \multicolumn{2}{c}{\bf Evolution} & \multicolumn{2}{c}{\bf Rates + Evolution}
\\
& $\bm{\overline{R}_{\text{mod}}}$ & \bf RAA & $\bm{\overline{R}_{\text{mod}}}$ & \bf RAA& $\bm{\overline{R}_{\text{mod}}}$ & \bf RAA
\\
\hline
\bf HM
&
$0.936 {}^{ + 0.024 }_{ - 0.023 }
$
&
$2.5\,\sigma$
&
$0.933 {}^{ + 0.025 }_{ - 0.024 }
$
&
$2.6\,\sigma$
&
$0.930 {}^{ + 0.024 }_{ - 0.023 }
$
&
$2.8\,\sigma$
\\
\bf EF
&
$0.960 {}^{ + 0.033 }_{ - 0.031 }
$
&
$1.2\,\sigma$
&
$0.975 {}^{ + 0.032 }_{ - 0.030 }
$
&
$0.8\,\sigma$
&
$0.975 {}^{ + 0.032 }_{ - 0.030 }
$
&
$0.8\,\sigma$
\\
\bf HKSS
&
$0.925 {}^{ + 0.025 }_{ - 0.023 }
$
&
$2.9\,\sigma$
&
$0.925 {}^{ + 0.026 }_{ - 0.024 }
$
&
$2.8\,\sigma$
&
$0.922 {}^{ + 0.024 }_{ - 0.023 }
$
&
$3.0\,\sigma$
\\
\bf KI
&
$0.975 {}^{ + 0.022 }_{ - 0.021 }
$
&
$1.1\,\sigma$
&
$0.973 {}^{ + 0.023 }_{ - 0.022 }
$
&
$1.2\,\sigma$
&
$0.970 \pm 0.021
$
&
$1.4\,\sigma$
\\
\bf HKSS-KI
&
$0.964 {}^{ + 0.023 }_{ - 0.022 }
$
&
$1.5\,\sigma$
&
$0.955 {}^{ + 0.024 }_{ - 0.023 }
$
&
$1.9\,\sigma$
&
$0.960 {}^{ + 0.022 }_{ - 0.021 }
$
&
$1.8\,\sigma$
\end{tabular}
\end{ruledtabular}
\caption{\label{tab.Rave}
Average ratio $\overline{R}_{\text{mod}}$
obtained from the least-squares analysis of the
reactor rates in Table~\ref{tab.rates}
and of the
Daya Bay~\protect\cite{DayaBay:2017jkb}
and
RENO~\protect\cite{RENO:2018pwo}
evolution data for the IBD yields of the five models in Table~\ref{tab.model_2020}.
The RAA columns give the corresponding statistical significance of the
reactor antineutrino anomaly.
}
\end{table*}

%%%%%%%%%%%%%%%%%%%%%%%%%%%%%%%%%%%%%%%%%%%%%%%%%%%%%%%%
%%%%%%%%%%%%%%%%%%%%%%%%%%%%%%%%%%%%%%%%%%%%%%%%%%%%%%%%

We performed the fit of the reactor rates
considering a constant new physics suppression $R_{\text{NP}}^{a}=\overline{R}$
in the least-squares function (\ref{genchi}).
Figure~\ref{fig.ratio_L}
shows the value of $\Delta\chi^2=\chi^2-\chi^2_{\text{min}}$ as a function of the average ratio $\overline{R}$
obtained in our least-squares analysis for the five models listed in Section~\ref{sec.model},
considering the IBD yields given in Table~\ref{tab.model_2020}.
The best-fit value of $\overline{R}$ for each of the five models
and the corresponding statistical significance of the
reactor antineutrino anomaly
are given in Table~\ref{tab.Rave}.

%%%%%%%%%%%%%%%%%%%%%%%%%%%%%%%%%%%%%%%%%%%%%%%%%%%%%%%%
%%%%%%%%%%%%%%%%%%%%%%%%%%%%%%%%%%%%%%%%%%%%%%%%%%%%%%%%

\begin{figure*}
\centering
\includegraphics*[width=0.7\linewidth]{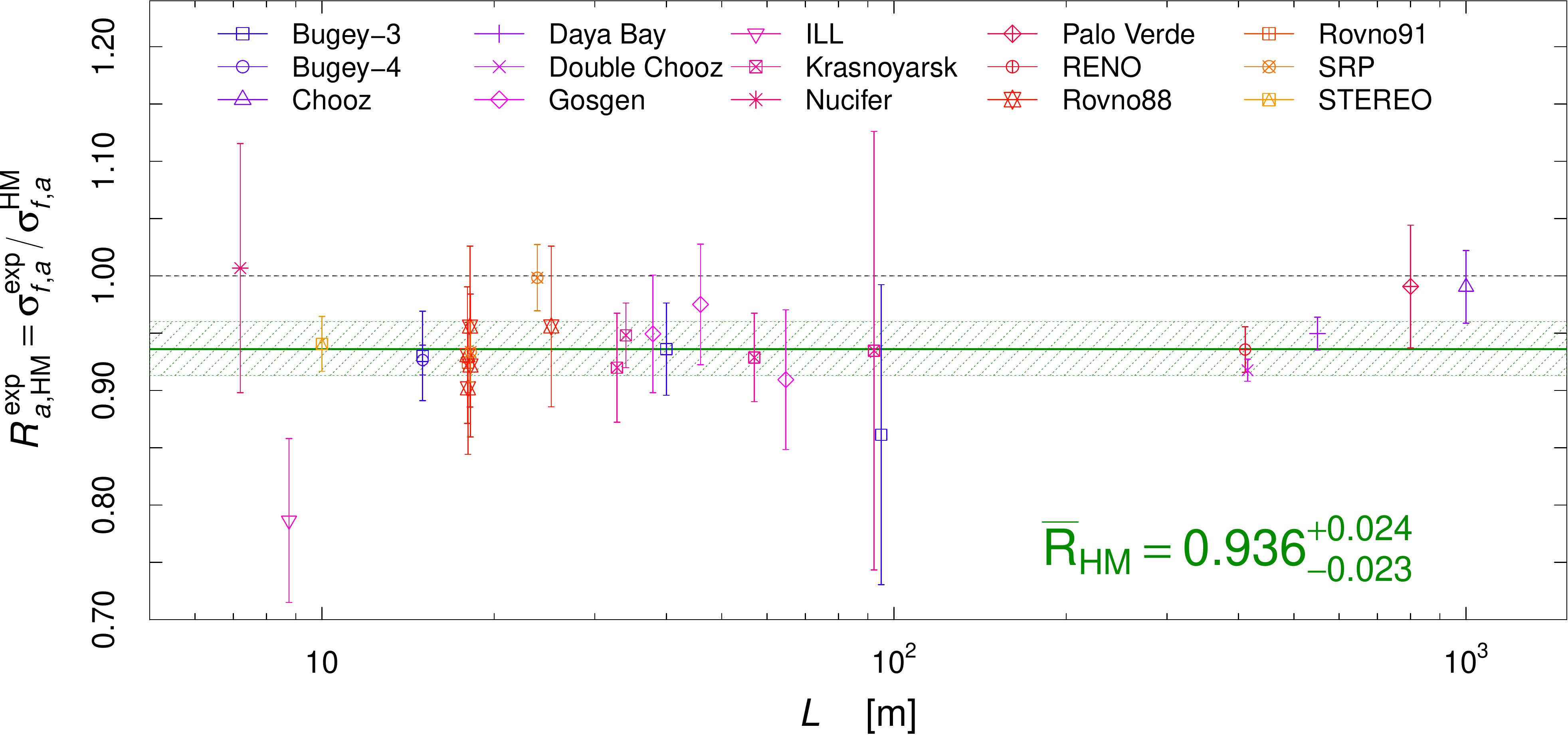}
\caption{ \label{fig.ratio_HM}
Ratio $R_{a,\text{HM}}^{\text{exp}}$ of measured and expected IBD yields
for the reactor experiments considered in our analysis as
a function of the reactor-detector distance $L$ for
the HM model.
The error bars show the experimental uncertainties.
The horizontal green band shows the average ratio $\overline{R}_{\text{HM}}$ and its uncertainty,
that gives a $\protect\input{wrk/rea-new-HM/rave/dat/rave-sig_one.dat}\,\sigma$ RAA.
}
\end{figure*}

\begin{figure*}
\centering
\setlength{\tabcolsep}{0pt}
\begin{tabular}{cc}
\subfigure[~EF model~\protect\cite{Estienne:2019ujo}: no RAA ($\protect\input{wrk/rea-new-EF/rave/dat/rave-sig_one.dat}\,\sigma$).]{\label{fig.ratio_EF}
\begin{tabular}{c}
\includegraphics*[width=0.49\linewidth]{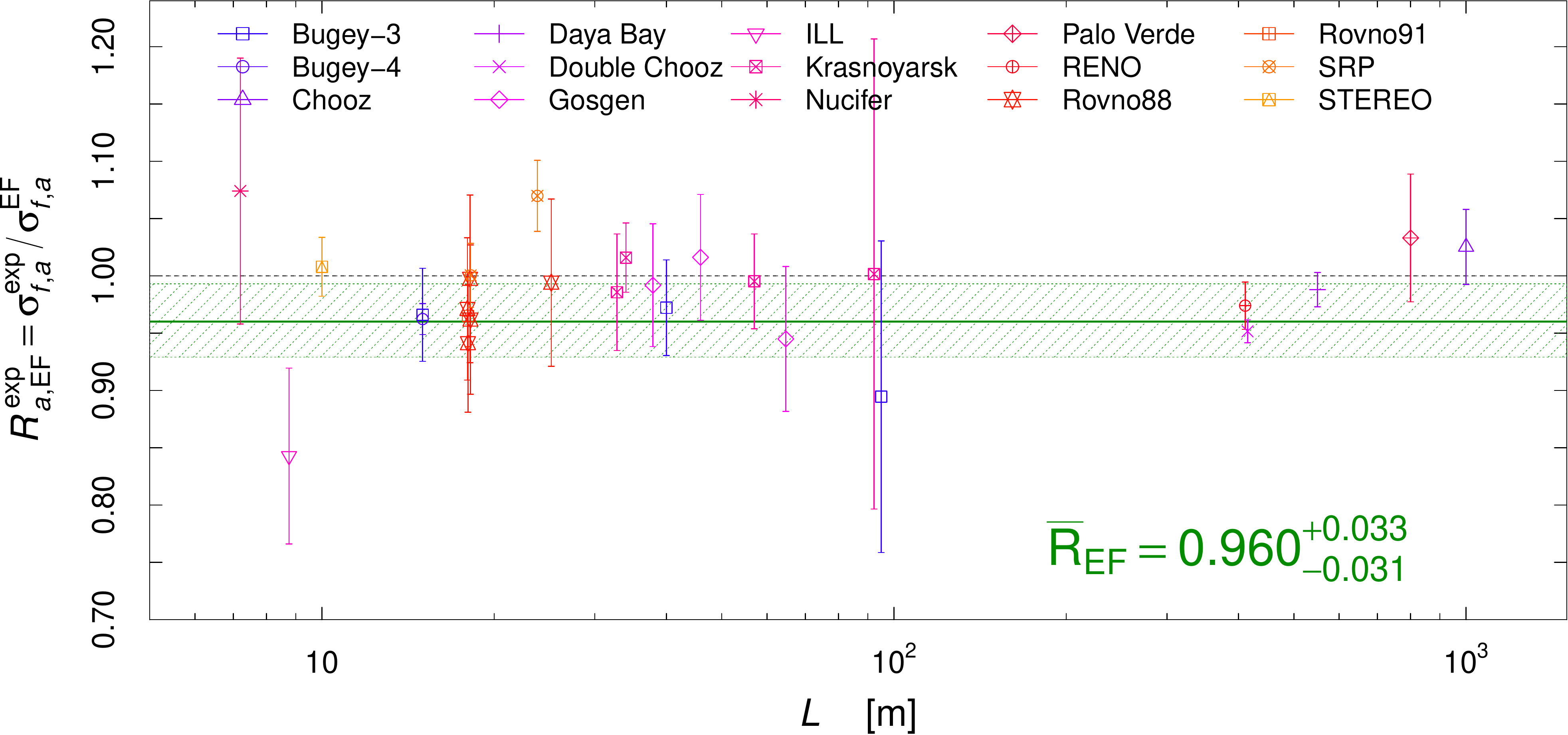}
\\
\end{tabular}
}
&
\subfigure[~HKSS model~\protect\cite{Hayen:2019eop}: RAA ($\protect\input{wrk/rea-new-HKSS/rave/dat/rave-sig_one.dat}\,\sigma$).]{\label{fig.ratio_HKSS}
\begin{tabular}{c}
\includegraphics*[width=0.49\linewidth]{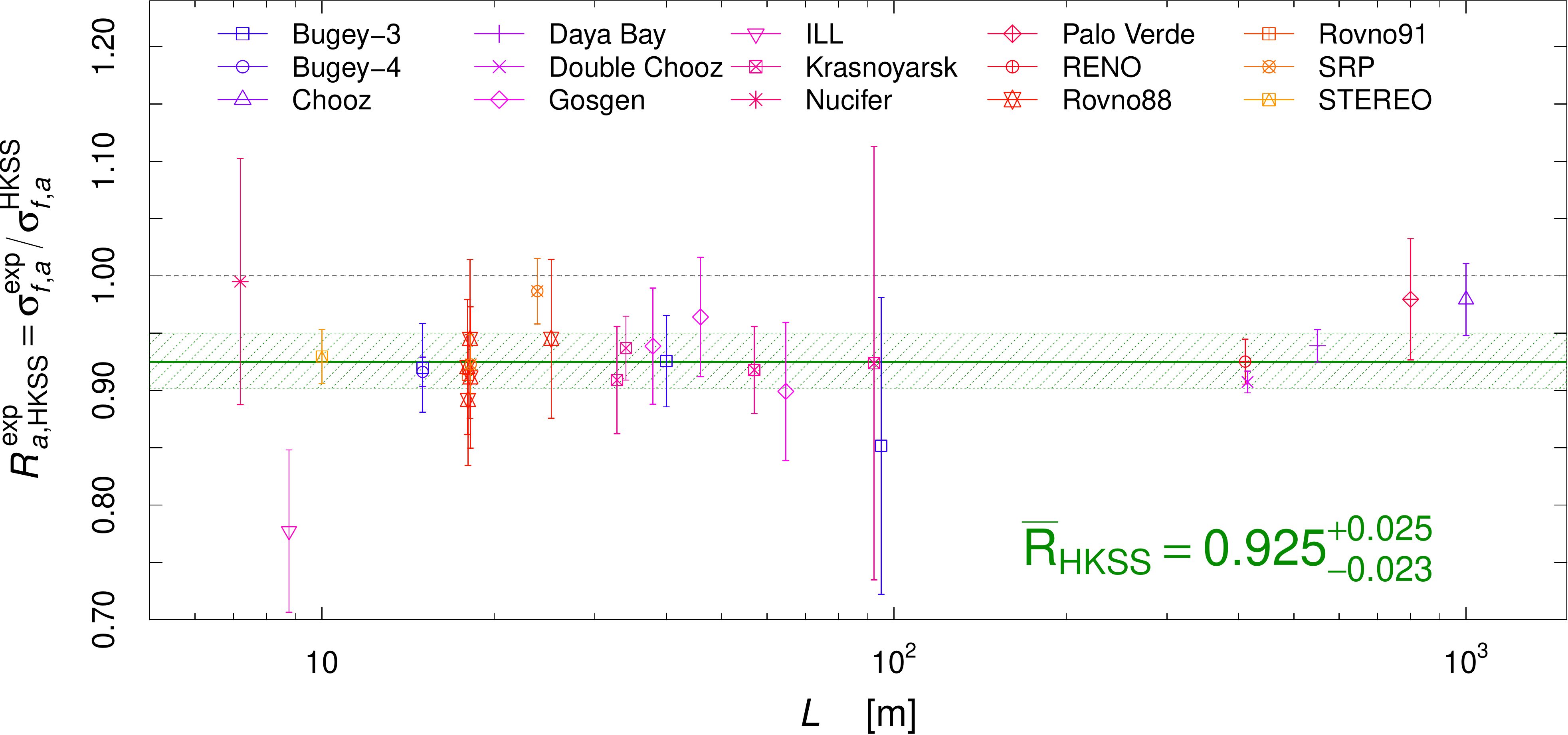}
\\
\end{tabular}
}
\\
\subfigure[~KI model~\protect\cite{Kopeikin:2021ugh}: no RAA ($\protect\input{wrk/rea-new-KI/rave/dat/rave-sig_one.dat}\,\sigma$).]{\label{fig.ratio_KI}
\begin{tabular}{c}
\includegraphics*[width=0.49\linewidth]{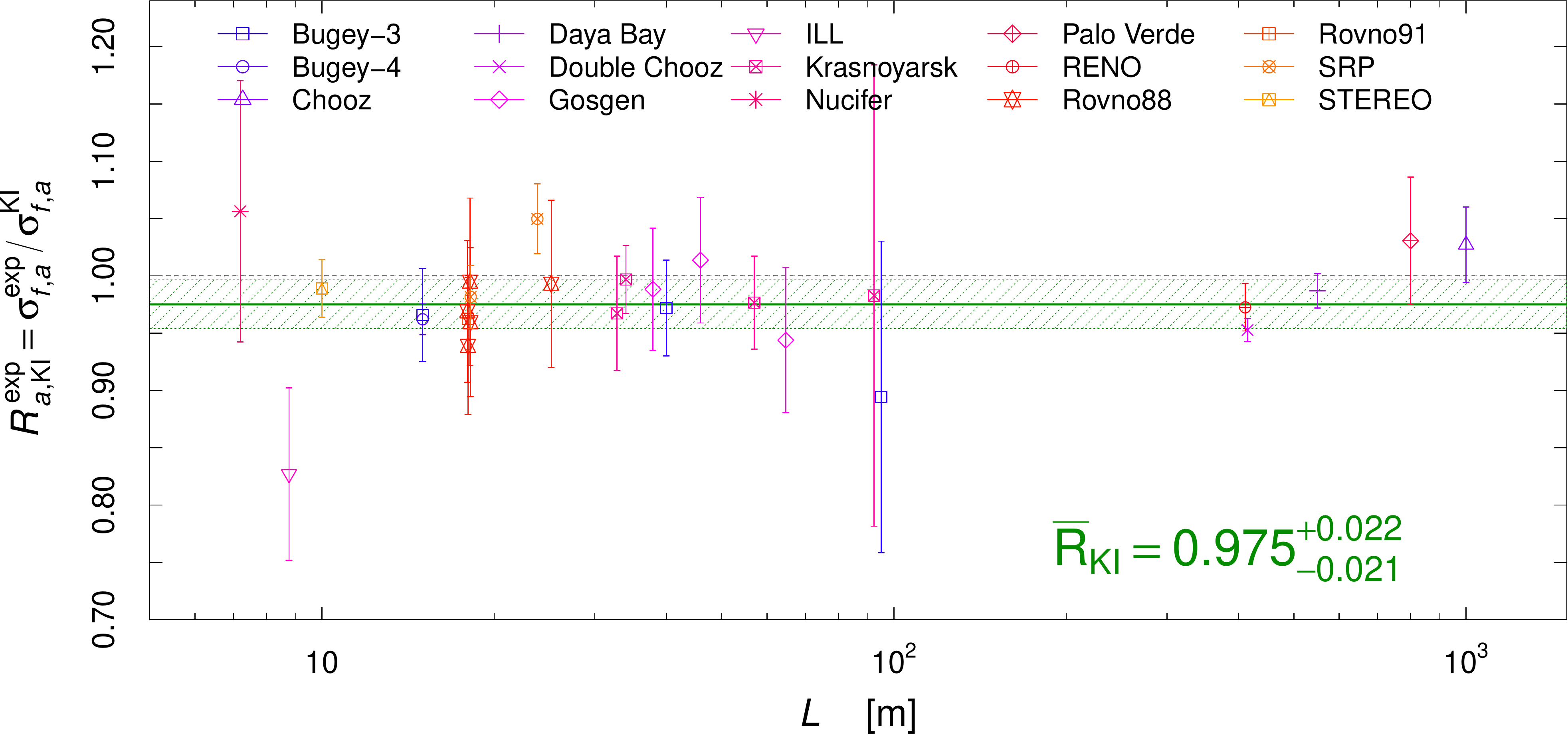}
\\
\end{tabular}
}
&
\subfigure[~HKSS-KI model: no RAA ($\protect\input{wrk/rea-new-HKSS-KI/rave/dat/rave-sig_one.dat}\,\sigma$).]{\label{fig.ratio_HKSS_KI}
\begin{tabular}{c}
\includegraphics*[width=0.49\linewidth]{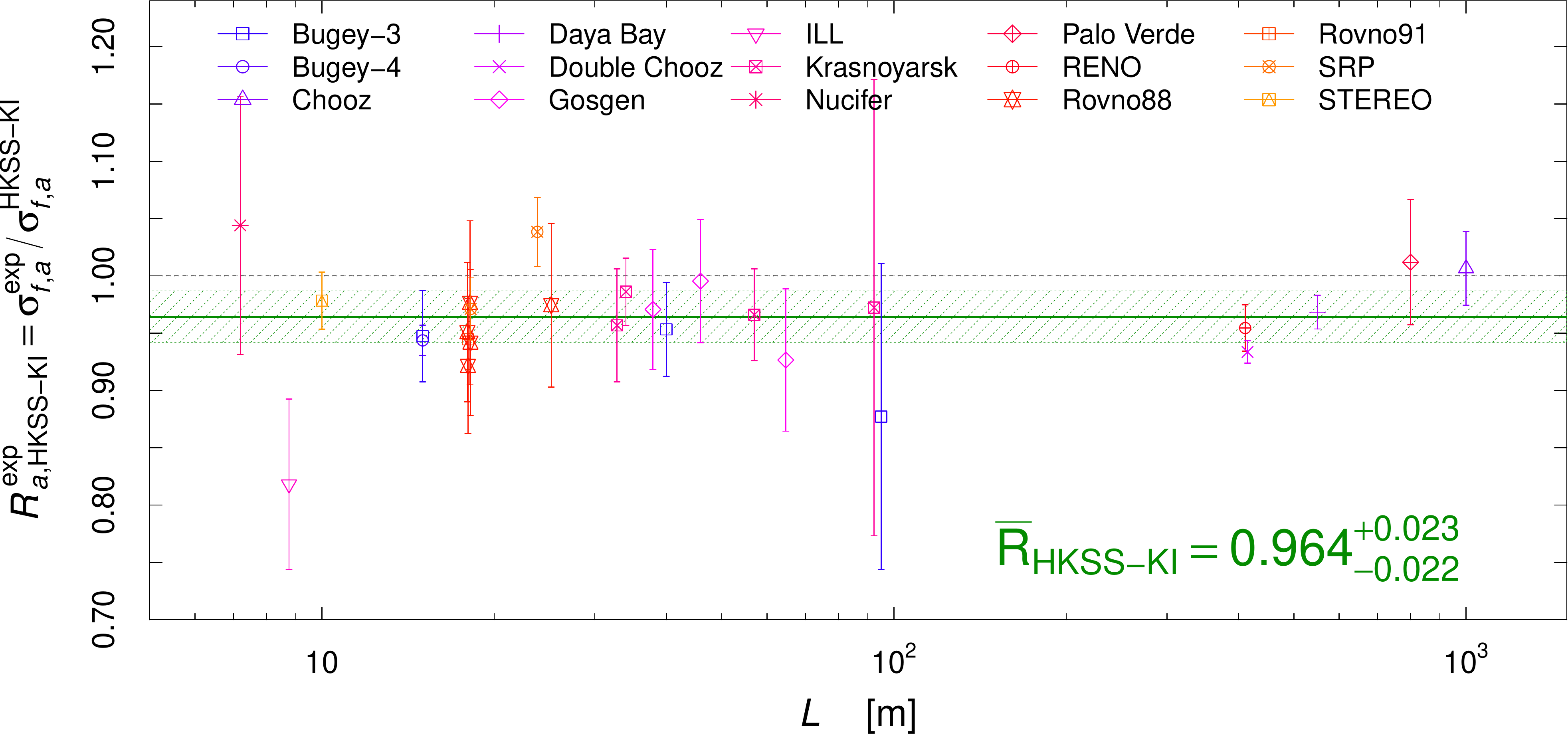}
\\
\end{tabular}
}
\end{tabular}
\caption{ \label{fig.ratios}
Ratio of measured and expected IBD yields
for the reactor experiments considered in our analysis as
a function of the reactor-detector distance $L$ for
the EF, HKSS, KI, and HKSS-KI models.
The error bars show the experimental uncertainties.
The horizontal green bands show the average ratio $\overline{R}$ and its uncertainty
that we obtained for each model.
}
\end{figure*}

\begin{figure*}
\centering
\setlength{\tabcolsep}{0pt}
\begin{tabular}{cc}
\subfigure[]{\label{fig.daya-lin-sig}
\begin{tabular}{c}
\includegraphics*[width=0.49\linewidth]{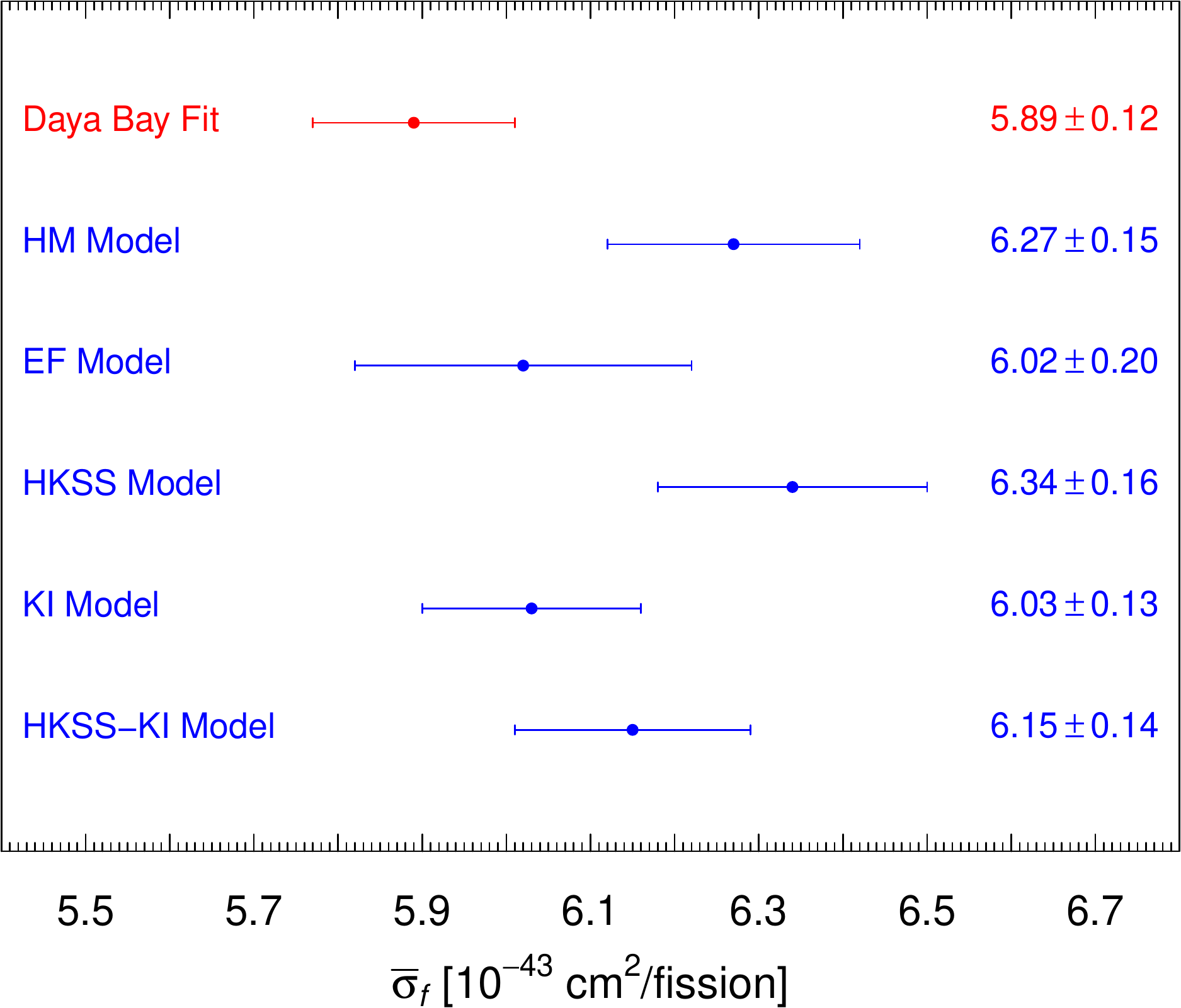}
\\
\end{tabular}
}
&
\subfigure[]{\label{fig.daya-lin-der}
\begin{tabular}{c}
\includegraphics*[width=0.49\linewidth]{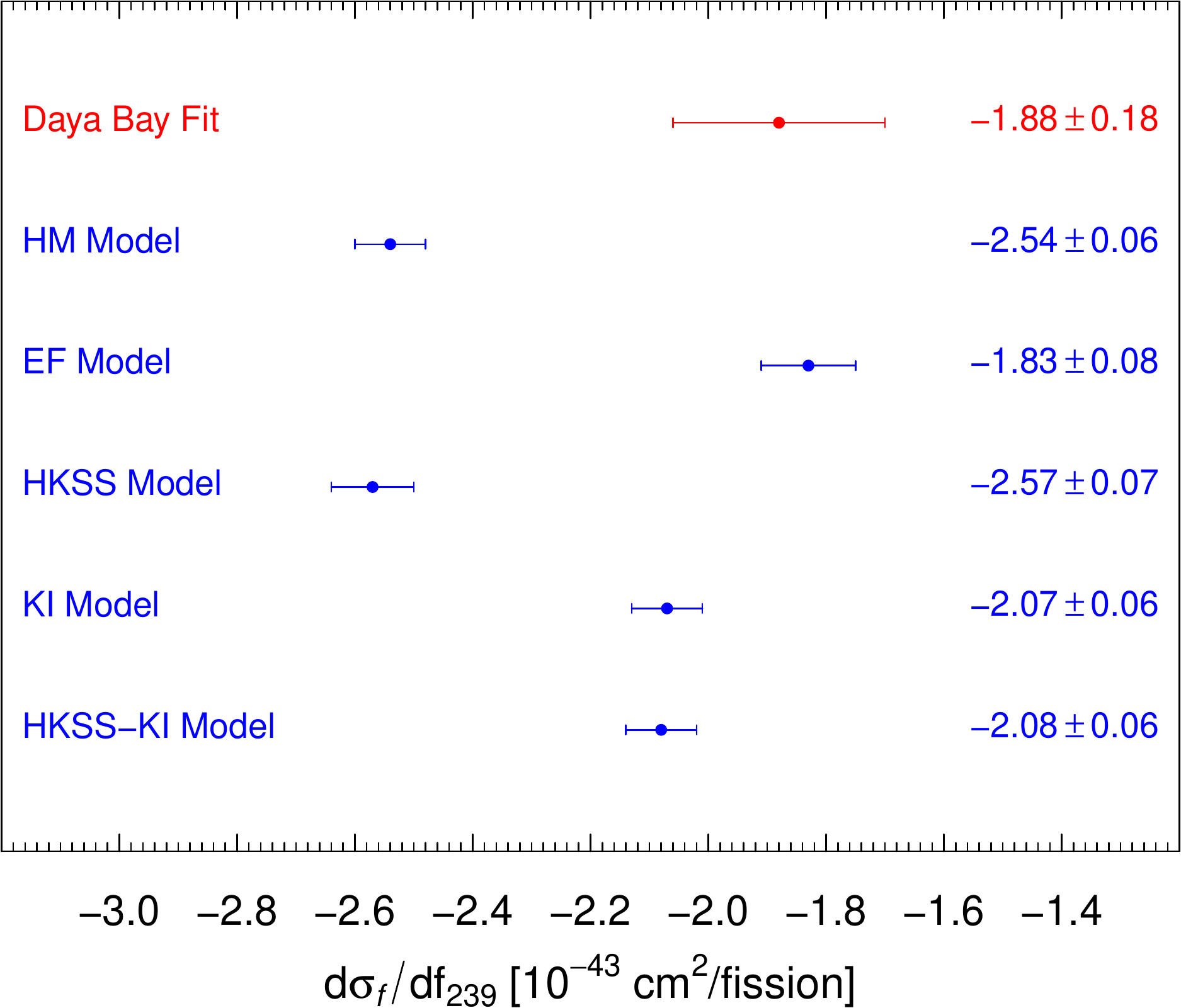}
\\
\end{tabular}
}
\\
\subfigure[]{\label{fig.reno-lin-sig}
\begin{tabular}{c}
\includegraphics*[width=0.49\linewidth]{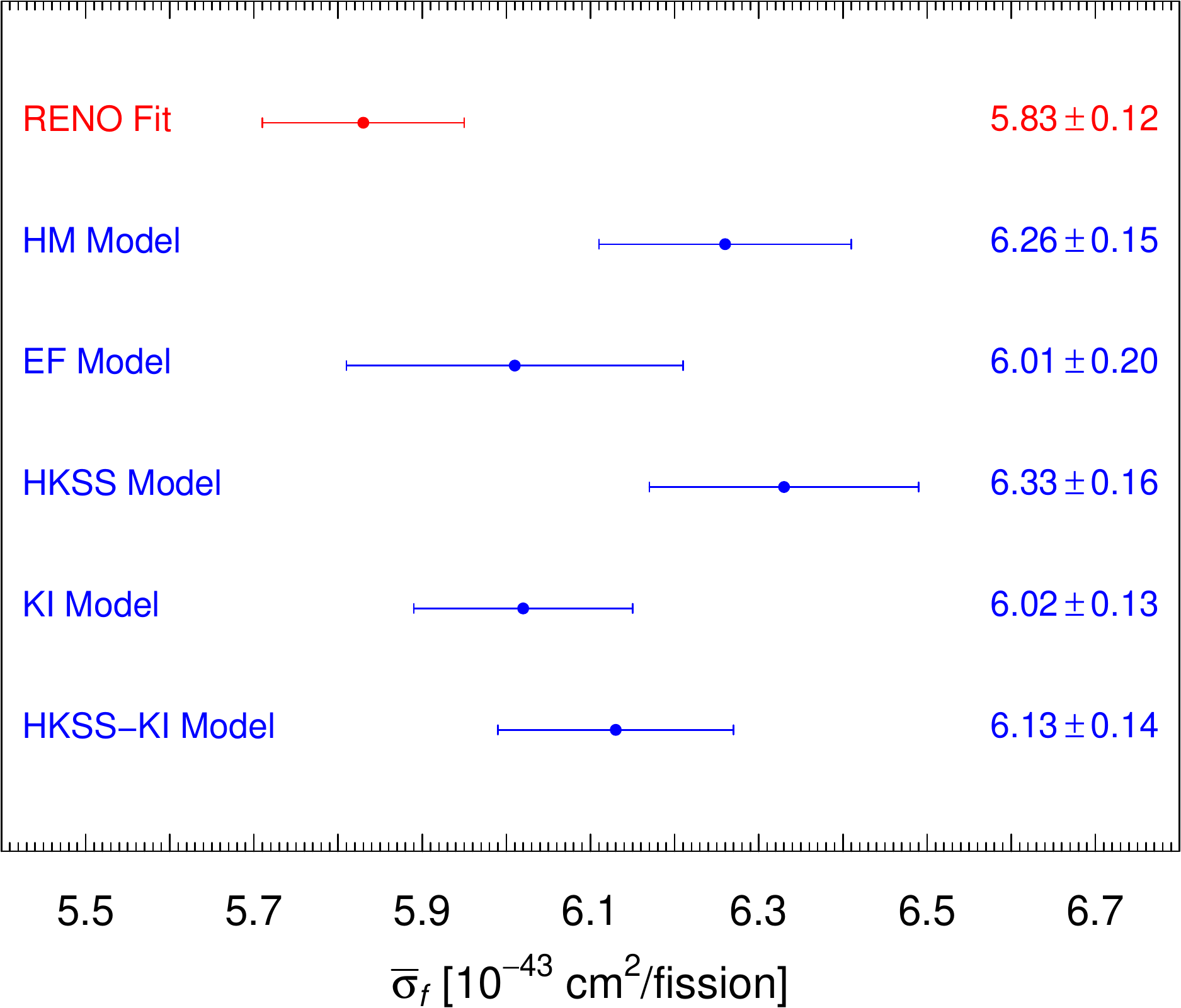}
\\
\end{tabular}
}
&
\subfigure[]{\label{fig.reno-lin-der}
\begin{tabular}{c}
\includegraphics*[width=0.49\linewidth]{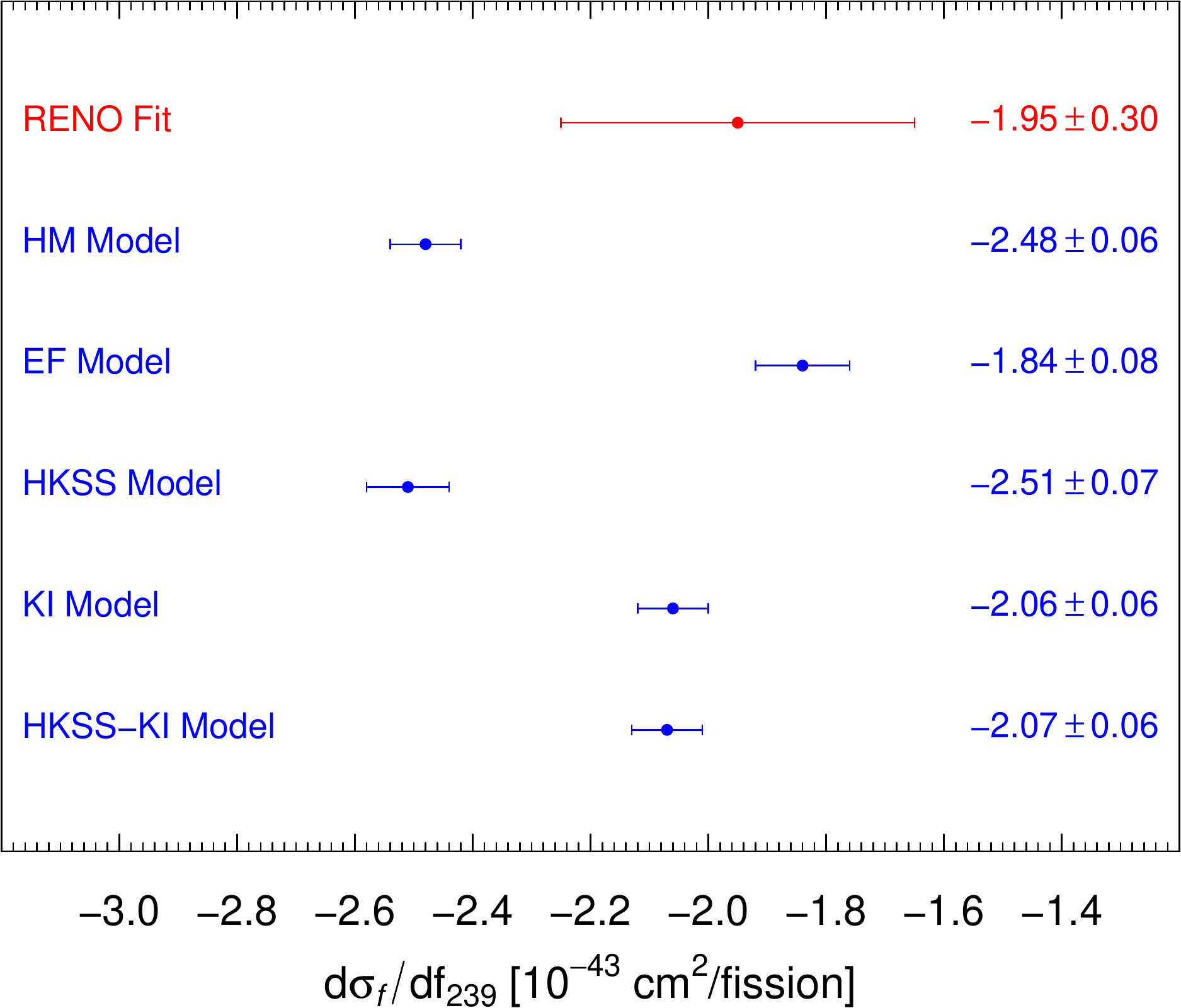}
\\
\end{tabular}
}
\end{tabular}
\caption{\label{fig.evo-lin}
Red:
results of the linear fits of
\subref{fig.daya-lin-sig}+\subref{fig.daya-lin-der}
the Daya Bay~\cite{DayaBay:2017jkb} evolution data
and
\subref{fig.reno-lin-sig}+\subref{fig.reno-lin-der}
the RENO~\cite{RENO:2018pwo} evolution data.
Blue: the corresponding
predictions of the five models in Table~\ref{tab.model_2020}.
}
\end{figure*}

%%%%%%%%%%%%%%%%%%%%%%%%%%%%%%%%%%%%%%%%%%%%%%%%%%%%%%%%
%%%%%%%%%%%%%%%%%%%%%%%%%%%%%%%%%%%%%%%%%%%%%%%%%%%%%%%%

Let us discuss first our results for the HM model,
that can be considered as the benchmark model,
since it is the original model that gave rise to the reactor antineutrino anomaly.
Figure~\ref{fig.ratio_HM}
shows the ratios of measured and expected rates
for the reactor experiments considered in our analysis
and the result for the average ratio
$\overline{R}_{\text{HM}}$
obtained with the HM model:
$ \overline{R}_{\text{HM}} = 0.936 {}^{ + 0.024 }_{ - 0.023 }
 $.
This value of $\overline{R}_{\text{HM}}$ shows that
the HM model implies a reactor antineutrino anomaly with a statistical significance of about
$ 2.5\,\sigma $.

The statistical significance of the HM RAA that we have obtained is similar to that obtained using
the original HM IBD yields in Table~\ref{tab.model},
which give
$ \overline{R}_{\text{HM}}^{\text{orig}} = 0.943 {}^{ + 0.021 }_{ - 0.020 }
 $,
corresponding to a
$ 2.6\,\sigma $
RAA.

The $ 2.5\,\sigma $
that we have obtained for the statistical significance of the HM RAA
is smaller than that obtained using the least-squares function
constructed with the method \textbf{(A)} described in Section~\ref{sec.method},
which gives
$ \overline{R}_{\text{HM}}^{\text{(A)}} = 0.922 \pm 0.024
 $,
corresponding to a
$ 3.3\,\sigma $
RAA.
The smallness of $ \overline{R}_{\text{HM}}^{\text{(A)}} $
is a manifestation of the PPP problem discussed in Section~\ref{sec.method}:
the average ratio is smaller than most of the experimental ratios
and
generates a misleading larger value of the statistical significance of the HM RAA.

Let us now consider the four new models EF, HKSS, KI, and HKSS-KI
discussed in Section~\ref{sec.model}.
The four panels in Figure~\ref{fig.ratios}
show the ratios of measured and expected rates
for the reactor experiments considered in our analysis
and the result for the average ratio
in each model.

Figure~\ref{fig.ratio_EF}
shows that with the EF calculation of the reactor antineutrino fluxes
there is practically no RAA,
since $ \overline{R}_{\text{EF}} $ differs from unity by only
$1.2\,\sigma$.
This difference is almost equal to that
%$\input{wrk/rea-org-EF/rave/dat/rave-sig_one.dat}\,\sigma$
given by the original EF IBD yields in Table~\ref{tab.model}.
Note that using the least-squares function
constructed with the method \textbf{(A)} described in Section~\ref{sec.method}
one would get a very misleading result:
$ \overline{R}_{\text{EF}}^{\text{(A)}} = 0.911 \pm 0.023
 $,
corresponding to a spurious indication of a
$ 3.9\,\sigma $
RAA.
This strong effect of PPP is due to the large uncertainties of the EF fluxes
that generate a strongly correlated covariance matrix
when they are added in quadrature to the experimental uncertainties
according to the prescription of method \textbf{(A)}.

On the other hand,
Figure~\ref{fig.ratio_HKSS} shows that with the HKSS model we obtain
$ \overline{R}_{\text{HKSS}} = 0.925 {}^{ + 0.025 }_{ - 0.023 }
 $,
corresponding to a
$ 2.9\,\sigma $ RAA,
that is larger than the
$ 2.5\,\sigma $ RAA obtained with the HM model.
Also in this case the method \textbf{(A)} gives an unbalanced average value
$ \overline{R}_{\text{HKSS}}^{\text{(A)}} = 0.910 \pm 0.024
 $,
corresponding to an excessive
$ 3.7\,\sigma $ RAA.

Figure~\ref{fig.ratio_KI}
shows that the reduced KI $\U$ IBD yield leads to the practical disappearance of the RAA,
since $ \overline{R}_{\text{KI}} $ differs from unity by only
$1.1\,\sigma$.
From Figure~\ref{fig.ratio_HKSS_KI}
one can see that
the addition of the HKSS corrections due to forbidden $\beta$ decays
slightly increases the RAA to
$1.5\,\sigma$.
Also for these two models the estimation of the RAA would be misleading
by adopting the least-squares method \textbf{(A)},
that gives the unbalanced ratios
$ \overline{R}_{\text{KI}}^{\text{(A)}} = 0.962 \pm 0.022
 $
($1.8\,\sigma$)
and
$ \overline{R}_{\text{HKSS-KI}}^{\text{(A)}} = 0.950 \pm 0.022
 $
($2.2\,\sigma$).

Summarizing the results of this Section,
we have shown in a quantitative way that the HM and HKSS conversion models of the reactor antineutrino fluxes
give a reactor antineutrino anomaly that is, however, not very large:
$\protect 2.5\,\sigma$
and
$\protect 2.9\,\sigma$,
respectively.
On the other hand,
we do not have a significant anomaly if we assume the EF reactor antineutrino fluxes
obtained from the nuclear databases with the summation method.
Also the KI reduction of the $\U$ IBD yield obtained with the conversion method
leads to the practical disappearance of the reactor antineutrino anomaly,
especially without the HKSS corrections.
Therefore, there is an approximate agreement between the results of the summation method implemented in the EF model
and the conversion method with the KI reduction of the $\U$ IBD yield.
There is, however, some concern regarding the problem of the 5 MeV bump,
that is not fitted by the EF summation spectra and the KI conversion spectra.
Since the HKSS corrections give only a partial explanation of the 5 MeV bump,
the problem of the calculation of the reactor antineutrino fluxes needs further studies.
We also think that the KI reduction of the $\U$ IBD yield
should be checked at least by an independent experiment.

%%%%%%%%%%%%%%%%%%%%%%%%%%%%%%%%%%%%%%%%%%%%%%%%%%%%%%%%
%%%%%%%%%%%%%%%%%%%%%%%%%%%%%%%%%%%%%%%%%%%%%%%%%%%%%%%%

\section{Fit of reactor fuel evolution data}
\label{sec.evolution}

The Daya Bay~\cite{DayaBay:2017jkb} and RENO~\cite{RENO:2018pwo}
collaborations have published measurements of the IBD yield during the evolution of the reactor fuel in multiple cycles.
These data provide important information on the reactor fluxes and can constrain reactor flux models and new physics~\cite{Giunti:2017nww,Hayes:2017res,Giunti:2017yid,Gebre:2017vmm,Giunti:2019qlt,Berryman:2020agd}.

To compare the flux evolution data with the different model predictions,
we first fit the evolution data 
with a linear function describing the IBD yield as a function of $f_{239}$,
as done by the Daya Bay~\cite{DayaBay:2017jkb} and RENO~\cite{RENO:2018pwo}
collaborations:
\begin{equation}
\sigma_{f,a}^{\text{lin}}
=
\bar{\sigma}_f
+
\dfrac{d \sigma_f}{d f_{239}}
\left(f_{239}^a-\bar{f}_{239} \right)
,
\label{eq.siglin}
\end{equation}
where $\overline{\sigma}_f$ is the average IBD yield
and $d \sigma_f / d f_{239}$
is the change in the IBD yield per unit change in the $\Pu$ fission fraction.
The linear approximation is justified by the approximately linear
variations of the $\U$, $\Um$, and $\Pum$ fission fractions
as function of the $\Pu$ fission fraction in
Daya Bay (see Figure~1 of Ref.~\cite{DayaBay:2017jkb})
and RENO (see Figure~1 of Ref.~\cite{RENO:2018pwo}).
The model predictions for the average IBD yields and $d \sigma_f / d f_{239}$ are obtained from the uncertainty propagation taking into account the experimental fission fraction variations and theoretical model correlations.  

We performed the linear analysis of the
Daya Bay and RENO evolution data
with the least-squares function
\begin{equation}
\chi^2_{\text{lin}}
=
\sum_{a,b}
\left( \sigma_{f,a}^{\text{exp}} - \sigma_{f,a}^{\text{lin}} \right)
\left(V^{\rm exp}\right)_{ab}^{-1}
\left( \sigma_{f,b}^{\text{exp}} - \sigma_{f,b}^{\text{lin}} \right)
,
\label{eq.chi2lin}
\end{equation}
where $V^{\rm exp}$
is the experimental covariance matrix
with statistical and systematic uncertainties added in quadrature.
The results are shown in Figure~\ref{fig.evo-lin},
in comparison with the predictions of the five models in Table~\ref{tab.model_2020}.

One can see from Figure~\ref{fig.evo-lin}
that the HM and HKSS models give values of
$\bar{\sigma}_f$
and
$d \sigma_f / d f_{239}$
that are in tension with the linear fits of the evolution data.
The discrepancies in $\bar{\sigma}_f$ are equivalent to the
average suppressions of the Daya Bay and RENO IBD yields
in the reactor antineutrino anomaly.
The discrepancy of $d \sigma_f / d f_{239}$
is an additional information that tells us that there are different suppressions
of the $\U$, $\Um$, $\Pu$ and $\Pum$ fluxes with respect to those
predicted by the HM and HKSS models.
The determination of the different suppressions has been discussed in several
papers~\cite{DayaBay:2017jkb,Giunti:2017nww,Hayes:2017res,Giunti:2017yid,Gebre:2017vmm,RENO:2018pwo,Giunti:2019qlt,Berryman:2020agd}.
Here we note that the $d \sigma_f / d f_{239}$ tension for the HM and HKSS models
is especially strong for the Daya Bay evolution data,
as shown in Figure~\ref{fig.daya-lin-der}:
$3.5\,\sigma$
for the HM model
and
$3.6\,\sigma$
for the HKSS model.
Therefore,
the HM and HKSS models are disfavored by the evolution data.

On the other hand,
Figure~\ref{fig.evo-lin} shows that the
EF, KI and HKSS-KI models give values of
$\bar{\sigma}_f$
and
$d \sigma_f / d f_{239}$
that agree with the fit of the evolution data within the uncertainties.
Therefore these models are preferred by the evolution data not only for the
absence of the reactor antineutrino anomaly,
but also for a better fit of the slope $d \sigma_f / d f_{239}$.

It is also interesting to fit the evolution data with the
least-squares function in Eq.~\eqref{genchi}
and a constant new physics suppression $R_{\text{NP}}^{a}=\overline{R}$
as we have done in Section~\ref{sec.rates} for the reactor rates.
Table~\ref{tab.Rave} shows the results for the five models under consideration,
by comparing the values of the fits of the reactor rates, the evolution data
and the combined rates and evolution data.
The combined fit is done by considering the
Daya Bay and RENO evolution data
instead of the corresponding rates in Table~\ref{tab.rates}.
One can see that the inclusion in the analysis of the
evolution data confirms the existence of a reactor antineutrino anomaly
for the HM
($2.8\,\sigma$)
and HKSS
($3.0\,\sigma$)
models
and the absence of the anomaly for the EF model
(only $0.8\,\sigma$).
This is due to the good fit of the evolution data by the EF model,
that has the smallest $\U$ IBD yield,
with a ratio of about 0.940 with respect to that in the original HM model
in Table~\ref{tab.model}.
This ratio is close to the best-fit ratio of
$ 0.925 \pm 0.015 $~\cite{Giunti:2019qlt},
obtained assuming the HM model for the other IBD yields,
which, however, are not very different in the EF and original HM model
(in the EF model, for the other important $\Pu$ IBD yield there is only the very small increase of 0.5\% with respect to the original HM model).

The KI and HKSS-KI models do not fit the evolution data as well as the EF model,
because they have larger $\U$ IBD yields and similar $\Pu$ IBD yields.
Therefore,
the inclusion in the analysis of the evolution data
increases the differences of
$ \overline{R}_{\text{KI}} $
and
$ \overline{R}_{\text{HKSS-KI}} $
from unity.
However, the resulting
$1.4\,\sigma$
for the KI model
and
$1.8\,\sigma$
for the HKSS-KI model
are still too small to claim an anomaly.

\section{Best-fit reactor flux model}
\label{sec.bestfit}

It is useful to establish which of the five models in Table~\ref{tab.model_2020}
provides the best fit of the reactor rates and the evolution data.
We have already seen in Sections~\ref{sec.rates} and~\ref{sec.evolution}
that the predictions of the HM and HKSS models
are in tension with the reactor rates and with
the Daya Bay and RENO evolution data.
The EF models appears to be favorite as the best-fit model,
with a difference of
$ \overline{R}_{\text{EF}} $ from unity
of only
$0.8\,\sigma$
considering the reactor rates and evolution data.
Among the conversion models the KI model
seems to be favorite,
with a difference of
$ \overline{R}_{\text{KI}} $ from unity
of
$1.4\,\sigma$,
and the addition of the HKSS corrections are disfavored by the
$1.8\,\sigma$
difference of
$ \overline{R}_{\text{HKSS-KI}} $ from unity.
In this section we apply goodness of fit tests to the reactor rates and evolution data
with the aims to check these indications and to select the best-fit model.

If the fluctuations of the data with respect to the model prediction are Gaussian,
the probability distribution function is given by
\begin{equation}
p(\sigma_{f,1}^{\text{exp}}, \ldots, \sigma_{f,N}^{\text{exp}})
=
\dfrac{ e^{ - \chi^2_{\text{tot}} / 2 } }{ \sqrt{ (2\pi)^N |V^{\text{tot}}| } }
,
\label{gauspdf}
\end{equation}
with
\begin{equation}
\chi^2_{\text{tot}}
=
\sum_{a,b}
\left( \sigma_{f,a}^{\text{exp}} - \sigma_{f,a}^{\text{mod}} \right)
\left( V^{\text{tot}} \right)^{-1}_{ab}
\left( \sigma_{f,b}^{\text{exp}} - \sigma_{f,b}^{\text{mod}} \right)
,
\label{chi2tot}
\end{equation}
where $N$ is the number of data points
and
$V^{\text{tot}}$ is the total covariance matrix
with experimental and theoretical uncertainties added in quadrature.
Note that Eq.~\eqref{chi2tot} considered as a least-squares function
would correspond to method \textbf{(A)}
discussed in Section~\ref{sec.method},
which suffers of the PPP problem in the determination of the average
suppression of the data with respect to the model predictions.
Since here we discuss the goodness of fit of the data for fixed model predictions,
there is no PPP problem and it is appropriate to consider
the Gaussian probability distribution function
in Eqs.~\eqref{gauspdf} and~\eqref{chi2tot}.

Table~\ref{tab.gof}
shows that the standard $\chi^2$ goodness of fit test is not able to reject
any of the five models under consideration if we consider the usual minimum $p$-value
of 5\% corresponding to a confidence level of 95\%.
However,
if the distribution of the data is Gaussian, there should be an approximately equal number of positive and negative deviations of the data from the model predictions.
This characteristic is not taken into account
by the standard $\chi^2$ goodness of fit test,
that is not sensitive to the signs of the deviations of the data
with respect to the model predictions.
From Figures~\ref{fig.ratio_HM} and \ref{fig.ratio_HKSS}
one can see that most of the reactor rates are smaller than the corresponding
predictions of the HM and HKSS models
(corresponding to $R_{a,\text{mod}}^{\text{exp}}=1$),
whereas
Figures~\ref{fig.ratio_EF}, \ref{fig.ratio_KI}, and \ref{fig.ratio_HKSS_KI}
show that the signs of the fluctuations of the reactor rates are
more balanced for the EF, KI, and HKSS-KI models,
especially for the EF model.
Therefore statistical tests that are sensitive to the signs of the deviations
should disfavor the HM and HKSS models
with respect to the EF, KI, and HKSS-KI models
and hopefully indicate which of these models provides the best fit of the data.

In order to perform statistical tests that probe the Gaussian distribution of the data
taking into account the signs of the deviations from the model predictions,
we need to transform the data to a univariate normal distribution.
This is achieved by considering the transformed data
\begin{equation}
x_{a}^{\text{mod}}
=
\sum_{b}
\left( V^{\text{tot}} \right)^{-1/2}_{ab}
\left( \sigma_{f,b}^{\text{exp}} - \sigma_{f,b}^{\text{mod}} \right)
,
\label{xamod}
\end{equation}
that should have a Gaussian distribution with
zero mean and unit standard deviation.

We first checked with the Shapiro-Wilk test (SW)
that the transformed data $x_{a}^{\text{mod}}$
have an empirical Gaussian distribution
around the sample mean.
Table~\ref{tab.gof}
shows that the transformed data pass this test
with a $p$-value larger than 5\% 
for all the models under consideration,
except the reactor rates for the EF model,
that have a $p$-value of
$4\%$.
This is somewhat puzzling,
because the other tests discussed below
confirm the expectation that the EF model is currently the best one.

We applied the following statistical tests
that are sensitive to the sign and size of the
deviations of the transformed data
with respect to the Gaussian distribution with
zero mean and unit standard deviation:

%%%%%%%%%%%%%%%%%%%%%%%%%%%%%%%%%%%%%%%%%%%%%%%%%%%%%%%%
%%%%%%%%%%%%%%%%%%%%%%%%%%%%%%%%%%%%%%%%%%%%%%%%%%%%%%%%

\begin{table*}
\begin{minipage}{0.5\linewidth}
\centering
\begin{ruledtabular}

\begin{tabular}{cccccc}
\bf Test & \bf HM & \bf EF & \bf HKSS & \bf KI & \bf HKSS-KI
\\
\hline
&
\multicolumn{5}{c}{\bf Rates}
\\
$\bm{\chi^2}$
&
$0.21$
&
$0.08$
&
$0.12$
&
$0.43$
&
$0.21$
\\
\bf SW
&
$0.14$
&
$0.04$
&
$0.13$
&
$0.20$
&
$0.13$
\\
\bf sign
&
$0.01$
&
$0.50$
&
$<10^{-3}$
&
$0.22$
&
$0.12$
\\
\bf KS
&
$0.01$
&
$0.77$
&
$0.004$
&
$0.36$
&
$0.19$
\\
\bf CVM
&
$0.01$
&
$0.74$
&
$0.005$
&
$0.37$
&
$0.17$
\\
\bf AD
&
$0.02$
&
$0.50$
&
$0.006$
&
$0.39$
&
$0.17$
\\
$\bm{Z_{\textbf{K}}}$
&
$<10^{-3}$
&
$0.004$
&
$<10^{-3}$
&
$0.06$
&
$0.01$
\\
$\bm{Z_{\textbf{C}}}$
&
$0.01$
&
$0.02$
&
$0.005$
&
$0.41$
&
$0.06$
\\
$\bm{Z_{\textbf{A}}}$
&
$0.02$
&
$0.13$
&
$0.009$
&
$0.38$
&
$0.12$
\\
\begin{tabular}{c}
\bf weighted
\\[-0.2cm]
\bf average
\end{tabular}
&
$0.06$
&
$0.29$
&
$0.04$
&
$0.45$
&
$0.16$
\\
\hline
&
\multicolumn{5}{c}{\bf Evolution}
\\
$\bm{\chi^2}$
&
$0.08$
&
$0.86$
&
$0.05$
&
$0.86$
&
$0.71$
\\
\bf SW
&
$0.70$
&
$0.94$
&
$0.66$
&
$0.19$
&
$0.20$
\\
\bf sign
&
$0.11$
&
$0.23$
&
$0.04$
&
$0.40$
&
$0.11$
\\
\bf KS
&
$0.13$
&
$0.45$
&
$0.06$
&
$0.46$
&
$0.21$
\\
\bf CVM
&
$0.05$
&
$0.56$
&
$0.02$
&
$0.37$
&
$0.10$
\\
\bf AD
&
$0.03$
&
$0.64$
&
$0.01$
&
$0.36$
&
$0.09$
\\
$\bm{Z_{\textbf{K}}}$
&
$<10^{-3}$
&
$0.02$
&
$<10^{-3}$
&
$0.003$
&
$<10^{-3}$
\\
$\bm{Z_{\textbf{C}}}$
&
$0.03$
&
$0.54$
&
$0.01$
&
$0.19$
&
$0.08$
\\
$\bm{Z_{\textbf{A}}}$
&
$0.05$
&
$0.63$
&
$0.02$
&
$0.16$
&
$0.05$
\\
\begin{tabular}{c}
\bf weighted
\\[-0.2cm]
\bf average
\end{tabular}
&
$0.08$
&
$0.50$
&
$0.05$
&
$0.27$
&
$0.11$
\\
\hline
&
\multicolumn{5}{c}{\bf Rates + Evolution}
\\
$\bm{\chi^2}$
&
$0.13$
&
$0.22$
&
$0.08$
&
$0.68$
&
$0.44$
\\
\bf SW
&
$0.32$
&
$0.13$
&
$0.35$
&
$0.59$
&
$0.41$
\\
\bf sign
&
$0.03$
&
$0.38$
&
$0.006$
&
$0.38$
&
$0.11$
\\
\bf KS
&
$0.04$
&
$0.84$
&
$0.02$
&
$0.39$
&
$0.20$
\\
\bf CVM
&
$0.02$
&
$0.67$
&
$0.006$
&
$0.38$
&
$0.14$
\\
\bf AD
&
$0.02$
&
$0.57$
&
$0.006$
&
$0.40$
&
$0.13$
\\
$\bm{Z_{\textbf{K}}}$
&
$<10^{-3}$
&
$0.05$
&
$<10^{-3}$
&
$0.05$
&
$0.008$
\\
$\bm{Z_{\textbf{C}}}$
&
$0.02$
&
$0.11$
&
$0.005$
&
$0.55$
&
$0.15$
\\
$\bm{Z_{\textbf{A}}}$
&
$0.03$
&
$0.20$
&
$0.01$
&
$0.41$
&
$0.12$
\\
\begin{tabular}{c}
\bf weighted
\\[-0.2cm]
\bf average
\end{tabular}
&
$0.05$
&
$0.35$
&
$0.03$
&
$0.42$
&
$0.16$
\end{tabular}
\end{ruledtabular}
\caption{\label{tab.gof}
$p$-values of goodness of fit tests of the
reactor rates in Table~\ref{tab.rates}
and of the
Daya Bay~\protect\cite{DayaBay:2017jkb}
and
RENO~\protect\cite{RENO:2018pwo}
evolution data for the five models in Table~\ref{tab.model_2020}.
The tests are:
$\chi^2$,
Shapiro-Wilk (SW),
sign test (sign),
Kolmogorov-Smirnov test (KS),
Cramer-von Mises test (CVM),
Anderson-Darling test (AD), and the
$Z_{\text{K}}$,
$Z_{\text{C}}$, and
$Z_{\text{A}}$ tests~\protect\cite{Zhang-JRSSB-2002}. The weighted averages are calculated by averaging for each model the $p$-values of the different tests normalized to unity.}
\end{minipage}
\end{table*}

\begin{figure}
\centering
\setlength{\tabcolsep}{0pt}
\begin{tabular}{c}
\subfigure[Reactor rates.]{\label{fig.ecdf-rea}
\begin{tabular}{c}
\includegraphics*[width=\linewidth]{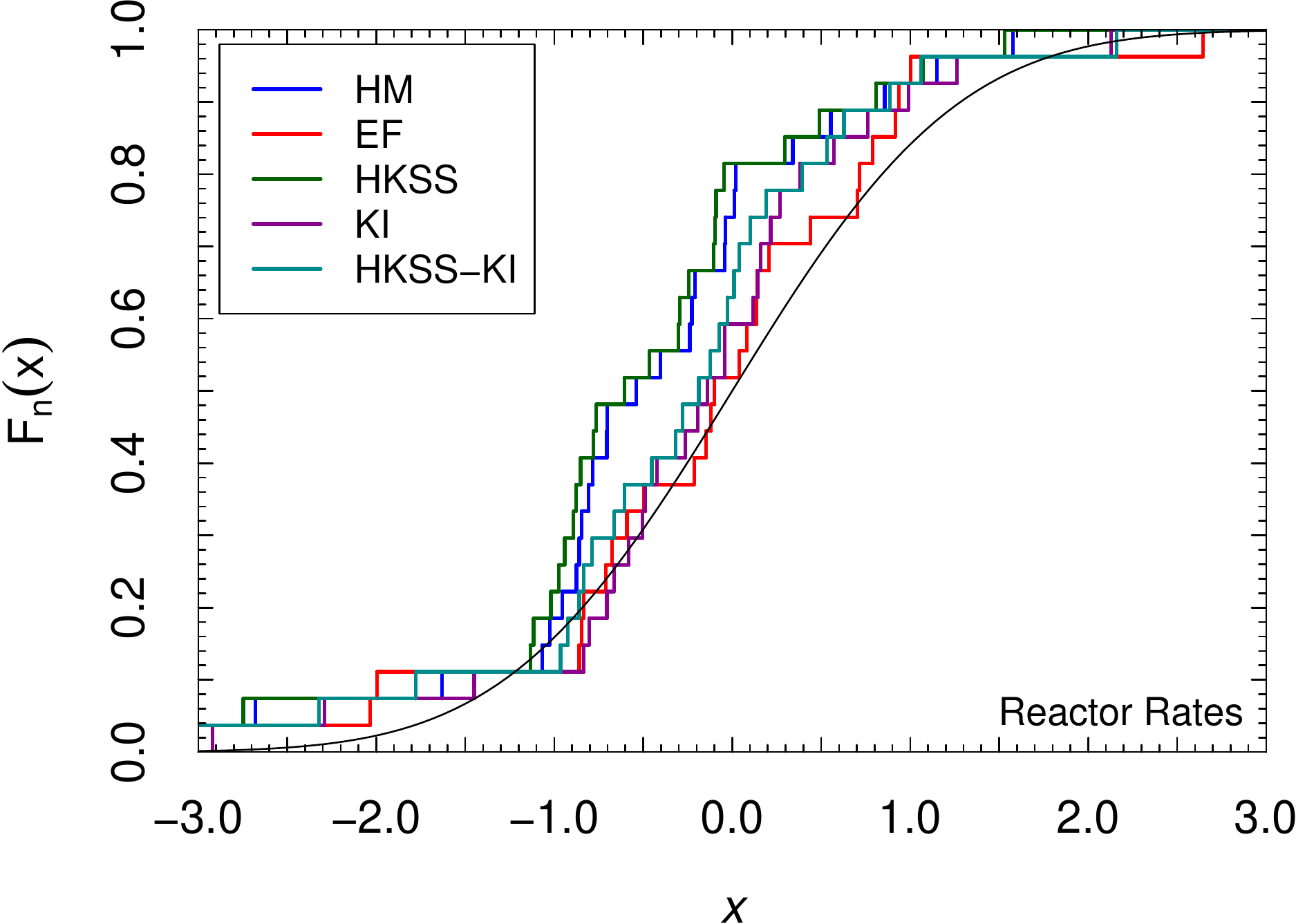}
\\
\end{tabular}
}
\\
\subfigure[Daya Bay and RENO evolution.]{\label{fig.ecdf-evo}
\begin{tabular}{c}
\includegraphics*[width=\linewidth]{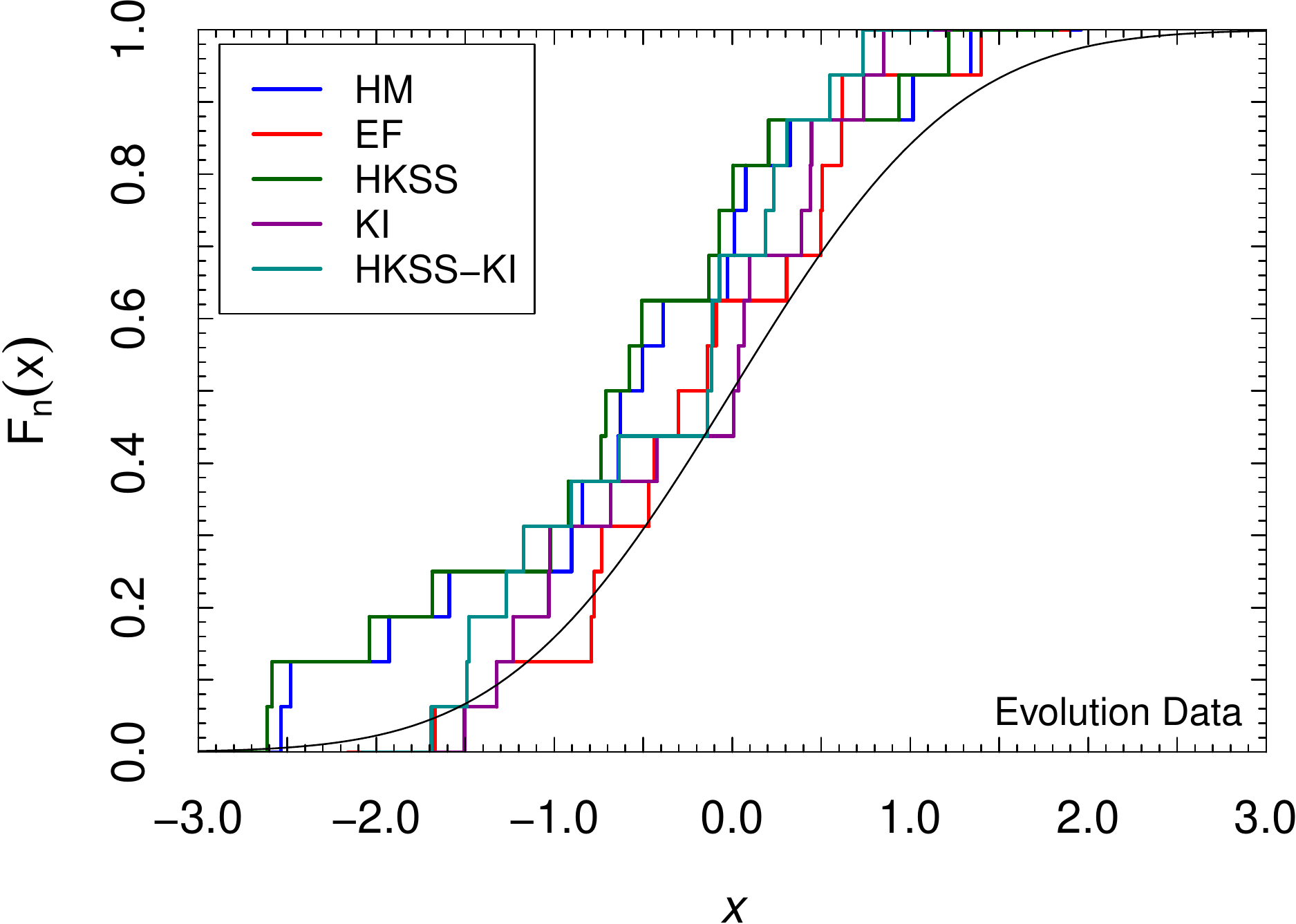}
\\
\end{tabular}
}
\\
\subfigure[Rates + evolution.]{\label{fig.ecdf-all}
\begin{tabular}{c}
\includegraphics*[width=\linewidth]{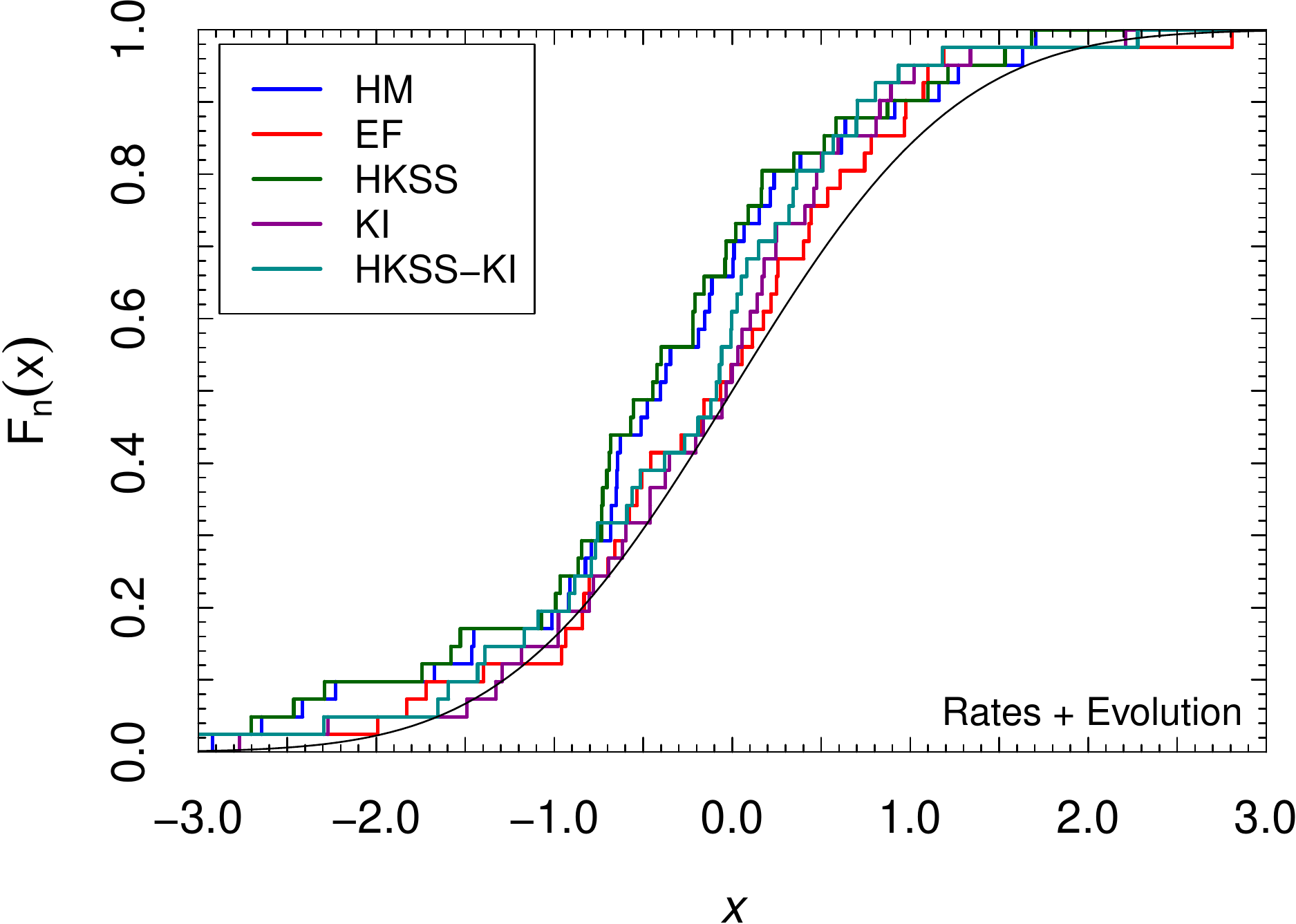}
\\
\end{tabular}
}
\end{tabular}
\caption{\label{fig.ecdf}
Empirical cumulative distribution functions
of
\subref{fig.ecdf-rea} the reactor rates in Table~\ref{tab.rates},
\subref{fig.ecdf-evo}
the
Daya Bay~\protect\cite{DayaBay:2017jkb}
and
RENO~\protect\cite{RENO:2018pwo}
evolution data, and
\subref{fig.ecdf-all}
the rates and evolution data,
assuming Gaussian distributions with respect to the
predictions of the five models in Table~\ref{tab.model_2020}.
The black curves shows the cumulative distribution function
of the Gaussian distribution with
zero mean and unit standard deviation.
}
\end{figure}

%%%%%%%%%%%%%%%%%%%%%%%%%%%%%%%%%%%%%%%%%%%%%%%%%%%%%%%%
%%%%%%%%%%%%%%%%%%%%%%%%%%%%%%%%%%%%%%%%%%%%%%%%%%%%%%%%

\begin{description}

\item[Sign test]
This classical test depends on the quantities of positive and negative
deviations of the transformed data
with respect to the Gaussian distribution with
zero mean and unit standard deviation, that should be approximately equal.
Table~\ref{tab.gof} shows that
the sign test
excludes the HM and HKSS models,
that have $p$-values smaller than 5\%
for the reactor rates
and in the combined analysis of the rates and evolution data.
The HKSS model has also a $p$-value smaller than 5\%
in the analysis of the evolution data alone.
The other models perform well under this test,
with the EF model preferred by the reactor rates,
the KI model preferred by the evolution data
and equal $p$-values for the two models
in the combined analysis of the rates and evolution data.

\item[Kolmogorov-Smirnov test (KS)]
This classical test depends on the
maximum distance between
the empirical cumulative distribution function of the sample
and
the theoretical cumulative distribution function
(see Figure~\ref{fig.ecdf}).
Since the KS test does not probe the whole differences of the two
cumulative distribution functions it is not very powerful.
However,
as one can see from Table~\ref{tab.gof},
the KS test disfavors the HM and HKSS model,
that have $p$-values smaller than 5\%
for the reactor rates
and in the combined analysis of the rates and evolution data.
The reactor rates and
the combined analysis of the rates and evolution data
favor the EF model.
However,
the EF and KI models have practically equal $p$-values
in the analysis of the the evolution data alone.

\item[Cramer-von Mises test (CVM)]
This classical test depends on the integral difference between
the empirical cumulative distribution function of the sample
and
the theoretical cumulative distribution function
(see Figure~\ref{fig.ecdf}).
Table~\ref{tab.gof} shows that,
considering $p$-values smaller than 5\%,
the CVM test
disfavors the HM and HKSS models
in the analysis of the reactor rates and
in the combined analysis of the rates and evolution data.
The evolution data alone disfavor the HKSS model,
with the HM model having the borderline $p$-value of 5\%.
The other models pass the test, but the EF model is favorite
by all the three data analyses.

\item[Anderson-Darling test (AD)]
Also this classical test depends on the integral difference between
the empirical cumulative distribution function of the sample
and
the theoretical cumulative distribution function
(see Figure~\ref{fig.ecdf}).
It differs from the CVM test by
placing more weight on observations in the tails of the distribution.
As one can see from Table~\ref{tab.gof},
this test disfavors the HM and HKSS models,
that have $p$-values smaller than 5\%,
and favor the EF model
in all the three data analyses.

\item[$Z_{\text{K}}$, $Z_{\text{C}}$, and $Z_{\text{A}}$ tests]
These tests~\cite{Zhang-JRSSB-2002}
are relatively new tests that are similar, respectively,
to the classical KS, CVM, and AD tests,
with more powerful measures,
based on the likelihood ratio,
of the difference between
the empirical cumulative distribution function of the sample
and
the theoretical cumulative distribution function.
One can see from Table~\ref{tab.gof}
that in particular the $Z_{\text{K}}$ test is very powerful
for rejecting the HM and HKSS models, that have
$p$-values smaller than 0.1\%
in all the three data analyses.
The $Z_{\text{K}}$ test disfavors also the HKSS-KI model,
with a $p$-value of 0.1\% for the reactor rates
and smaller $p$-value for the other two data analyses.
For the reactor rates,
the three tests favor the KI model,
with the EF model performing very badly in the $Z_{\text{K}}$ test
($0.4\%$ $p$-value)
and badly in the $Z_{\text{C}}$ test
($2\%$ $p$-value).
On the other hand,
for the evolution data the three tests prefer the EF model,
although it has a $Z_{\text{K}}$ $p$-value of only
$2\%$.
The $Z_{\text{K}}$ $p$-value of the KI model is worse:
$0.3\%$.
Both the EF and KI models are accepted by the three tests
in the combined analysis of the rates and evolution data,
with a preference for the KI model in the $Z_{\text{C}}$ and $Z_{\text{A}}$ tests.

\end{description}

%%%%%%%%%%%%%%%%%%%%%%%%%%%%%%%%%%%%%%%%%%%%%%%%%%%%%%%%
%%%%%%%%%%%%%%%%%%%%%%%%%%%%%%%%%%%%%%%%%%%%%%%%%%%%%%%%

\begin{figure*}
\centering
\setlength{\tabcolsep}{0pt}
\begin{tabular}{cc}
\subfigure[]{\label{fig.osc-new-rat}
\begin{tabular}{c}
\includegraphics*[width=0.49\linewidth]{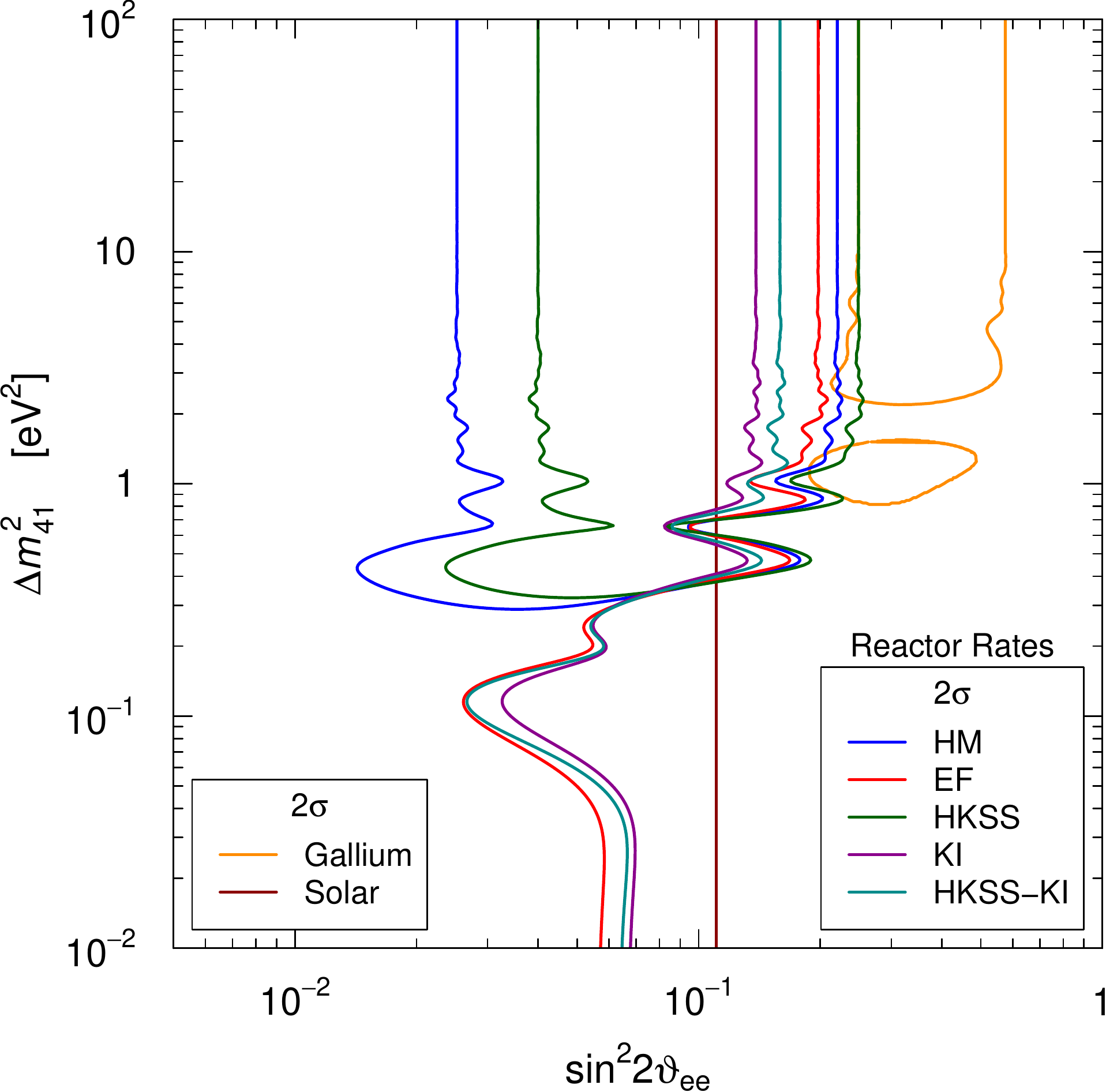}
\\
\end{tabular}
}
&
\subfigure[]{\label{fig.osc-new-rat+evo}
\begin{tabular}{c}
\includegraphics*[width=0.49\linewidth]{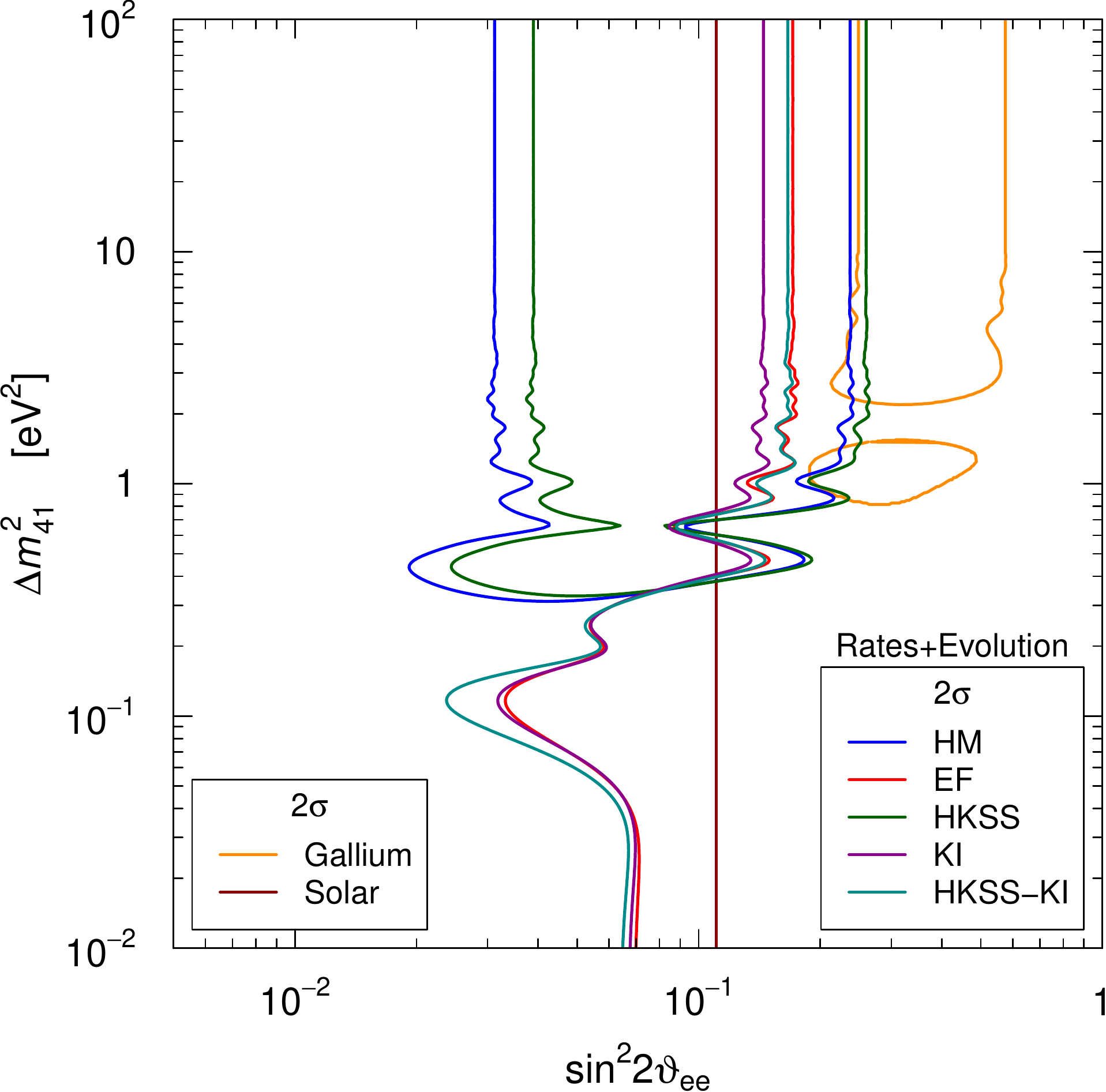}
\\
\end{tabular}
}
\end{tabular}
\caption{\label{fig.osc-rat}
Contours of the $2\sigma$ allowed regions in the
($\sin^2\!2\vartheta_{ee},\Delta{m}^2_{41}$)
plane obtained
from the neutrino oscillation fit of
\subref{fig.osc-new-rat}
the reactor rates in Table~\ref{tab.rates}
and
\subref{fig.osc-new-rat+evo}
the combined fit of the reactor rates
and the
Daya Bay~\protect\cite{DayaBay:2017jkb}
and
RENO~\protect\cite{RENO:2018pwo}
evolution data.
The
blue, red, green, magenta, and cyan curves correspond, respectively,
to the HM, EF, HKSS, KI, and HKSS-KI models in Table~\ref{tab.model_2020}.
Also shown are the contour of the $2\sigma$ allowed regions of the Gallium anomaly
obtained in Ref.~\cite{Barinov:2021asz} from the combined analysis of
the GALLEX, SAGE and BEST data (orange curve),
and the $2\sigma$ bound obtained from the analysis of solar neutrino data
in Ref.~\cite{Goldhagen:2021kxe} (dark red vertical line).
}
\end{figure*}

%%%%%%%%%%%%%%%%%%%%%%%%%%%%%%%%%%%%%%%%%%%%%%%%%%%%%%%%
%%%%%%%%%%%%%%%%%%%%%%%%%%%%%%%%%%%%%%%%%%%%%%%%%%%%%%%%

Considering the overall performance of the five models under consideration
under the goodness of fit tests that we have performed,
we can conclude that the HM and HKSS models are excluded by the data
with a high confidence level.
Note that this conclusion is supported not only by the powerful relatively new
$Z_{\text{K}}$ test, but also by the classical sign test and,
especially for the HKSS model,
by the classical KS, CVM and AD tests.
The HM and HKSS models are rejected with these classical tests
by the analysis of the reactor rates and
the combined analysis of reactor rates and evolution data.
The evolution data alone are less powerful in rejecting these models,
because they are less numerous,
they have a smaller variability of the fuel fractions, and
they have large correlated systematic uncertainties.

We have seen that among the other models,
the HKSS-KI model is not the best one for any test
and it is strongly disfavored by the $Z_{\text{K}}$ test applied
to the combined analysis of reactor rates and evolution data.

The EF and KI models are alternatively the best for some test
in each of the three data analyses.
In order to try to establish a ranking,
we consider the weighted averages of the $p$-values of the tests
for each of the three analyses in Table~\ref{tab.gof}.
These weighted averages have been obtained by averaging for each model the $p$-values of the different
tests normalized to unity.
Such normalization is necessary in order to weight equally the different tests,
including those that give small $p$-values for all the models\footnote{Without the normalization of the rows in Table~\ref{tab.gof}, the average would be dominated by the tests that give large $p$-values and the tests that give small $p$-values for all the models (as the $Z_{\text{K}}$ test) would be irrelevant.}.
Let us however emphasize that these weighted averages do not have a statistical meaning
and are only helpful for establishing an indicative ranking of the models.
From the values of the weighted averages reported in Table~\ref{tab.gof}
one can see that the KI model is favored by the reactor rates
and
the EF model is favored by the evolution data.
The combined analysis of reactor rates and evolution data
slightly favors the KI model,
with an average $p$-value of
$42\%$,
but the EF model has the almost as good average $p$-value of
$35\%$.

In conclusion of this attempt to find the best-fit reactor flux model,
since the EF and KI models are preferred by different tests
and  sets of data,
with similar overall performances in
the combined analysis of reactor rates and evolution data,
it is fair to consider both as favorite,
specifying that they have been obtained with different methods:
the EF model has been obtained entirely with the summation method,
and the KI model has been obtained entirely with the conversion method.
The strongly disfavored HM and HKSS models
can be considered in practice as conversion models,
because only the minor $\Um$ neutrino flux has been calculated with the summation method.
Since we obtained the HKSS-KI model by reducing the
$\U$ IBD yield of the HKSS model by the KI factor and
keeping the same $\Um$, $\Pu$, and $\Pum$ neutrino fluxes of the HKSS model
(see Section~\ref{sec.model}),
also the HKSS-KI model can be considered in practice as a conversion model.
Therefore,
from our analysis we conclude that
the KI model is the best among the conversion models
and
the only summation model that we considered\footnote{
We considered only the EF summation model because it is the only
recent one and it is obviously not useful
to consider previous summation models obtained
with outdated versions of the nuclear databases.
},
the EF model,
is practically equally good.

\section{Neutrino oscillations}
\label{sec.oscillations}

In this section we discuss the implications of the reactor neutrino flux models
for the neutrino oscillation analysis of the short-baseline reactor neutrino data.
We consider the effective 3+1 survival probability
of electron neutrinos and antineutrinos in Eq.~\eqref{Pee}.
However,
since the reactor neutrino measurements can probe only the disappearance of
electron antineutrinos the results apply to any model of neutrino masses and mixing that predicts an effective two-neutrino-like survival probability
of the type in Eq.~\eqref{Pee},
that depends only on an effective mixing angle $\vartheta_{ee}$
and a squared-mass difference $\Delta m_{41}^2$.

Figure~\ref{fig.osc-new-rat}
shows the contours of the $2\sigma$ allowed regions in the
($\sin^2\!2\vartheta_{ee},\Delta{m}^2_{41}$)
plane obtained from
the neutrino oscillation fit of
the reactor rates in Table~\ref{tab.rates}
considering the IBD yields of the HM, EF, HKSS, KI, and HKSS-KI models in Table~\ref{tab.model_2020}.
One can see that there is an indication in favor of neutrino oscillations
only for the HM and HKSS models
that give a significant reactor rate anomaly
above $2\sigma$, as discussed in Section~\ref{sec.rates}
(see Table~\ref{tab.Rave}).
Considering the EF, KI, and HKSS-KI models,
for which the reactor rate anomaly is smaller than $2\sigma$,
we obtained the $2\sigma$ exclusion curves shown in Figure~\ref{fig.osc-new-rat},
that allow only small values of $\sin^2 2\vartheta_{ee}$,
including $\sin^2 2\vartheta_{ee}=0$,
that corresponds to the absence of short-baseline oscillations.

For all the reactor flux models,
there are upper bounds for the value of the mixing parameter $\sin^2 2\vartheta_{ee}$
that depend on the value of $\Delta m_{41}^2$.
For $\Delta m_{41}^2 \gtrsim 2 \, \text{eV}^2$
the upper bounds for $\sin^2 2\vartheta_{ee}$
are between
$0.14$
and
$0.25$
for all the reactor flux models.

In Figure~\ref{fig.osc-new-rat} we also reproduced the
contours of the regions that are allowed at $2\sigma$
by the Gallium anomaly, according to the
combined analysis in Ref.~\cite{Barinov:2021asz}
of the new data of the BEST experiment and the old data of the
GALLEX~\cite{Kaether:2010ag}
and
SAGE~\cite{Abdurashitov:2005tb}
experiments.
One can see that the Gallium anomaly region
lies at rather large values of
$\sin^2 2\vartheta_{ee}$ and is in tension with the
reactor upper bounds on $\sin^2 2\vartheta_{ee}$
for all the reactor flux models.
The Gallium anomaly region is also in strong tension
with the more stringent $2\sigma$ solar electron neutrino
upper bound
$\sin^2 2\vartheta_{ee}<0.11$~\cite{Goldhagen:2021kxe}
shown in Figure~\ref{fig.osc-new-rat}
(see Refs.~\cite{Palazzo:2011rj,Palazzo:2012yf,Giunti:2012tn,Kopp:2013vaa,Gariazzo:2017fdh,Dentler:2018sju,Giunti:2021iti} for previous analyses that constrained $\sin^2 2\vartheta_{ee}$
using the solar electron neutrino data).

Figure~\ref{fig.osc-new-rat+evo}
shows the contours of the $2\sigma$ allowed regions in the
($\sin^2\!2\vartheta_{ee},\Delta{m}^2_{41}$)
plane obtained from
the combined neutrino oscillation fit of
the reactor rates
and the
Daya Bay~\cite{DayaBay:2017jkb}
and
RENO~\cite{RENO:2018pwo}
evolution data.
One can see that the allowed regions are similar
to those obtained in Figure~\ref{fig.osc-new-rat}
without the evolution data.
The reason is that the main information on neutrino oscillations
coming from the Daya Bay and RENO evolution data
is given by the comparison of the measured average rates with the model predictions,
that is already taken into account in the rate analysis.

Considering the EF and KI models that are favored
according to the discussion in Section~\ref{sec.bestfit}
assuming the absence of neutrino oscillations,
the $2\sigma$ upper bounds on $\sin^2 2\vartheta_{ee}$
for $\Delta m_{41}^2 \gtrsim 2 \, \text{eV}^2$
are, respectively,
$0.20$
and
$0.14$
from the rates analysis,
and
$0.17$
and
$0.14$
from the combined rates and evolution analysis.
Therefore,
there is a significant tension with the $2\sigma$ allowed region of the Gallium anomaly,
that lies above $\sin^2 2\vartheta_{ee} \approx 0.2$.

The tension between the Gallium anomaly obtained in Ref.~\cite{Barinov:2021asz}
and the exclusion curves of the
DANSS~\cite{Danilov:2020ucs},
PROSPECT~\cite{PROSPECT:2020sxr}, and
STEREO~\cite{STEREO:2019ztb},
experiments is shown in Figure~2 of Ref.~\cite{Barinov:2021mjj},
where one can see that,
however,
there is a partial compatibility for $\Delta m_{41}^2 \approx 9 \, \text{eV}^2$.
The exclusion curves of these experiments are obtained comparing
the reactor neutrino induced spectrum at different distances around 10 m.
They disappear for $\Delta m_{41}^2 \gtrsim 10 \, \text{eV}^2$.
Therefore,
these experiments cannot probe the high-$\Delta m_{41}^2$ part of the
Gallium anomaly region.
The results of our analyses of the reactor rates and the reactor evolution data
fills this gap and
extends the tension between the reactor neutrino data and the Gallium anomaly
to large values of $\Delta m_{41}^2$.

%%%%%%%%%%%%%%%%%%%%%%%%%%%%%%%%%%%%%%%%%%%%%%%%%%%%%%%%
%%%%%%%%%%%%%%%%%%%%%%%%%%%%%%%%%%%%%%%%%%%%%%%%%%%%%%%%

\section{Summary and conclusions}
\label{sec.conclusions}

In this paper we revisited the reactor antineutrino anomaly
in light of the recent reactor antineutrino flux calculations.

We have first performed, in Section~\ref{sec.model}, an improved calculation of
the IBD yields of five reactor antineutrino flux models:
the HM~\cite{Mueller:2011nm,Huber:2011wv},
HKSS~\cite{Hayen:2019eop},
KI~\cite{Kopeikin:2021ugh}, and
HKSS-KI conversion models
and the
EF~\cite{Estienne:2019ujo} summation model.
The HM model has been considered the standard one since 2011
and is the basis of the reactor antineutrino
anomaly~\cite{Mention:2011rk,Abazajian:2012ys}.
The HKSS model was proposed in 2019 as an improvement of the HM model
that gives a partial explanation of the 5 MeV bump
obtained by taking into account forbidden $\beta$ decays.
The KI model, proposed in 2021,
is based on a decrease of the $\U$ antineutrino flux
obtained with the conversion method
motivated by new relative measurements of the ratio of the $\U$ and $\Pu$ $\beta$ spectra
at the Kurchatov Institute.
The HKSS-KI model is a model that we obtained by applying to the
HKSS model the KI suppression of the $\U$ antineutrino flux.
The EF model is the most updated summation model,
that was published in 2019.
The improved IBD yields that we obtained for the EF model agree very well with the original ones~\cite{Estienne:2019ujo},
whereas those of the HM, HKSS, and KI model are slightly larger than the original ones
(see Tables~\ref{tab.model} and~\ref{tab.model_2020}).

Then, after a discussion in Section~\ref{sec.method}
of the proper statistical method for the analysis of the data,
we calculated the suppression of the reactor antineutrino flux
predicted by each of the 5 models
that is necessary for fitting the rates measured by the experiments
listed in Table~\ref{tab.rates} (Section~\ref{sec.rates}).
We found that,
the reactor antineutrino anomaly
is larger than $2\sigma$ only for the HM and HKSS models.
The difference between the data and the predictions of the EF and KI models
is slightly larger than $1\sigma$
(see Table~\ref{tab.Rave}).
It slightly increases to
$1.5\,\sigma$
for the HKSS-KI model.
Therefore there is practically no anomaly for the EF, KI, and HKSS-KI models.
The addition of the
Daya Bay~\cite{DayaBay:2017jkb} and RENO~\cite{RENO:2018pwo}
evolution data discussed in Section~\ref{sec.evolution}
confirms these conclusions,
but establish a more pronounced ranking of the models,
with the EF model having the smallest deviation from the data
($0.8\,\sigma$),
followed by the KI model
($1.4\,\sigma$), and
the HKSS-KI model
($1.8\,\sigma$).

In Section~\ref{sec.bestfit}
we further explored the question of which is the best-fit model
by applying several goodness of fit tests.
This is an important question,
because although there is an approximate agreement of the new
EF summation model and the new KI and HKSS-KI conversion models
on the demise of the reactor antineutrino anomaly,
the models are different and have different IBD yields
(see Table~\ref{tab.model_2020}).
We found that the EF and KI models are favored,
with the EF model fitting better the evolution data
and the KI model fitting better the reactor rates.
Therefore, we can consider EF as the best summation model
and KI as the best conversion model,
leaving the decision of a clear preference between the two models
to future studies with more data.

We finally discussed the implications of the five models
for short-baseline neutrino oscillations
due to active-sterile neutrino mixing.
We have shown that the five models imply upper bounds that are
not very different for the mixing
generating the oscillations.
Hence,
this is a robust constraint given by reactor antineutrino data.

The reactor bound on active-sterile neutrino mixing
agrees with the solar bound~\cite{Goldhagen:2021kxe}
and both bounds are
in tension with the large mixing~\cite{Barinov:2021asz}
required to explain the anomaly of the
GALLEX~\cite{Kaether:2010ag},
SAGE~\cite{Abdurashitov:2005tb}, and
BEST~\cite{Barinov:2021asz}
Gallium experiments with short baseline neutrino oscillations.
This is a puzzling new development in the phenomenology
of short-baseline neutrino oscillations
that may require an explanation that goes beyond
the simplest model of active-sterile neutrino mixing.

%%%%%%%%%%%%%%%%%%%%%%%%%%%%%%%%%%%%%%%%%%%%%%%%%%%%%%%%
%%%%%%%%%%%%%%%%%%%%%%%%%%%%%%%%%%%%%%%%%%%%%%%%%%%%%%%%

\begin{acknowledgments}
The work of C. Giunti and C.A. Ternes was supported by the research grant "The Dark Universe: A Synergic Multimessenger Approach" number 2017X7X85K under the program PRIN 2017 funded by the Ministero dell'Istruzione, Universit\`a e della Ricerca (MIUR).
The work of Y.F.Li and Z.Xin was supported by National Natural Science Foundation of China under Grant Nos.~12075255 and 11835013, by Beijing Natural Science Foundation under Grant No.~1192019, by the Key Research Program of the Chinese Academy of Sciences under Grant No.~XDPB15.
Y.F. Li is also grateful for the
support by the CAS Center for Excellence in Particle Physics (CCEPP).
\end{acknowledgments}

%%%%%%%%%%%%%%%%%%%%%%%%%%%%%%%%%%%%%%%%%%%%%%%%%%%%%%%%
%%%%%%%%%%%%%%%%%%%%%%%%%%%%%%%%%%%%%%%%%%%%%%%%%%%%%%%%

\bibliographystyle{apsrev4-1}
\bibliography{main}

\begin{thebibliography}{76}%
\makeatletter
\providecommand \@ifxundefined [1]{%
 \@ifx{#1\undefined}
}%
\providecommand \@ifnum [1]{%
 \ifnum #1\expandafter \@firstoftwo
 \else \expandafter \@secondoftwo
 \fi
}%
\providecommand \@ifx [1]{%
 \ifx #1\expandafter \@firstoftwo
 \else \expandafter \@secondoftwo
 \fi
}%
\providecommand \natexlab [1]{#1}%
\providecommand \enquote  [1]{``#1''}%
\providecommand \bibnamefont  [1]{#1}%
\providecommand \bibfnamefont [1]{#1}%
\providecommand \citenamefont [1]{#1}%
\providecommand \href@noop [0]{\@secondoftwo}%
\providecommand \href [0]{\begingroup \@sanitize@url \@href}%
\providecommand \@href[1]{\@@startlink{#1}\@@href}%
\providecommand \@@href[1]{\endgroup#1\@@endlink}%
\providecommand \@sanitize@url [0]{\catcode `\\12\catcode `\$12\catcode
  `\&12\catcode `\#12\catcode `\^12\catcode `\_12\catcode `\%12\relax}%
\providecommand \@@startlink[1]{}%
\providecommand \@@endlink[0]{}%
\providecommand \url  [0]{\begingroup\@sanitize@url \@url }%
\providecommand \@url [1]{\endgroup\@href {#1}{\urlprefix }}%
\providecommand \urlprefix  [0]{URL }%
\providecommand \Eprint [0]{\href }%
\providecommand \doibase [0]{http://dx.doi.org/}%
\providecommand \selectlanguage [0]{\@gobble}%
\providecommand \bibinfo  [0]{\@secondoftwo}%
\providecommand \bibfield  [0]{\@secondoftwo}%
\providecommand \translation [1]{[#1]}%
\providecommand \BibitemOpen [0]{}%
\providecommand \bibitemStop [0]{}%
\providecommand \bibitemNoStop [0]{.\EOS\space}%
\providecommand \EOS [0]{\spacefactor3000\relax}%
\providecommand \BibitemShut  [1]{\csname bibitem#1\endcsname}%
\let\auto@bib@innerbib\@empty
%</preamble>
\bibitem [{\citenamefont {Zyla}\ \emph {et~al.}(2020)\citenamefont {Zyla} \emph
  {et~al.}}]{ParticleDataGroup:2020ssz}%
  \BibitemOpen
  \bibfield  {author} {\bibinfo {author} {\bibfnamefont {P.}~\bibnamefont
  {Zyla}} \emph {et~al.} (\bibinfo {collaboration} {Particle Data Group}),\
  }\href {\doibase 10.1093/ptep/ptaa104} {\bibfield  {journal} {\bibinfo
  {journal} {PTEP}\ }\textbf {\bibinfo {volume} {2020}},\ \bibinfo {pages}
  {083C01} (\bibinfo {year} {2020})}\BibitemShut {NoStop}%
\bibitem [{\citenamefont {Bemporad}\ \emph {et~al.}(2002)\citenamefont
  {Bemporad}, \citenamefont {Gratta},\ and\ \citenamefont
  {Vogel}}]{Bemporad:2001qy}%
  \BibitemOpen
  \bibfield  {author} {\bibinfo {author} {\bibfnamefont {C.}~\bibnamefont
  {Bemporad}}, \bibinfo {author} {\bibfnamefont {G.}~\bibnamefont {Gratta}}, \
  and\ \bibinfo {author} {\bibfnamefont {P.}~\bibnamefont {Vogel}},\
  }\href@noop {} {\bibfield  {journal} {\bibinfo  {journal} {Rev. Mod. Phys.}\
  }\textbf {\bibinfo {volume} {74}},\ \bibinfo {pages} {297} (\bibinfo {year}
  {2002})},\ \Eprint {http://arxiv.org/abs/hep-ph/0107277} {hep-ph/0107277}
  \BibitemShut {NoStop}%
\bibitem [{\citenamefont {Huber}(2016)}]{Huber:2016fkt}%
  \BibitemOpen
  \bibfield  {author} {\bibinfo {author} {\bibfnamefont {P.}~\bibnamefont
  {Huber}},\ }\href@noop {} {\bibfield  {journal} {\bibinfo  {journal} {Nucl.
  Phys.}\ }\textbf {\bibinfo {volume} {B908}},\ \bibinfo {pages} {268}
  (\bibinfo {year} {2016})},\ \Eprint {http://arxiv.org/abs/arXiv:1602.01499}
  {arXiv:1602.01499 [hep-ph]} \BibitemShut {NoStop}%
\bibitem [{\citenamefont {Hayes}\ and\ \citenamefont
  {Vogel}(2016)}]{Hayes:2016qnu}%
  \BibitemOpen
  \bibfield  {author} {\bibinfo {author} {\bibfnamefont {A.~C.}\ \bibnamefont
  {Hayes}}\ and\ \bibinfo {author} {\bibfnamefont {P.}~\bibnamefont {Vogel}},\
  }\href@noop {} {\bibfield  {journal} {\bibinfo  {journal}
  {Ann.Rev.Nucl.Part.Sci.}\ }\textbf {\bibinfo {volume} {66}},\ \bibinfo
  {pages} {219} (\bibinfo {year} {2016})},\ \Eprint
  {http://arxiv.org/abs/arXiv:1605.02047} {arXiv:1605.02047 [hep-ph]}
  \BibitemShut {NoStop}%
\bibitem [{\citenamefont {Reines}\ \emph {et~al.}(1960)\citenamefont {Reines},
  \citenamefont {Cowan}, \citenamefont {Harrison}, \citenamefont {McGuire},\
  and\ \citenamefont {Kruse}}]{Reines:1960pr}%
  \BibitemOpen
  \bibfield  {author} {\bibinfo {author} {\bibfnamefont {F.}~\bibnamefont
  {Reines}}, \bibinfo {author} {\bibfnamefont {C.~L.}\ \bibnamefont {Cowan}},
  \bibinfo {author} {\bibfnamefont {F.~B.}\ \bibnamefont {Harrison}}, \bibinfo
  {author} {\bibfnamefont {A.~D.}\ \bibnamefont {McGuire}}, \ and\ \bibinfo
  {author} {\bibfnamefont {H.~W.}\ \bibnamefont {Kruse}},\ }\href@noop {}
  {\bibfield  {journal} {\bibinfo  {journal} {Phys. Rev.}\ }\textbf {\bibinfo
  {volume} {117}},\ \bibinfo {pages} {159} (\bibinfo {year}
  {1960})}\BibitemShut {NoStop}%
\bibitem [{\citenamefont {Davis}\ \emph {et~al.}(1979)\citenamefont {Davis},
  \citenamefont {Vogel}, \citenamefont {Mann},\ and\ \citenamefont
  {Schenter}}]{Davis:1979gg}%
  \BibitemOpen
  \bibfield  {author} {\bibinfo {author} {\bibfnamefont {B.~R.}\ \bibnamefont
  {Davis}}, \bibinfo {author} {\bibfnamefont {P.}~\bibnamefont {Vogel}},
  \bibinfo {author} {\bibfnamefont {F.~M.}\ \bibnamefont {Mann}}, \ and\
  \bibinfo {author} {\bibfnamefont {R.~E.}\ \bibnamefont {Schenter}},\
  }\href@noop {} {\bibfield  {journal} {\bibinfo  {journal} {Phys. Rev.}\
  }\textbf {\bibinfo {volume} {C19}},\ \bibinfo {pages} {2259} (\bibinfo {year}
  {1979})}\BibitemShut {NoStop}%
\bibitem [{\citenamefont {Mueller}\ \emph {et~al.}(2011)\citenamefont {Mueller}
  \emph {et~al.}}]{Mueller:2011nm}%
  \BibitemOpen
  \bibfield  {author} {\bibinfo {author} {\bibfnamefont {T.~A.}\ \bibnamefont
  {Mueller}} \emph {et~al.},\ }\href@noop {} {\bibfield  {journal} {\bibinfo
  {journal} {Phys. Rev.}\ }\textbf {\bibinfo {volume} {C83}},\ \bibinfo {pages}
  {054615} (\bibinfo {year} {2011})},\ \Eprint
  {http://arxiv.org/abs/arXiv:1101.2663} {arXiv:1101.2663 [hep-ex]}
  \BibitemShut {NoStop}%
\bibitem [{\citenamefont {Huber}(2011)}]{Huber:2011wv}%
  \BibitemOpen
  \bibfield  {author} {\bibinfo {author} {\bibfnamefont {P.}~\bibnamefont
  {Huber}},\ }\href@noop {} {\bibfield  {journal} {\bibinfo  {journal} {Phys.
  Rev.}\ }\textbf {\bibinfo {volume} {C84}},\ \bibinfo {pages} {024617}
  (\bibinfo {year} {2011})},\ \Eprint {http://arxiv.org/abs/arXiv:1106.0687}
  {arXiv:1106.0687 [hep-ph]} \BibitemShut {NoStop}%
\bibitem [{\citenamefont {Mention}\ \emph {et~al.}(2011)\citenamefont {Mention}
  \emph {et~al.}}]{Mention:2011rk}%
  \BibitemOpen
  \bibfield  {author} {\bibinfo {author} {\bibfnamefont {G.}~\bibnamefont
  {Mention}} \emph {et~al.},\ }\href@noop {} {\bibfield  {journal} {\bibinfo
  {journal} {Phys. Rev.}\ }\textbf {\bibinfo {volume} {D83}},\ \bibinfo {pages}
  {073006} (\bibinfo {year} {2011})},\ \Eprint
  {http://arxiv.org/abs/arXiv:1101.2755} {arXiv:1101.2755 [hep-ex]}
  \BibitemShut {NoStop}%
\bibitem [{\citenamefont {Von~Feilitzsch}\ \emph {et~al.}(1982)\citenamefont
  {Von~Feilitzsch}, \citenamefont {Hahn},\ and\ \citenamefont
  {Schreckenbach}}]{VonFeilitzsch:1982jw}%
  \BibitemOpen
  \bibfield  {author} {\bibinfo {author} {\bibfnamefont {F.}~\bibnamefont
  {Von~Feilitzsch}}, \bibinfo {author} {\bibfnamefont {A.~A.}\ \bibnamefont
  {Hahn}}, \ and\ \bibinfo {author} {\bibfnamefont {K.}~\bibnamefont
  {Schreckenbach}},\ }\href@noop {} {\bibfield  {journal} {\bibinfo  {journal}
  {Phys. Lett.}\ }\textbf {\bibinfo {volume} {B118}},\ \bibinfo {pages} {162}
  (\bibinfo {year} {1982})}\BibitemShut {NoStop}%
\bibitem [{\citenamefont {Schreckenbach}\ \emph {et~al.}(1985)\citenamefont
  {Schreckenbach}, \citenamefont {Colvin}, \citenamefont {Gelletly},\ and\
  \citenamefont {Von~Feilitzsch}}]{Schreckenbach:1985ep}%
  \BibitemOpen
  \bibfield  {author} {\bibinfo {author} {\bibfnamefont {K.}~\bibnamefont
  {Schreckenbach}}, \bibinfo {author} {\bibfnamefont {G.}~\bibnamefont
  {Colvin}}, \bibinfo {author} {\bibfnamefont {W.}~\bibnamefont {Gelletly}}, \
  and\ \bibinfo {author} {\bibfnamefont {F.}~\bibnamefont {Von~Feilitzsch}},\
  }\href@noop {} {\bibfield  {journal} {\bibinfo  {journal} {Phys. Lett.}\
  }\textbf {\bibinfo {volume} {B160}},\ \bibinfo {pages} {325} (\bibinfo {year}
  {1985})}\BibitemShut {NoStop}%
\bibitem [{\citenamefont {Hahn}\ \emph {et~al.}(1989)\citenamefont {Hahn} \emph
  {et~al.}}]{Hahn:1989zr}%
  \BibitemOpen
  \bibfield  {author} {\bibinfo {author} {\bibfnamefont {A.~A.}\ \bibnamefont
  {Hahn}} \emph {et~al.},\ }\href@noop {} {\bibfield  {journal} {\bibinfo
  {journal} {Phys. Lett.}\ }\textbf {\bibinfo {volume} {B218}},\ \bibinfo
  {pages} {365} (\bibinfo {year} {1989})}\BibitemShut {NoStop}%
\bibitem [{\citenamefont {Haag}\ \emph {et~al.}(2014)\citenamefont {Haag},
  \citenamefont {Gutlein}, \citenamefont {Hofmann}, \citenamefont {Oberauer},
  \citenamefont {Potzel} \emph {et~al.}}]{Haag:2013raa}%
  \BibitemOpen
  \bibfield  {author} {\bibinfo {author} {\bibfnamefont {N.}~\bibnamefont
  {Haag}}, \bibinfo {author} {\bibfnamefont {A.}~\bibnamefont {Gutlein}},
  \bibinfo {author} {\bibfnamefont {M.}~\bibnamefont {Hofmann}}, \bibinfo
  {author} {\bibfnamefont {L.}~\bibnamefont {Oberauer}}, \bibinfo {author}
  {\bibfnamefont {W.}~\bibnamefont {Potzel}},  \emph {et~al.},\ }\href@noop {}
  {\bibfield  {journal} {\bibinfo  {journal} {Phys. Rev. Lett.}\ }\textbf
  {\bibinfo {volume} {112}},\ \bibinfo {pages} {122501} (\bibinfo {year}
  {2014})},\ \Eprint {http://arxiv.org/abs/arXiv:1312.5601} {arXiv:1312.5601
  [nucl-ex]} \BibitemShut {NoStop}%
\bibitem [{\citenamefont {Vogel}(2007)}]{Vogel:2007du}%
  \BibitemOpen
  \bibfield  {author} {\bibinfo {author} {\bibfnamefont {P.}~\bibnamefont
  {Vogel}},\ }\href@noop {} {\bibfield  {journal} {\bibinfo  {journal} {Phys.
  Rev.}\ }\textbf {\bibinfo {volume} {C76}},\ \bibinfo {pages} {025504}
  (\bibinfo {year} {2007})},\ \Eprint {http://arxiv.org/abs/arXiv:0708.0556}
  {arXiv:0708.0556 [hep-ph]} \BibitemShut {NoStop}%
\bibitem [{\citenamefont {Choi}\ \emph {et~al.}(2016)\citenamefont {Choi} \emph
  {et~al.}}]{RENO:2015ksa}%
  \BibitemOpen
  \bibfield  {author} {\bibinfo {author} {\bibfnamefont {J.}~\bibnamefont
  {Choi}} \emph {et~al.} (\bibinfo {collaboration} {RENO}),\ }\href@noop {}
  {\bibfield  {journal} {\bibinfo  {journal} {Phys. Rev. Lett.}\ }\textbf
  {\bibinfo {volume} {116}},\ \bibinfo {pages} {211801} (\bibinfo {year}
  {2016})},\ \Eprint {http://arxiv.org/abs/arXiv:1511.05849} {arXiv:1511.05849
  [hep-ex]} \BibitemShut {NoStop}%
\bibitem [{\citenamefont {Abe}\ \emph {et~al.}(2014)\citenamefont {Abe} \emph
  {et~al.}}]{DoubleChooz:2014kuw}%
  \BibitemOpen
  \bibfield  {author} {\bibinfo {author} {\bibfnamefont {Y.}~\bibnamefont
  {Abe}} \emph {et~al.} (\bibinfo {collaboration} {Double Chooz}),\ }\href
  {\doibase 10.1007/JHEP02(2015)074, 10.1007/JHEP10(2014)086} {\bibfield
  {journal} {\bibinfo  {journal} {JHEP}\ }\textbf {\bibinfo {volume} {10}},\
  \bibinfo {pages} {086} (\bibinfo {year} {2014})},\ \bibinfo {note} {[Erratum:
  JHEP 02, 074 (2015)]},\ \Eprint {http://arxiv.org/abs/arXiv:1406.7763}
  {arXiv:1406.7763 [hep-ex]} \BibitemShut {NoStop}%
\bibitem [{\citenamefont {An}\ \emph {et~al.}(2016)\citenamefont {An} \emph
  {et~al.}}]{DayaBay:2015lja}%
  \BibitemOpen
  \bibfield  {author} {\bibinfo {author} {\bibfnamefont {F.~P.}\ \bibnamefont
  {An}} \emph {et~al.} (\bibinfo {collaboration} {Daya Bay}),\ }\href@noop {}
  {\bibfield  {journal} {\bibinfo  {journal} {Phys. Rev. Lett.}\ }\textbf
  {\bibinfo {volume} {116}},\ \bibinfo {pages} {061801} (\bibinfo {year}
  {2016})},\ \Eprint {http://arxiv.org/abs/arXiv:1508.04233} {arXiv:1508.04233
  [hep-ex]} \BibitemShut {NoStop}%
\bibitem [{\citenamefont {Estienne}\ \emph {et~al.}(2019)\citenamefont
  {Estienne}, \citenamefont {Fallot} \emph {et~al.}}]{Estienne:2019ujo}%
  \BibitemOpen
  \bibfield  {author} {\bibinfo {author} {\bibfnamefont {M.}~\bibnamefont
  {Estienne}}, \bibinfo {author} {\bibfnamefont {M.}~\bibnamefont {Fallot}},
  \emph {et~al.},\ }\href {\doibase 10.1103/PhysRevLett.123.022502} {\bibfield
  {journal} {\bibinfo  {journal} {Phys. Rev. Lett.}\ }\textbf {\bibinfo
  {volume} {123}},\ \bibinfo {pages} {022502} (\bibinfo {year} {2019})},\
  \Eprint {http://arxiv.org/abs/arXiv:1904.09358} {arXiv:1904.09358 [nucl-ex]}
  \BibitemShut {NoStop}%
\bibitem [{\citenamefont {Silaeva}\ and\ \citenamefont
  {Sinev}()}]{Silaeva:2020msh}%
  \BibitemOpen
  \bibfield  {author} {\bibinfo {author} {\bibfnamefont {S.~V.}\ \bibnamefont
  {Silaeva}}\ and\ \bibinfo {author} {\bibfnamefont {V.~V.}\ \bibnamefont
  {Sinev}},\ }\href@noop {} {\ }\Eprint {http://arxiv.org/abs/arXiv:2012.09917}
  {arXiv:2012.09917 [nucl-ex]} \BibitemShut {NoStop}%
\bibitem [{\citenamefont {Fang}\ \emph {et~al.}(2021)\citenamefont {Fang},
  \citenamefont {Li},\ and\ \citenamefont {Zhang}}]{Fang:2020emq}%
  \BibitemOpen
  \bibfield  {author} {\bibinfo {author} {\bibfnamefont {D.-L.}\ \bibnamefont
  {Fang}}, \bibinfo {author} {\bibfnamefont {Y.-F.}\ \bibnamefont {Li}}, \ and\
  \bibinfo {author} {\bibfnamefont {D.}~\bibnamefont {Zhang}},\ }\href@noop {}
  {\bibfield  {journal} {\bibinfo  {journal} {Phys.Lett.}\ }\textbf {\bibinfo
  {volume} {B813}},\ \bibinfo {pages} {136067} (\bibinfo {year} {2021})},\
  \Eprint {http://arxiv.org/abs/arXiv:2001.01689} {arXiv:2001.01689 [hep-ph]}
  \BibitemShut {NoStop}%
\bibitem [{\citenamefont {An}\ \emph {et~al.}(2017{\natexlab{a}})\citenamefont
  {An} \emph {et~al.}}]{DayaBay:2016ssb}%
  \BibitemOpen
  \bibfield  {author} {\bibinfo {author} {\bibfnamefont {F.}~\bibnamefont {An}}
  \emph {et~al.} (\bibinfo {collaboration} {Daya Bay}),\ }\href@noop {}
  {\bibfield  {journal} {\bibinfo  {journal} {Chin.Phys.}\ }\textbf {\bibinfo
  {volume} {C41}},\ \bibinfo {pages} {013002} (\bibinfo {year}
  {2017}{\natexlab{a}})},\ \Eprint {http://arxiv.org/abs/arXiv:1607.05378}
  {arXiv:1607.05378 [hep-ex]} \BibitemShut {NoStop}%
\bibitem [{\citenamefont {Hayen}\ \emph {et~al.}(2019)\citenamefont {Hayen},
  \citenamefont {Kostensalo}, \citenamefont {Severijns},\ and\ \citenamefont
  {Suhonen}}]{Hayen:2019eop}%
  \BibitemOpen
  \bibfield  {author} {\bibinfo {author} {\bibfnamefont {L.}~\bibnamefont
  {Hayen}}, \bibinfo {author} {\bibfnamefont {J.}~\bibnamefont {Kostensalo}},
  \bibinfo {author} {\bibfnamefont {N.}~\bibnamefont {Severijns}}, \ and\
  \bibinfo {author} {\bibfnamefont {J.}~\bibnamefont {Suhonen}},\ }\href@noop
  {} {\bibfield  {journal} {\bibinfo  {journal} {Phys.Rev.}\ }\textbf {\bibinfo
  {volume} {C100}},\ \bibinfo {pages} {054323} (\bibinfo {year} {2019})},\
  \Eprint {http://arxiv.org/abs/arXiv:1908.08302} {arXiv:1908.08302 [nucl-th]}
  \BibitemShut {NoStop}%
\bibitem [{\citenamefont {Li}\ and\ \citenamefont {Zhang}(2019)}]{Li:2019quv}%
  \BibitemOpen
  \bibfield  {author} {\bibinfo {author} {\bibfnamefont {Y.-F.}\ \bibnamefont
  {Li}}\ and\ \bibinfo {author} {\bibfnamefont {D.}~\bibnamefont {Zhang}},\
  }\href@noop {} {\bibfield  {journal} {\bibinfo  {journal} {Phys.Rev.}\
  }\textbf {\bibinfo {volume} {D100}},\ \bibinfo {pages} {053005} (\bibinfo
  {year} {2019})},\ \Eprint {http://arxiv.org/abs/arXiv:1904.07791}
  {arXiv:1904.07791 [hep-ph]} \BibitemShut {NoStop}%
\bibitem [{\citenamefont {Kopeikin}\ \emph {et~al.}(2021)\citenamefont
  {Kopeikin}, \citenamefont {Skorokhvatov},\ and\ \citenamefont
  {Titov}}]{Kopeikin:2021ugh}%
  \BibitemOpen
  \bibfield  {author} {\bibinfo {author} {\bibfnamefont {V.}~\bibnamefont
  {Kopeikin}}, \bibinfo {author} {\bibfnamefont {M.}~\bibnamefont
  {Skorokhvatov}}, \ and\ \bibinfo {author} {\bibfnamefont {O.}~\bibnamefont
  {Titov}},\ }\href {\doibase 10.1103/PhysRevD.104.L071301} {\bibfield
  {journal} {\bibinfo  {journal} {Phys. Rev. D}\ }\textbf {\bibinfo {volume}
  {104}},\ \bibinfo {pages} {L071301} (\bibinfo {year} {2021})},\ \Eprint
  {http://arxiv.org/abs/2103.01684} {arXiv:2103.01684 [nucl-ex]} \BibitemShut
  {NoStop}%
\bibitem [{\citenamefont {An}\ \emph {et~al.}(2017{\natexlab{b}})\citenamefont
  {An} \emph {et~al.}}]{DayaBay:2017jkb}%
  \BibitemOpen
  \bibfield  {author} {\bibinfo {author} {\bibfnamefont {F.~P.}\ \bibnamefont
  {An}} \emph {et~al.} (\bibinfo {collaboration} {Daya Bay}),\ }\href@noop {}
  {\bibfield  {journal} {\bibinfo  {journal} {Phys.Rev.Lett.}\ }\textbf
  {\bibinfo {volume} {118}},\ \bibinfo {pages} {251801} (\bibinfo {year}
  {2017}{\natexlab{b}})},\ \Eprint {http://arxiv.org/abs/arXiv:1704.01082}
  {arXiv:1704.01082 [physics]} \BibitemShut {NoStop}%
\bibitem [{\citenamefont {Adey}\ \emph {et~al.}(2019)\citenamefont {Adey} \emph
  {et~al.}}]{DayaBay:2019yxq}%
  \BibitemOpen
  \bibfield  {author} {\bibinfo {author} {\bibfnamefont {D.}~\bibnamefont
  {Adey}} \emph {et~al.} (\bibinfo {collaboration} {Daya Bay}),\ }\href@noop {}
  {\bibfield  {journal} {\bibinfo  {journal} {Phys.Rev.Lett.}\ }\textbf
  {\bibinfo {volume} {123}},\ \bibinfo {pages} {111801} (\bibinfo {year}
  {2019})},\ \Eprint {http://arxiv.org/abs/arXiv:1904.07812} {arXiv:1904.07812
  [hep-ex]} \BibitemShut {NoStop}%
\bibitem [{\citenamefont {Bak}\ \emph {et~al.}(2019)\citenamefont {Bak} \emph
  {et~al.}}]{RENO:2018pwo}%
  \BibitemOpen
  \bibfield  {author} {\bibinfo {author} {\bibfnamefont {G.}~\bibnamefont
  {Bak}} \emph {et~al.} (\bibinfo {collaboration} {RENO}),\ }\href@noop {}
  {\bibfield  {journal} {\bibinfo  {journal} {Phys.Rev.Lett.}\ }\textbf
  {\bibinfo {volume} {122}},\ \bibinfo {pages} {232501} (\bibinfo {year}
  {2019})},\ \Eprint {http://arxiv.org/abs/arXiv:1806.00574} {arXiv:1806.00574
  [hep-ex]} \BibitemShut {NoStop}%
\bibitem [{\citenamefont {An}\ \emph {et~al.}()\citenamefont {An} \emph
  {et~al.}}]{An:2021tyg}%
  \BibitemOpen
  \bibfield  {author} {\bibinfo {author} {\bibfnamefont {F.~P.}\ \bibnamefont
  {An}} \emph {et~al.} (\bibinfo {collaboration} {Daya Bay, PROSPECT}),\
  }\href@noop {} {\ }\Eprint {http://arxiv.org/abs/arXiv:2106.12251}
  {arXiv:2106.12251 [nucl-ex]} \BibitemShut {NoStop}%
\bibitem [{\citenamefont {Almazan}\ \emph {et~al.}()\citenamefont {Almazan}
  \emph {et~al.}}]{Prospect:2021lbs}%
  \BibitemOpen
  \bibfield  {author} {\bibinfo {author} {\bibfnamefont {H.}~\bibnamefont
  {Almazan}} \emph {et~al.} (\bibinfo {collaboration} {PROSPECT, STEREO}),\
  }\href@noop {} {\ }\Eprint {http://arxiv.org/abs/arXiv:2107.03371}
  {arXiv:2107.03371 [nucl-ex]} \BibitemShut {NoStop}%
\bibitem [{\citenamefont {Almazan~Molina}\ \emph
  {et~al.}(2020{\natexlab{a}})\citenamefont {Almazan~Molina} \emph
  {et~al.}}]{STEREO:2020fvd}%
  \BibitemOpen
  \bibfield  {author} {\bibinfo {author} {\bibfnamefont {H.}~\bibnamefont
  {Almazan~Molina}} \emph {et~al.} (\bibinfo {collaboration} {STEREO}),\
  }\href@noop {} {\bibfield  {journal} {\bibinfo  {journal} {Phys.Rev.Lett.}\
  }\textbf {\bibinfo {volume} {125}},\ \bibinfo {pages} {201801} (\bibinfo
  {year} {2020}{\natexlab{a}})},\ \Eprint
  {http://arxiv.org/abs/arXiv:2004.04075} {arXiv:2004.04075 [hep-ex]}
  \BibitemShut {NoStop}%
\bibitem [{\citenamefont {Giunti}(2017{\natexlab{a}})}]{Giunti:2016elf}%
  \BibitemOpen
  \bibfield  {author} {\bibinfo {author} {\bibfnamefont {C.}~\bibnamefont
  {Giunti}},\ }\href@noop {} {\bibfield  {journal} {\bibinfo  {journal}
  {Phys.Lett.}\ }\textbf {\bibinfo {volume} {B764}},\ \bibinfo {pages} {145}
  (\bibinfo {year} {2017}{\natexlab{a}})},\ \Eprint
  {http://arxiv.org/abs/arXiv:1608.04096} {arXiv:1608.04096 [hep-ph]}
  \BibitemShut {NoStop}%
\bibitem [{\citenamefont {Giunti}(2017{\natexlab{b}})}]{Giunti:2017nww}%
  \BibitemOpen
  \bibfield  {author} {\bibinfo {author} {\bibfnamefont {C.}~\bibnamefont
  {Giunti}},\ }\href@noop {} {\bibfield  {journal} {\bibinfo  {journal}
  {Phys.Rev.}\ }\textbf {\bibinfo {volume} {D96}},\ \bibinfo {pages} {033005}
  (\bibinfo {year} {2017}{\natexlab{b}})},\ \Eprint
  {http://arxiv.org/abs/arXiv:1704.02276} {arXiv:1704.02276 [hep-ph]}
  \BibitemShut {NoStop}%
\bibitem [{\citenamefont {Hayes}\ \emph {et~al.}(2018)\citenamefont {Hayes}
  \emph {et~al.}}]{Hayes:2017res}%
  \BibitemOpen
  \bibfield  {author} {\bibinfo {author} {\bibfnamefont {A.}~\bibnamefont
  {Hayes}} \emph {et~al.},\ }\href@noop {} {\bibfield  {journal} {\bibinfo
  {journal} {Phys.Rev.Lett.}\ }\textbf {\bibinfo {volume} {120}},\ \bibinfo
  {pages} {022503} (\bibinfo {year} {2018})},\ \Eprint
  {http://arxiv.org/abs/arXiv:1707.07728} {arXiv:1707.07728 [nucl-th]}
  \BibitemShut {NoStop}%
\bibitem [{\citenamefont {Giunti}\ \emph {et~al.}(2017)\citenamefont {Giunti},
  \citenamefont {Ji}, \citenamefont {Laveder}, \citenamefont {Li},\ and\
  \citenamefont {Littlejohn}}]{Giunti:2017yid}%
  \BibitemOpen
  \bibfield  {author} {\bibinfo {author} {\bibfnamefont {C.}~\bibnamefont
  {Giunti}}, \bibinfo {author} {\bibfnamefont {X.~P.}\ \bibnamefont {Ji}},
  \bibinfo {author} {\bibfnamefont {M.}~\bibnamefont {Laveder}}, \bibinfo
  {author} {\bibfnamefont {Y.~F.}\ \bibnamefont {Li}}, \ and\ \bibinfo {author}
  {\bibfnamefont {B.~R.}\ \bibnamefont {Littlejohn}},\ }\href@noop {}
  {\bibfield  {journal} {\bibinfo  {journal} {JHEP}\ }\textbf {\bibinfo
  {volume} {1710}},\ \bibinfo {pages} {143} (\bibinfo {year} {2017})},\ \Eprint
  {http://arxiv.org/abs/arXiv:1708.01133} {arXiv:1708.01133 [hep-ph]}
  \BibitemShut {NoStop}%
\bibitem [{\citenamefont {Gebre}\ \emph {et~al.}(2018)\citenamefont {Gebre},
  \citenamefont {Littlejohn},\ and\ \citenamefont {Surukuchi}}]{Gebre:2017vmm}%
  \BibitemOpen
  \bibfield  {author} {\bibinfo {author} {\bibfnamefont {Y.}~\bibnamefont
  {Gebre}}, \bibinfo {author} {\bibfnamefont {B.~R.}\ \bibnamefont
  {Littlejohn}}, \ and\ \bibinfo {author} {\bibfnamefont {P.~T.}\ \bibnamefont
  {Surukuchi}},\ }\href@noop {} {\bibfield  {journal} {\bibinfo  {journal}
  {Phys.Rev.}\ }\textbf {\bibinfo {volume} {D97}},\ \bibinfo {pages} {013003}
  (\bibinfo {year} {2018})},\ \Eprint {http://arxiv.org/abs/arXiv:1709.10051}
  {arXiv:1709.10051 [hep-ph]} \BibitemShut {NoStop}%
\bibitem [{\citenamefont {Giunti}\ \emph {et~al.}(2019)\citenamefont {Giunti},
  \citenamefont {Li}, \citenamefont {Littlejohn},\ and\ \citenamefont
  {Surukuchi}}]{Giunti:2019qlt}%
  \BibitemOpen
  \bibfield  {author} {\bibinfo {author} {\bibfnamefont {C.}~\bibnamefont
  {Giunti}}, \bibinfo {author} {\bibfnamefont {Y.~F.}\ \bibnamefont {Li}},
  \bibinfo {author} {\bibfnamefont {B.~R.}\ \bibnamefont {Littlejohn}}, \ and\
  \bibinfo {author} {\bibfnamefont {P.~T.}\ \bibnamefont {Surukuchi}},\
  }\href@noop {} {\bibfield  {journal} {\bibinfo  {journal} {Phys.Rev.}\
  }\textbf {\bibinfo {volume} {D99}},\ \bibinfo {pages} {073005} (\bibinfo
  {year} {2019})},\ \Eprint {http://arxiv.org/abs/arXiv:1901.01807}
  {arXiv:1901.01807 [hep-ph]} \BibitemShut {NoStop}%
\bibitem [{\citenamefont {Berryman}\ and\ \citenamefont
  {Huber}(2021)}]{Berryman:2020agd}%
  \BibitemOpen
  \bibfield  {author} {\bibinfo {author} {\bibfnamefont {J.~M.}\ \bibnamefont
  {Berryman}}\ and\ \bibinfo {author} {\bibfnamefont {P.}~\bibnamefont
  {Huber}},\ }\href@noop {} {\bibfield  {journal} {\bibinfo  {journal} {JHEP}\
  }\textbf {\bibinfo {volume} {2101}},\ \bibinfo {pages} {167} (\bibinfo {year}
  {2021})},\ \Eprint {http://arxiv.org/abs/arXiv:2005.01756} {arXiv:2005.01756
  [hep-ph]} \BibitemShut {NoStop}%
\bibitem [{\citenamefont {Gariazzo}\ \emph {et~al.}(2016)\citenamefont
  {Gariazzo}, \citenamefont {Giunti}, \citenamefont {Laveder}, \citenamefont
  {Li},\ and\ \citenamefont {Zavanin}}]{Gariazzo:2015rra}%
  \BibitemOpen
  \bibfield  {author} {\bibinfo {author} {\bibfnamefont {S.}~\bibnamefont
  {Gariazzo}}, \bibinfo {author} {\bibfnamefont {C.}~\bibnamefont {Giunti}},
  \bibinfo {author} {\bibfnamefont {M.}~\bibnamefont {Laveder}}, \bibinfo
  {author} {\bibfnamefont {Y.~F.}\ \bibnamefont {Li}}, \ and\ \bibinfo {author}
  {\bibfnamefont {E.}~\bibnamefont {Zavanin}},\ }\href@noop {} {\bibfield
  {journal} {\bibinfo  {journal} {J. Phys.}\ }\textbf {\bibinfo {volume}
  {G43}},\ \bibinfo {pages} {033001} (\bibinfo {year} {2016})},\ \Eprint
  {http://arxiv.org/abs/arXiv:1507.08204} {arXiv:1507.08204 [hep-ph]}
  \BibitemShut {NoStop}%
\bibitem [{\citenamefont {Giunti}\ and\ \citenamefont
  {Lasserre}(2019)}]{Giunti:2019aiy}%
  \BibitemOpen
  \bibfield  {author} {\bibinfo {author} {\bibfnamefont {C.}~\bibnamefont
  {Giunti}}\ and\ \bibinfo {author} {\bibfnamefont {T.}~\bibnamefont
  {Lasserre}},\ }\href {\doibase 10.1146/annurev-nucl-101918-023755} {\bibfield
   {journal} {\bibinfo  {journal} {Ann. Rev. Nucl. Part. Sci.}\ }\textbf
  {\bibinfo {volume} {69}},\ \bibinfo {pages} {163} (\bibinfo {year} {2019})},\
  \Eprint {http://arxiv.org/abs/arXiv:1901.08330} {arXiv:1901.08330 [hep-ph]}
  \BibitemShut {NoStop}%
\bibitem [{\citenamefont {Diaz}\ \emph {et~al.}(2020)\citenamefont {Diaz},
  \citenamefont {Arguelles}, \citenamefont {Collin}, \citenamefont {Conrad},\
  and\ \citenamefont {Shaevitz}}]{Diaz:2019fwt}%
  \BibitemOpen
  \bibfield  {author} {\bibinfo {author} {\bibfnamefont {A.}~\bibnamefont
  {Diaz}}, \bibinfo {author} {\bibfnamefont {C.}~\bibnamefont {Arguelles}},
  \bibinfo {author} {\bibfnamefont {G.}~\bibnamefont {Collin}}, \bibinfo
  {author} {\bibfnamefont {J.}~\bibnamefont {Conrad}}, \ and\ \bibinfo {author}
  {\bibfnamefont {M.}~\bibnamefont {Shaevitz}},\ }\href@noop {} {\bibfield
  {journal} {\bibinfo  {journal} {Phys.Rept.}\ }\textbf {\bibinfo {volume}
  {884}},\ \bibinfo {pages} {1} (\bibinfo {year} {2020})},\ \Eprint
  {http://arxiv.org/abs/arXiv:1906.00045} {arXiv:1906.00045 [hep-ex]}
  \BibitemShut {NoStop}%
\bibitem [{\citenamefont {Boser}\ \emph {et~al.}(2020)\citenamefont {Boser},
  \citenamefont {Buck}, \citenamefont {Giunti}, \citenamefont {Lesgourgues},
  \citenamefont {Ludhova}, \citenamefont {Mertens}, \citenamefont {Schukraft},\
  and\ \citenamefont {Wurm}}]{Boser:2019rta}%
  \BibitemOpen
  \bibfield  {author} {\bibinfo {author} {\bibfnamefont {S.}~\bibnamefont
  {Boser}}, \bibinfo {author} {\bibfnamefont {C.}~\bibnamefont {Buck}},
  \bibinfo {author} {\bibfnamefont {C.}~\bibnamefont {Giunti}}, \bibinfo
  {author} {\bibfnamefont {J.}~\bibnamefont {Lesgourgues}}, \bibinfo {author}
  {\bibfnamefont {L.}~\bibnamefont {Ludhova}}, \bibinfo {author} {\bibfnamefont
  {S.}~\bibnamefont {Mertens}}, \bibinfo {author} {\bibfnamefont
  {A.}~\bibnamefont {Schukraft}}, \ and\ \bibinfo {author} {\bibfnamefont
  {M.}~\bibnamefont {Wurm}},\ }\href@noop {} {\bibfield  {journal} {\bibinfo
  {journal} {Prog.Part.Nucl.Phys.}\ }\textbf {\bibinfo {volume} {111}},\
  \bibinfo {pages} {103736} (\bibinfo {year} {2020})},\ \Eprint
  {http://arxiv.org/abs/arXiv:1906.01739} {arXiv:1906.01739 [hep-ex]}
  \BibitemShut {NoStop}%
\bibitem [{\citenamefont {Dasgupta}\ and\ \citenamefont
  {Kopp}(2021)}]{Dasgupta:2021ies}%
  \BibitemOpen
  \bibfield  {author} {\bibinfo {author} {\bibfnamefont {B.}~\bibnamefont
  {Dasgupta}}\ and\ \bibinfo {author} {\bibfnamefont {J.}~\bibnamefont
  {Kopp}},\ }\href@noop {} {\bibfield  {journal} {\bibinfo  {journal}
  {Phys.Rept.}\ }\textbf {\bibinfo {volume} {928}},\ \bibinfo {pages} {63}
  (\bibinfo {year} {2021})},\ \Eprint {http://arxiv.org/abs/arXiv:2106.05913}
  {arXiv:2106.05913 [hep-ph]} \BibitemShut {NoStop}%
\bibitem [{\citenamefont {de~Salas}\ \emph {et~al.}(2020)\citenamefont
  {de~Salas}, \citenamefont {Forero}, \citenamefont {Gariazzo}, \citenamefont
  {Martinez-Mirave}, \citenamefont {Mena}, \citenamefont {Ternes},
  \citenamefont {Tortola},\ and\ \citenamefont {Valle}}]{deSalas:2020pgw}%
  \BibitemOpen
  \bibfield  {author} {\bibinfo {author} {\bibfnamefont {P.~F.}\ \bibnamefont
  {de~Salas}}, \bibinfo {author} {\bibfnamefont {D.~V.}\ \bibnamefont
  {Forero}}, \bibinfo {author} {\bibfnamefont {S.}~\bibnamefont {Gariazzo}},
  \bibinfo {author} {\bibfnamefont {P.}~\bibnamefont {Martinez-Mirave}},
  \bibinfo {author} {\bibfnamefont {O.}~\bibnamefont {Mena}}, \bibinfo {author}
  {\bibfnamefont {C.~A.}\ \bibnamefont {Ternes}}, \bibinfo {author}
  {\bibfnamefont {M.}~\bibnamefont {Tortola}}, \ and\ \bibinfo {author}
  {\bibfnamefont {J.~W.~F.}\ \bibnamefont {Valle}},\ }\href@noop {} {\bibfield
  {journal} {\bibinfo  {journal} {JHEP}\ }\textbf {\bibinfo {volume} {2021}},\
  \bibinfo {pages} {071} (\bibinfo {year} {2020})},\ \Eprint
  {http://arxiv.org/abs/arXiv:2006.11237} {arXiv:2006.11237 [hep-ph]}
  \BibitemShut {NoStop}%
\bibitem [{\citenamefont {Esteban}\ \emph {et~al.}(2020)\citenamefont
  {Esteban}, \citenamefont {Gonzalez-Garcia}, \citenamefont {Maltoni},
  \citenamefont {Schwetz},\ and\ \citenamefont {Zhou}}]{Esteban:2020cvm}%
  \BibitemOpen
  \bibfield  {author} {\bibinfo {author} {\bibfnamefont {I.}~\bibnamefont
  {Esteban}}, \bibinfo {author} {\bibfnamefont {M.}~\bibnamefont
  {Gonzalez-Garcia}}, \bibinfo {author} {\bibfnamefont {M.}~\bibnamefont
  {Maltoni}}, \bibinfo {author} {\bibfnamefont {T.}~\bibnamefont {Schwetz}}, \
  and\ \bibinfo {author} {\bibfnamefont {A.}~\bibnamefont {Zhou}},\ }\href
  {\doibase 10.1007/JHEP09(2020)178} {\bibfield  {journal} {\bibinfo  {journal}
  {JHEP}\ }\textbf {\bibinfo {volume} {09}},\ \bibinfo {pages} {178} (\bibinfo
  {year} {2020})},\ \Eprint {http://arxiv.org/abs/arXiv:2007.14792}
  {arXiv:2007.14792 [hep-ph]} \BibitemShut {NoStop}%
\bibitem [{\citenamefont {Capozzi}\ \emph {et~al.}()\citenamefont {Capozzi},
  \citenamefont {Di~Valentino}, \citenamefont {Lisi}, \citenamefont {Marrone},
  \citenamefont {Melchiorri},\ and\ \citenamefont {Palazzo}}]{Capozzi:2021fjo}%
  \BibitemOpen
  \bibfield  {author} {\bibinfo {author} {\bibfnamefont {F.}~\bibnamefont
  {Capozzi}}, \bibinfo {author} {\bibfnamefont {E.}~\bibnamefont
  {Di~Valentino}}, \bibinfo {author} {\bibfnamefont {E.}~\bibnamefont {Lisi}},
  \bibinfo {author} {\bibfnamefont {A.}~\bibnamefont {Marrone}}, \bibinfo
  {author} {\bibfnamefont {A.}~\bibnamefont {Melchiorri}}, \ and\ \bibinfo
  {author} {\bibfnamefont {A.}~\bibnamefont {Palazzo}},\ }\href@noop {} {\
  }\Eprint {http://arxiv.org/abs/arXiv:2107.00532} {arXiv:2107.00532 [hep-ph]}
  \BibitemShut {NoStop}%
\bibitem [{\citenamefont {Abazajian}\ \emph {et~al.}()\citenamefont {Abazajian}
  \emph {et~al.}}]{Abazajian:2012ys}%
  \BibitemOpen
  \bibfield  {author} {\bibinfo {author} {\bibfnamefont {K.~N.}\ \bibnamefont
  {Abazajian}} \emph {et~al.},\ }\href@noop {} {\ }\Eprint
  {http://arxiv.org/abs/arXiv:1204.5379} {arXiv:1204.5379 [hep-ph]}
  \BibitemShut {NoStop}%
\bibitem [{\citenamefont {Strumia}\ and\ \citenamefont
  {Vissani}(2003)}]{Strumia:2003zx}%
  \BibitemOpen
  \bibfield  {author} {\bibinfo {author} {\bibfnamefont {A.}~\bibnamefont
  {Strumia}}\ and\ \bibinfo {author} {\bibfnamefont {F.}~\bibnamefont
  {Vissani}},\ }\href@noop {} {\bibfield  {journal} {\bibinfo  {journal} {Phys.
  Lett.}\ }\textbf {\bibinfo {volume} {B564}},\ \bibinfo {pages} {42} (\bibinfo
  {year} {2003})},\ \Eprint {http://arxiv.org/abs/astro-ph/0302055}
  {astro-ph/0302055} \BibitemShut {NoStop}%
\bibitem [{\citenamefont {Vogel}\ and\ \citenamefont
  {Beacom}(1999)}]{Vogel:1999zy}%
  \BibitemOpen
  \bibfield  {author} {\bibinfo {author} {\bibfnamefont {P.}~\bibnamefont
  {Vogel}}\ and\ \bibinfo {author} {\bibfnamefont {J.~F.}\ \bibnamefont
  {Beacom}},\ }\href@noop {} {\bibfield  {journal} {\bibinfo  {journal} {Phys.
  Rev.}\ }\textbf {\bibinfo {volume} {D60}},\ \bibinfo {pages} {053003}
  (\bibinfo {year} {1999})},\ \Eprint {http://arxiv.org/abs/hep-ph/9903554}
  {hep-ph/9903554} \BibitemShut {NoStop}%
\bibitem [{\citenamefont {Llewellyn~Smith}(1972)}]{LlewellynSmith:1971zm}%
  \BibitemOpen
  \bibfield  {author} {\bibinfo {author} {\bibfnamefont {C.~H.}\ \bibnamefont
  {Llewellyn~Smith}},\ }\href@noop {} {\bibfield  {journal} {\bibinfo
  {journal} {Phys. Rep.}\ }\textbf {\bibinfo {volume} {3}},\ \bibinfo {pages}
  {261} (\bibinfo {year} {1972})}\BibitemShut {NoStop}%
\bibitem [{\citenamefont {Kurylov}\ \emph {et~al.}(2003)\citenamefont
  {Kurylov}, \citenamefont {Ramsey-Musolf},\ and\ \citenamefont
  {Vogel}}]{Kurylov:2002vj}%
  \BibitemOpen
  \bibfield  {author} {\bibinfo {author} {\bibfnamefont {A.}~\bibnamefont
  {Kurylov}}, \bibinfo {author} {\bibfnamefont {M.}~\bibnamefont
  {Ramsey-Musolf}}, \ and\ \bibinfo {author} {\bibfnamefont {P.}~\bibnamefont
  {Vogel}},\ }\href@noop {} {\bibfield  {journal} {\bibinfo  {journal} {Phys.
  Rev.}\ }\textbf {\bibinfo {volume} {C67}},\ \bibinfo {pages} {035502}
  (\bibinfo {year} {2003})},\ \Eprint {http://arxiv.org/abs/hep-ph/0211306}
  {hep-ph/0211306} \BibitemShut {NoStop}%
\bibitem [{\citenamefont {Kozlov}\ \emph {et~al.}(2000)\citenamefont {Kozlov},
  \citenamefont {Khalturtsev}, \citenamefont {Machulin}, \citenamefont
  {Martemyanov}, \citenamefont {Martemyanov}, \citenamefont {Sukhotin},
  \citenamefont {Tarasenkov}, \citenamefont {Turbin},\ and\ \citenamefont
  {Vyrodov}}]{Kozlov:1999ct}%
  \BibitemOpen
  \bibfield  {author} {\bibinfo {author} {\bibfnamefont {Y.~V.}\ \bibnamefont
  {Kozlov}}, \bibinfo {author} {\bibfnamefont {S.~V.}\ \bibnamefont
  {Khalturtsev}}, \bibinfo {author} {\bibfnamefont {I.~N.}\ \bibnamefont
  {Machulin}}, \bibinfo {author} {\bibfnamefont {A.~V.}\ \bibnamefont
  {Martemyanov}}, \bibinfo {author} {\bibfnamefont {V.~P.}\ \bibnamefont
  {Martemyanov}}, \bibinfo {author} {\bibfnamefont {S.~V.}\ \bibnamefont
  {Sukhotin}}, \bibinfo {author} {\bibfnamefont {V.~G.}\ \bibnamefont
  {Tarasenkov}}, \bibinfo {author} {\bibfnamefont {E.~V.}\ \bibnamefont
  {Turbin}}, \ and\ \bibinfo {author} {\bibfnamefont {V.~N.}\ \bibnamefont
  {Vyrodov}},\ }\href {\doibase 10.1134/1.855742} {\bibfield  {journal}
  {\bibinfo  {journal} {Phys. Atom. Nucl.}\ }\textbf {\bibinfo {volume} {63}},\
  \bibinfo {pages} {1016} (\bibinfo {year} {2000})},\ \Eprint
  {http://arxiv.org/abs/hep-ex/9912047} {hep-ex/9912047 [hep-ex]} \BibitemShut
  {NoStop}%
\bibitem [{\citenamefont {Wilks}(1938)}]{Wilks:1938dza}%
  \BibitemOpen
  \bibfield  {author} {\bibinfo {author} {\bibfnamefont {S.~S.}\ \bibnamefont
  {Wilks}},\ }\href {\doibase 10.1214/aoms/1177732360} {\bibfield  {journal}
  {\bibinfo  {journal} {Annals Math. Statist.}\ }\textbf {\bibinfo {volume}
  {9}},\ \bibinfo {pages} {60} (\bibinfo {year} {1938})}\BibitemShut {NoStop}%
\bibitem [{\citenamefont {Gariazzo}\ \emph {et~al.}(2017)\citenamefont
  {Gariazzo}, \citenamefont {Giunti}, \citenamefont {Laveder},\ and\
  \citenamefont {Li}}]{Gariazzo:2017fdh}%
  \BibitemOpen
  \bibfield  {author} {\bibinfo {author} {\bibfnamefont {S.}~\bibnamefont
  {Gariazzo}}, \bibinfo {author} {\bibfnamefont {C.}~\bibnamefont {Giunti}},
  \bibinfo {author} {\bibfnamefont {M.}~\bibnamefont {Laveder}}, \ and\
  \bibinfo {author} {\bibfnamefont {Y.~F.}\ \bibnamefont {Li}},\ }\href@noop {}
  {\bibfield  {journal} {\bibinfo  {journal} {JHEP}\ }\textbf {\bibinfo
  {volume} {1706}},\ \bibinfo {pages} {135} (\bibinfo {year} {2017})},\ \Eprint
  {http://arxiv.org/abs/arXiv:1703.00860} {arXiv:1703.00860 [hep-ph]}
  \BibitemShut {NoStop}%
\bibitem [{\citenamefont {Zhang}\ \emph {et~al.}(2013)\citenamefont {Zhang},
  \citenamefont {Qian},\ and\ \citenamefont {Vogel}}]{Zhang:2013ela}%
  \BibitemOpen
  \bibfield  {author} {\bibinfo {author} {\bibfnamefont {C.}~\bibnamefont
  {Zhang}}, \bibinfo {author} {\bibfnamefont {X.}~\bibnamefont {Qian}}, \ and\
  \bibinfo {author} {\bibfnamefont {P.}~\bibnamefont {Vogel}},\ }\href@noop {}
  {\bibfield  {journal} {\bibinfo  {journal} {Phys. Rev.}\ }\textbf {\bibinfo
  {volume} {D87}},\ \bibinfo {pages} {073018} (\bibinfo {year} {2013})},\
  \Eprint {http://arxiv.org/abs/arXiv:1303.0900} {arXiv:1303.0900 [nucl-ex]}
  \BibitemShut {NoStop}%
\bibitem [{\citenamefont {Berryman}\ and\ \citenamefont
  {Huber}(2020)}]{Berryman:2019hme}%
  \BibitemOpen
  \bibfield  {author} {\bibinfo {author} {\bibfnamefont {J.}~\bibnamefont
  {Berryman}}\ and\ \bibinfo {author} {\bibfnamefont {P.}~\bibnamefont
  {Huber}},\ }\href@noop {} {\bibfield  {journal} {\bibinfo  {journal}
  {Phys.Rev.}\ }\textbf {\bibinfo {volume} {D101}},\ \bibinfo {pages} {015008}
  (\bibinfo {year} {2020})},\ \Eprint {http://arxiv.org/abs/arXiv:1909.09267}
  {arXiv:1909.09267 [hep-ph]} \BibitemShut {NoStop}%
\bibitem [{Pee(1987)}]{Peelle:1987}%
  \BibitemOpen
  \href@noop {} {\  (\bibinfo {year} {1987})},\ \bibinfo {note} {{R.W. Peelle,
  Oak Ridge National Laboratory Informal Memorandum, 1987}}\BibitemShut
  {NoStop}%
\bibitem [{\citenamefont {Chiba}\ and\ \citenamefont
  {Smith}(1994)}]{Chiba-Smith-JNST-1994}%
  \BibitemOpen
  \bibfield  {author} {\bibinfo {author} {\bibfnamefont {S.}~\bibnamefont
  {Chiba}}\ and\ \bibinfo {author} {\bibfnamefont {D.~L.}\ \bibnamefont
  {Smith}},\ }\href {\doibase 10.1080/18811248.1994.9735223} {\bibfield
  {journal} {\bibinfo  {journal} {Journal of Nuclear Science and Technology}\
  }\textbf {\bibinfo {volume} {31}},\ \bibinfo {pages} {770} (\bibinfo {year}
  {1994})}\BibitemShut {NoStop}%
\bibitem [{\citenamefont {D'Agostini}(1994)}]{DAgostini:1993arp}%
  \BibitemOpen
  \bibfield  {author} {\bibinfo {author} {\bibfnamefont {G.}~\bibnamefont
  {D'Agostini}},\ }\href {\doibase 10.1016/0168-9002(94)90719-6} {\bibfield
  {journal} {\bibinfo  {journal} {Nucl. Instrum. Meth.}\ }\textbf {\bibinfo
  {volume} {A346}},\ \bibinfo {pages} {306} (\bibinfo {year}
  {1994})}\BibitemShut {NoStop}%
\bibitem [{\citenamefont {Giunti}\ \emph {et~al.}(2020)\citenamefont {Giunti},
  \citenamefont {Li},\ and\ \citenamefont {Zhang}}]{Giunti:2019fcj}%
  \BibitemOpen
  \bibfield  {author} {\bibinfo {author} {\bibfnamefont {C.}~\bibnamefont
  {Giunti}}, \bibinfo {author} {\bibfnamefont {Y.}~\bibnamefont {Li}}, \ and\
  \bibinfo {author} {\bibfnamefont {Y.}~\bibnamefont {Zhang}},\ }\href@noop {}
  {\bibfield  {journal} {\bibinfo  {journal} {JHEP}\ }\textbf {\bibinfo
  {volume} {2005}},\ \bibinfo {pages} {061} (\bibinfo {year} {2020})},\ \Eprint
  {http://arxiv.org/abs/arXiv:1912.12956} {arXiv:1912.12956 [hep-ph]}
  \BibitemShut {NoStop}%
\bibitem [{\citenamefont {de~Kerret}\ \emph {et~al.}(2020)\citenamefont
  {de~Kerret} \emph {et~al.}}]{DoubleChooz:2019qbj}%
  \BibitemOpen
  \bibfield  {author} {\bibinfo {author} {\bibfnamefont {H.}~\bibnamefont
  {de~Kerret}} \emph {et~al.} (\bibinfo {collaboration} {Double Chooz}),\
  }\href {\doibase 10.1038/s41567-020-0831-y} {\bibfield  {journal} {\bibinfo
  {journal} {Nature Phys.}\ }\textbf {\bibinfo {volume} {16}},\ \bibinfo
  {pages} {558} (\bibinfo {year} {2020})},\ \Eprint
  {http://arxiv.org/abs/1901.09445} {arXiv:1901.09445 [hep-ex]} \BibitemShut
  {NoStop}%
\bibitem [{\citenamefont {Atif}\ \emph {et~al.}()\citenamefont {Atif} \emph
  {et~al.}}]{RENO:2020dxd}%
  \BibitemOpen
  \bibfield  {author} {\bibinfo {author} {\bibfnamefont {Z.}~\bibnamefont
  {Atif}} \emph {et~al.} (\bibinfo {collaboration} {RENO}),\ }\href@noop {} {\
  }\Eprint {http://arxiv.org/abs/arXiv:2010.14989} {arXiv:2010.14989 [hep-ex]}
  \BibitemShut {NoStop}%
\bibitem [{\citenamefont {Zhang}(64)}]{Zhang-JRSSB-2002}%
  \BibitemOpen
  \bibfield  {author} {\bibinfo {author} {\bibfnamefont {J.}~\bibnamefont
  {Zhang}},\ }\href {\doibase 10.1111/1467-9868.00337} {\bibfield  {journal}
  {\bibinfo  {journal} {Journal of the Royal Statistical Society Series B}\ ,\
  \bibinfo {pages} {281}} (\bibinfo {year} {64})}\BibitemShut {NoStop}%
\bibitem [{\citenamefont {Barinov}\ \emph {et~al.}()\citenamefont {Barinov}
  \emph {et~al.}}]{Barinov:2021asz}%
  \BibitemOpen
  \bibfield  {author} {\bibinfo {author} {\bibfnamefont {V.}~\bibnamefont
  {Barinov}} \emph {et~al.},\ }\href@noop {} {\ }\Eprint
  {http://arxiv.org/abs/arXiv:2109.11482} {arXiv:2109.11482 [nucl-ex]}
  \BibitemShut {NoStop}%
\bibitem [{\citenamefont {Goldhagen}\ \emph {et~al.}()\citenamefont
  {Goldhagen}, \citenamefont {Maltoni}, \citenamefont {Reichard},\ and\
  \citenamefont {Schwetz}}]{Goldhagen:2021kxe}%
  \BibitemOpen
  \bibfield  {author} {\bibinfo {author} {\bibfnamefont {K.}~\bibnamefont
  {Goldhagen}}, \bibinfo {author} {\bibfnamefont {M.}~\bibnamefont {Maltoni}},
  \bibinfo {author} {\bibfnamefont {S.}~\bibnamefont {Reichard}}, \ and\
  \bibinfo {author} {\bibfnamefont {T.}~\bibnamefont {Schwetz}},\ }\href@noop
  {} {\ }\Eprint {http://arxiv.org/abs/arXiv:2109.14898} {arXiv:2109.14898
  [hep-ph]} \BibitemShut {NoStop}%
\bibitem [{\citenamefont {Kaether}\ \emph {et~al.}(2010)\citenamefont
  {Kaether}, \citenamefont {Hampel}, \citenamefont {Heusser}, \citenamefont
  {Kiko},\ and\ \citenamefont {Kirsten}}]{Kaether:2010ag}%
  \BibitemOpen
  \bibfield  {author} {\bibinfo {author} {\bibfnamefont {F.}~\bibnamefont
  {Kaether}}, \bibinfo {author} {\bibfnamefont {W.}~\bibnamefont {Hampel}},
  \bibinfo {author} {\bibfnamefont {G.}~\bibnamefont {Heusser}}, \bibinfo
  {author} {\bibfnamefont {J.}~\bibnamefont {Kiko}}, \ and\ \bibinfo {author}
  {\bibfnamefont {T.}~\bibnamefont {Kirsten}},\ }\href@noop {} {\bibfield
  {journal} {\bibinfo  {journal} {Phys. Lett.}\ }\textbf {\bibinfo {volume}
  {B685}},\ \bibinfo {pages} {47} (\bibinfo {year} {2010})},\ \Eprint
  {http://arxiv.org/abs/arXiv:1001.2731} {arXiv:1001.2731 [hep-ex]}
  \BibitemShut {NoStop}%
\bibitem [{\citenamefont {Abdurashitov}\ \emph {et~al.}(2006)\citenamefont
  {Abdurashitov} \emph {et~al.}}]{Abdurashitov:2005tb}%
  \BibitemOpen
  \bibfield  {author} {\bibinfo {author} {\bibfnamefont {J.~N.}\ \bibnamefont
  {Abdurashitov}} \emph {et~al.} (\bibinfo {collaboration} {SAGE}),\
  }\href@noop {} {\bibfield  {journal} {\bibinfo  {journal} {Phys. Rev.}\
  }\textbf {\bibinfo {volume} {C73}},\ \bibinfo {pages} {045805} (\bibinfo
  {year} {2006})},\ \Eprint {http://arxiv.org/abs/nucl-ex/0512041}
  {nucl-ex/0512041} \BibitemShut {NoStop}%
\bibitem [{\citenamefont {Palazzo}(2011)}]{Palazzo:2011rj}%
  \BibitemOpen
  \bibfield  {author} {\bibinfo {author} {\bibfnamefont {A.}~\bibnamefont
  {Palazzo}},\ }\href@noop {} {\bibfield  {journal} {\bibinfo  {journal} {Phys.
  Rev.}\ }\textbf {\bibinfo {volume} {D83}},\ \bibinfo {pages} {113013}
  (\bibinfo {year} {2011})},\ \Eprint {http://arxiv.org/abs/arXiv:1105.1705}
  {arXiv:1105.1705 [hep-ph]} \BibitemShut {NoStop}%
\bibitem [{\citenamefont {Palazzo}(2012)}]{Palazzo:2012yf}%
  \BibitemOpen
  \bibfield  {author} {\bibinfo {author} {\bibfnamefont {A.}~\bibnamefont
  {Palazzo}},\ }\href@noop {} {\bibfield  {journal} {\bibinfo  {journal} {Phys.
  Rev.}\ }\textbf {\bibinfo {volume} {D85}},\ \bibinfo {pages} {077301}
  (\bibinfo {year} {2012})},\ \Eprint {http://arxiv.org/abs/arXiv:1201.4280}
  {arXiv:1201.4280 [hep-ph]} \BibitemShut {NoStop}%
\bibitem [{\citenamefont {Giunti}\ \emph {et~al.}(2012)\citenamefont {Giunti},
  \citenamefont {Laveder}, \citenamefont {Li}, \citenamefont {Liu},\ and\
  \citenamefont {Long}}]{Giunti:2012tn}%
  \BibitemOpen
  \bibfield  {author} {\bibinfo {author} {\bibfnamefont {C.}~\bibnamefont
  {Giunti}}, \bibinfo {author} {\bibfnamefont {M.}~\bibnamefont {Laveder}},
  \bibinfo {author} {\bibfnamefont {Y.~F.}\ \bibnamefont {Li}}, \bibinfo
  {author} {\bibfnamefont {Q.}~\bibnamefont {Liu}}, \ and\ \bibinfo {author}
  {\bibfnamefont {H.}~\bibnamefont {Long}},\ }\href@noop {} {\bibfield
  {journal} {\bibinfo  {journal} {Phys. Rev.}\ }\textbf {\bibinfo {volume}
  {D86}},\ \bibinfo {pages} {113014} (\bibinfo {year} {2012})},\ \Eprint
  {http://arxiv.org/abs/arXiv:1210.5715} {arXiv:1210.5715 [hep-ph]}
  \BibitemShut {NoStop}%
\bibitem [{\citenamefont {Kopp}\ \emph {et~al.}(2013)\citenamefont {Kopp},
  \citenamefont {Machado}, \citenamefont {Maltoni},\ and\ \citenamefont
  {Schwetz}}]{Kopp:2013vaa}%
  \BibitemOpen
  \bibfield  {author} {\bibinfo {author} {\bibfnamefont {J.}~\bibnamefont
  {Kopp}}, \bibinfo {author} {\bibfnamefont {P.~A.~N.}\ \bibnamefont
  {Machado}}, \bibinfo {author} {\bibfnamefont {M.}~\bibnamefont {Maltoni}}, \
  and\ \bibinfo {author} {\bibfnamefont {T.}~\bibnamefont {Schwetz}},\ }\href
  {\doibase 10.1007/JHEP05(2013)050} {\bibfield  {journal} {\bibinfo  {journal}
  {JHEP}\ }\textbf {\bibinfo {volume} {1305}},\ \bibinfo {pages} {050}
  (\bibinfo {year} {2013})},\ \Eprint {http://arxiv.org/abs/arXiv:1303.3011}
  {arXiv:1303.3011 [hep-ph]} \BibitemShut {NoStop}%
\bibitem [{\citenamefont {Dentler}\ \emph {et~al.}(2018)\citenamefont
  {Dentler}, \citenamefont {Hernandez-Cabezudo}, \citenamefont {Kopp},
  \citenamefont {Machado}, \citenamefont {Maltoni}, \citenamefont
  {Martinez-Soler},\ and\ \citenamefont {Schwetz}}]{Dentler:2018sju}%
  \BibitemOpen
  \bibfield  {author} {\bibinfo {author} {\bibfnamefont {M.}~\bibnamefont
  {Dentler}}, \bibinfo {author} {\bibfnamefont {A.}~\bibnamefont
  {Hernandez-Cabezudo}}, \bibinfo {author} {\bibfnamefont {J.}~\bibnamefont
  {Kopp}}, \bibinfo {author} {\bibfnamefont {P.~A.~N.}\ \bibnamefont
  {Machado}}, \bibinfo {author} {\bibfnamefont {M.}~\bibnamefont {Maltoni}},
  \bibinfo {author} {\bibfnamefont {I.}~\bibnamefont {Martinez-Soler}}, \ and\
  \bibinfo {author} {\bibfnamefont {T.}~\bibnamefont {Schwetz}},\ }\href@noop
  {} {\bibfield  {journal} {\bibinfo  {journal} {JHEP}\ }\textbf {\bibinfo
  {volume} {1808}},\ \bibinfo {pages} {010} (\bibinfo {year} {2018})},\ \Eprint
  {http://arxiv.org/abs/arXiv:1803.10661} {arXiv:1803.10661 [hep-ph]}
  \BibitemShut {NoStop}%
\bibitem [{\citenamefont {Giunti}\ \emph {et~al.}(2021)\citenamefont {Giunti},
  \citenamefont {Li}, \citenamefont {Ternes},\ and\ \citenamefont
  {Zhang}}]{Giunti:2021iti}%
  \BibitemOpen
  \bibfield  {author} {\bibinfo {author} {\bibfnamefont {C.}~\bibnamefont
  {Giunti}}, \bibinfo {author} {\bibfnamefont {Y.~F.}\ \bibnamefont {Li}},
  \bibinfo {author} {\bibfnamefont {C.~A.}\ \bibnamefont {Ternes}}, \ and\
  \bibinfo {author} {\bibfnamefont {Y.~Y.}\ \bibnamefont {Zhang}},\ }\href
  {\doibase 10.1016/j.physletb.2021.136214} {\bibfield  {journal} {\bibinfo
  {journal} {Phys. Lett. B}\ }\textbf {\bibinfo {volume} {816}},\ \bibinfo
  {pages} {136214} (\bibinfo {year} {2021})},\ \Eprint
  {http://arxiv.org/abs/2101.06785} {arXiv:2101.06785 [hep-ph]} \BibitemShut
  {NoStop}%
\bibitem [{\citenamefont {Danilov}(2021)}]{Danilov:2020ucs}%
  \BibitemOpen
  \bibfield  {author} {\bibinfo {author} {\bibfnamefont {M.}~\bibnamefont
  {Danilov}},\ }\href@noop {} {\bibfield  {journal} {\bibinfo  {journal} {PoS}\
  }\textbf {\bibinfo {volume} {ICHEP2020}},\ \bibinfo {pages} {121} (\bibinfo
  {year} {2021})},\ \Eprint {http://arxiv.org/abs/arXiv:2012.10255}
  {arXiv:2012.10255 [hep-ex]} \BibitemShut {NoStop}%
\bibitem [{\citenamefont {Andriamirado}\ \emph {et~al.}(2021)\citenamefont
  {Andriamirado} \emph {et~al.}}]{PROSPECT:2020sxr}%
  \BibitemOpen
  \bibfield  {author} {\bibinfo {author} {\bibfnamefont {M.}~\bibnamefont
  {Andriamirado}} \emph {et~al.} (\bibinfo {collaboration} {PROSPECT}),\
  }\href@noop {} {\bibfield  {journal} {\bibinfo  {journal} {Phys.Rev.}\
  }\textbf {\bibinfo {volume} {D103}},\ \bibinfo {pages} {032001} (\bibinfo
  {year} {2021})},\ \Eprint {http://arxiv.org/abs/arXiv:2006.11210}
  {arXiv:2006.11210 [hep-ex]} \BibitemShut {NoStop}%
\bibitem [{\citenamefont {Almazan~Molina}\ \emph
  {et~al.}(2020{\natexlab{b}})\citenamefont {Almazan~Molina} \emph
  {et~al.}}]{STEREO:2019ztb}%
  \BibitemOpen
  \bibfield  {author} {\bibinfo {author} {\bibfnamefont {H.}~\bibnamefont
  {Almazan~Molina}} \emph {et~al.} (\bibinfo {collaboration} {STEREO}),\
  }\href@noop {} {\bibfield  {journal} {\bibinfo  {journal} {Phys.Rev.}\
  }\textbf {\bibinfo {volume} {D102}},\ \bibinfo {pages} {052002} (\bibinfo
  {year} {2020}{\natexlab{b}})},\ \Eprint
  {http://arxiv.org/abs/arXiv:1912.06582} {arXiv:1912.06582 [hep-ex]}
  \BibitemShut {NoStop}%
\bibitem [{\citenamefont {Barinov}\ and\ \citenamefont
  {Gorbunov}()}]{Barinov:2021mjj}%
  \BibitemOpen
  \bibfield  {author} {\bibinfo {author} {\bibfnamefont {V.}~\bibnamefont
  {Barinov}}\ and\ \bibinfo {author} {\bibfnamefont {D.}~\bibnamefont
  {Gorbunov}},\ }\href@noop {} {\ }\Eprint
  {http://arxiv.org/abs/arXiv:2109.14654} {arXiv:2109.14654 [hep-ph]}
  \BibitemShut {NoStop}%
\end{thebibliography}%

\end{document}